\documentclass{aa}
\usepackage{graphicx}
\usepackage{subfig}
\usepackage{txfonts}
\usepackage{amstext}
\usepackage{color}
\usepackage{longtable}
\usepackage{xcolor}
\usepackage{diagbox}
\usepackage{hyperref}
\hypersetup{
 colorlinks   = true, 
 urlcolor     = blue, 
 linkcolor    = blue,
 citecolor    = blue
 }

\usepackage[labelfont=bf]{caption}

\begin{document}

\newcommand{\bu}{$\bullet$ }
\newcommand{\omicron}{o}
\newcommand{\ksi}{\xi}

\newcommand{\Mnom}{\ifmmode{\rm M}_\odot\else M$_{\odot}$\fi}
\newcommand{\Rnom}{\hbox{$\mathcal{R}^{\mathrm N}_\odot$}}
\newcommand{\Lnom}{\hbox{$\mathcal{L}^{\mathrm N}_\odot$}}
\newcommand{\GMnom}{\hbox{$\mathcal{GM}^{\mathrm N}_\odot$}}

\newcommand{\ubv}{\hbox{$U\kern-0.13em B\kern-0.01em V$}}
\newcommand{\ubvr}{\hbox{$U\kern-0.13em B\kern-0.01em V\kern-0.1em R$}}
\newcommand{\ubvror}{\hbox{$U\kern-0.13em B\kern-0.01em V\kern-0.1em(R)$}}
\newcommand{\vr}{\hbox{$V\!-\!R$}}
\newcommand{\bv}{\hbox{$B\!-\!V$}}
\newcommand{\ub}{\hbox{$U\!-\!B$}}
\newcommand{\hp}{$H_{\rm p}$}

\newcommand{\m}{\ifmmode^{\rm m}\!\!.\else $^{\rm m}\!\!.$\fi}
\newcommand{\D}{\ifmmode^{\rm d}\!\!.\else $^{\rm d}\!\!.$\fi}
\newcommand{\ks}{km~s$^{-1}$}
\newcommand{\ms}{M$_{\odot}$}
\newcommand{\oc}{$O-C$}
\newcommand{\ha}{H$_\alpha$ }
\newcommand{\hb}{H$_\beta$ }
\newcommand{\hg}{H$_\gamma$ }
\newcommand{\tef}{$T_{\rm eff}$ }

\def\jaavso{Journal of the American Association of Variable Star Observers}

\title{Refined reduction and standardisation of 53 years of \ubv\ photometry at Hvar
\thanks{To the memory of the following colleagues
who actively participated in Hvar observations
and are no longer among us:
Josef Havelka ($\dag$2009),
Ji\v{r}\'\i\ Horn ($\dag$1994),
Karel Juza ($\dag$1994),
V\'aclav Kocourek ($\dag$2026),
Svatopluk K\v{r}\'\i\v{z} ($\dag$2018),
Pavel Mayer ($\dag$2018),
Stanislav \v{S}tefl ($\dag$2014),
Miloslav Tlamicha ($\dag$2026).
}
\fnmsep
}

\subtitle{I. Hot emission-line stars and binaries}

\author{
H.~Bo\v{z}i\'c\inst{1}\and
P. Harmanec\inst{2}\and
M.~Bro\v{z}\inst{2}\and
A.~Opli\v{s}tilov\'a\inst{2,3}\and
P.~Koubsk\'y\inst{4}\and
P.~Hadrava\inst{5}\and
D.~Ru\v{z}djak\inst{1}\and
D.~Sudar\inst{1}\and
M.~Wolf\inst{2}\and
P.~Zasche\inst{2}\and
J.~Honsa\inst{4}\and
F.~\v{Z}\v{d}\!\'arsk\'y\inst{4}\and
A.~Harmanec\inst{2}\and
J.~Jon\'ak\inst{2,6}\and
I.~Piantschitsch\inst{7,8,9}\and
I.~Skoki\'c\inst{1}\and
J.~\v{S}vr\v{c}kov\'a\inst{2}\and
K.~Vitovsk\'y\inst{2,10}\and
D.~Vr\v{s}nak\inst{11}\and
M.~Zummer\inst{2}
}
\institute{
University of Zagreb, Faculty of Geodesy, Hvar Observatory, Ka\v ci\'ceva~26, HR-10000 Zagreb, Croatia
  \and
Charles University, Faculty of Mathematics and Physics, Astronomical Institute, V~Hole\v{s}ovi\v{c}k\'ach~2, CZ-180~00~Praha~8,\\ Czech Republic
  \and
Université de Liège, GAPHE, STAR, B5c, Allée du 6 Août 19c, B-4000 Sart Tilman, Liège, Belgium
  \and        
Astronomical Institute of the Czech Academy of Sciences, Fri\v{c}ova~298, CZ-251~65~Ond\v{r}ejov, Czech Republic
  \and
Astronomical Institute of the Czech Academy of Sciences, Bo\v{c}n\'\i~II~1401/1, CZ-141~00 Praha~4, Czech Republic
 \and
Université Côte d’Azur, Observatoire de la Côte d’Azur, CNRS, Laboratoire Lagrange, Bd de l’Observatoire, 06304 Nice, France 
 \and
International Centre for Neuroscience and Ethics (CINET),
Fundación Tatiana Pérez de Guzmán el Bueno, P.º del General Martínez Campos, 25,
28010 Madrid, Spain
 \and 
Universitat Autònoma de Barcelona (UAB), Department of Philosophy, Carrer de la
Fortuna, 08193 Bellaterra, Barcelona, Spain
 \and 
Institute of Applied Computing \& Community Code (IAC3),
Universitat de les Illes Balears (UIB), Carretera de Valldemossa, km 7.4,
07122 Palma de Mallorca, Spain
 \and
Heidelberger Institut für Theoretische Studien, Schloss-Wolfsbrunnenweg 35, 69118 Heidelberg, Germany
 \and
University of Zagreb, Faculty of Electrical Engineering and Computing, Unska~3, HR-10000 Zagreb, Croatia
}
\date{Received \today}

\abstract{
{
Emission-line stars classified as Be exhibit light and colour variability
on various timescales, ranging from days to decades.
Their evolution must be constrained by long-term observations
that are accurately calibrated and stable.
}
}
{
{
Here, we provide a new reduction of photoelectric \ubv\ observations
obtained at the Hvar observatory,
spanning more than 50 years (1972--2025).
This unique dataset is highly complementary to the
Transiting Exoplanet Survey Satellite,
which has been conducting observations since 2018,
not only in terms of the time baseline,
but also in providing fundamental constraints in the U and B bands.
}
}
{
{
We used new, non-linear reduction equations,
with temporally variable extinction over the course of the night,
which allowed us to achieve long-term accuracy of
0.008 to 0.016\,mag (1-$\sigma$ uncertainty),
as verified by the Johnson standards.
We then classified 59 Be stars into five classes,
based on their variability patterns; namely,
long-term envelope (LTE),
long-term cyclic (LTC),
binarity (BIN),
rapid low-amplitude (RLA),
and long-term quiescence (LTQ).
}
}
{
{
According to our observations,
the percentages of stars in the individual classes are
44\%,
24\%,
25\%,
66\%, and
19\%,
respectively.
We note that stars in the sample often exhibited more than one pattern.
At certain times, changes in the U and B bands were markedly different from those in~V
(e.g. for BU Tau, V744 Her, V923 Aql, and V1294 Aql).
We confirm that the LTE-positive variability is more common than the inverse
(20~vs~6);
in addition, two stars exhibited both types
($\zeta$~Tau and V1294~Aql).
According to our observations,
the LTC variability and the LTE-positive variability
are almost mutually exclusive.
Among 26 binary systems
with previously known orbital solutions,
circular orbits are more common than eccentric ones
(18~vs~8).
As for the brightness variations between different 
quiescent phases,
an increasing trend is less common than a decreasing one
(4~vs~7);
spanning from $-6.5$ to $+6.0\,{\rm mmag}\,{\rm yr}^{-1}$.
}
}
{
{
Our observations provide well-calibrated \ubv\ light curves spanning several decades, 
offering a valuable dataset for investigations of Be-star variability and
tests of various models, including the viscous decretion disc model.
Continuous monitoring is important for the most interesting objects, namely,
$\beta$~Lyr,
EW~Lac,
$\delta$~Sco,
$\gamma$~Cas, and
V1294~Aql.
}
}


\maketitle
\nolinenumbers

\section{Introduction}

The astronomical observatory at Hvar, Croatia
($43^\circ10'43''\,{\rm N}$,
$16^\circ 26'54''\,{\rm E}$,
${\sim}200\,{\rm m}$ above sea level)
is a site of historic importance due to its long-term observing programme,
focussed on hot emission-line stars
\citep{hb2013,mayer2013}.
These objects have exhibited a temporary presence of hydrogen lines in emission.
When this programme started, little was known about their light variability.
Over time, numerous papers based on Hvar \ubv\ photometry
have been published
\citep[e.g.][]{zarf6,kxand01,hvar4,zarf10,zarf11,zarf7b,zarf7a,zarf14,cxdrahorn,zarf17,hvar5d,hvar5,cxdrahvar,zarf24,zarf30,mourard2018,zarf31,broz2021,wolf2021,hec2022},
shedding light on the possible causes of Be-star variability. 

A classical result based on Hvar photometry was reported by \citet{hec83},
who explained the two basic kinds of photometric variability 
observed among Be stars:
positive and inverse correlations
between \bv\ and \ub\ colour indices.
He suggested that the behaviour is governed primarily 
by the inclination angle, that is, whether the envelope is viewed edge-on or pole-on. 
Specifically, pole-on Be stars are due to their discs being 'brighter-when-redder'
(and vice versa).
This distinguishes them from other intrinsically variable stars, 
which generally become 'brighter-when-bluer',
as recently confirmed by Gaia
\citep{Eyer_2019A&A...623A.110G}.

We monitored more than one hundred bright stars at Hvar,
including already known binaries,
which have exhibited Balmer emission lines in their recorded history. We made sure not
to restrict our study to only the sort of objects that are commonly referred to as `classical Be stars'
\citep{struve31,rivi2013}, defined as stars belonging to apparent luminosity classes III to V.
We remain of the opinion that hot emission-line objects
should be viewed in terms of their complexity.
This is supported, for instance, by the recent finding by \citet{alzbeta2025}
that even the B0Ia supergiant $\varepsilon$~Ori is not likely to be a genuine supergiant,
but a rapidly rotating and flattened object,
most likely surrounded with circumstellar matter
\citep{Krticka_2018A&A...617A.121K}.

\vskip\baselineskip

Ad Be phenomenon.
The Be phenomenon refers to the episodic ejection of matter 
from rapidly rotating B-type stars, leading to the formation of a viscous, 
Keplerian circumstellar decretion disc responsible for emission lines and 
characteristic variability.
Recent research on the Be phenomenon has been focussed on constraining
the standard viscous decretion disc model
\citep{rivi98,carciofi2011,baade2017,Rimulo_2018MNRAS.476.3555R},
describing the temporal evolution of gaseous, rotating discs
around classical Be stars,
but not the decretion or mass loss mechanism itself.

In this context,
the star $\omega$~CMa was studied in great detail
in a series of papers by
\citet{Ghoreyshi_2018MNRAS.479.2214G,ghore2021,ghore2023}.
Its long-term light curve exhibits four cycles of a decreasing cycle length (10.5 to 7\,yr),
with correlated brightenings, which are also referred to as 'outbursts' in the literature,
and quiescence durations.
According to the model,
the viscosity parameter,~$\alpha$,
\citep{Shakura_1973A&A....24..337S}
is very large, ranging from 0.1 to 1.0;
moreover, it evolves non-monotonically.
The minimum light in quiescence decreases,
whereas the overall minimum is interpreted as stellar flux.
However, Ghoreyshi et al. admit that their model exhibits tension between other observables:
the synthetic flux in J, H, K, and L is too high,
the equivalent width of H$_\alpha$ decreases too slowly, and
the degree of polarisation decreases too slowly.

\citet{Neiner_2020A&A...644A...9N} addressed a possible
internal mechanism launching the Be phenomenon ---
an excitation of g-modes by vigorous convection in the core, followed by
their propagation and damping by thermal diffusion, thereby
transporting the angular momentum outwards.
This subsequently results in unstable g-modes at the surface
and outflow of gas.
Since there is no cavity when the outer layers have been launched,
the p-modes disappear,
but after a relaxation of the surface,
they are expected to reappear.

\vskip\baselineskip

Ad binaries among the Be stars.
There are undoubtedly some Be stars
for which the Balmer emission arises from the accretion disc
formed during large-scale mass exchange
\citep[as suggested by][]{kh75, hk76}.
One prominent example is the $\beta$~Lyr system
\citep{mourard2018,broz2021,Vitovsky2025A&A...704A.189V}.
However, most of the known binaries among classical Be stars
\citep{struve31}
are observed in later evolutionary stages,
following the end of the mass transfer phase
\citep{pols91,Bodensteiner_2020A&A...641A..42B,rivi2025}.

Of the 57 Be stars observed by \citet{Klement_2019ApJ...885..147K},
at least 46\% have companions according to the SED turndown,
corresponding to truncated outer discs.
In at least 30 cases, they were confirmed by
direct spectroscopic or interferometric detections
\citep{wang2021,Wang_2023AJ....165..203W,Gies2023ApJ...942L...6G,klement2024}.
The companions are often hot, compact, low-mass, and
slowly rotating stars, in the He core-burning phase (sdO stars) or even
white dwarfs. Their Be primaries, on the contrary, are high-mass objects,
rotating near the critical limit, indicative of previous mass transfer, which
had led to substantial spin-up. 

The combined solutions available for spectroscopy and interferometry led to total masses
from 7.0 to 18.6\,\Mnom\ \citep{klement2024}.
This is highly comparable to some emission-line binaries
undergoing mass exchange, such as $\beta$~Lyr,
with a total mass of 13.0\,\Mnom\
\citep{mourard2018}.
This has been corroborated by finding binaries with stripped components,
such as
LB~1,
HR~6819,
HD~15124,
and V742~Cas;
\dots\
\citep{Chochol_2002ASPC..279..143C,Elbadry_2022MNRAS.516.3602E,rivi2025},
specifically, in the stage prior to a contraction to the helium main sequence.

In this context, \citet{Dodd_2024MNRAS.527.3076D}
analysed Gaia astrometry of B versus Be stars,
with a sensitivity down to 20\,mas. They found that the binary fraction is (counterintuitively)
lower for Be stars.
This is in agreement with one of the components being stripped,
and, hence, undetectable. Alternatively, in triple systems, the components might interact,
migrate,
resulting in a `hardening' and leading to the highly eccentric orbits of secondaries
\citep{Kervella_2022A&A...667A.111K,davidge2023,rast2024}.

\vskip\baselineskip

Ad Be stars photometric surveys.
To obtain properties characterising the population,
\citet{Labadie_2017AJ....153..252L}
analysed photometry from the small-aperture KELT survey
for 610 Be stars.
Their classification according to light curves showed that
37\% of objects had long-term variability,
36\% light brightenings,
24\% non-radial pulsations, and
only 2\% were eclipsing binaries
(similarly as \citet{Bernhard2018MNRAS.479.2909B}).
Some stars exhibited ${\sim}10\,{\rm d}$-long light brightenings,
with amplitudes up to $0.5$\,mag.

Subsequently,
\citet{bartz2022}
analysed photometry from the Transiting Exoplanet Survey Satellite (TESS)
for 420 Be stars.
Their classification according to periodograms showed that
87\% of objects had frequency groups,
18\% flickers (${\sim}10\,{\rm d}$ long),
25\% low-frequency stochasticity,
and 29\% isolated frequencies.
Some amplitudes were as low as $0.0001$\,mag.
Therefore, even small mass-loss events were detected.

From this perspective,
\citet{bartz2025} organised a~coordinated H$_\alpha$ spectroscopic monitoring for 13~stars,
measured the equivalent widths of the blue and red wings of the H$_\alpha$ profiles separately, and analysed 
their ratio for periodicity. They recovered cyclic variations with gradually decreasing
amplitudes on typical timescales from 0.5 to $2\,{\rm d}^{-1}$,
corresponding to a~near-surface orbital motion.
The observed profiles correspond to localised, asymmetric ejection and later (after five to ten cycles) 
became symmetric again due to the gradual dispersion of the matter.

\vskip\baselineskip

Hereinafter, we provide long-term context and constraints for such Be star models,
by extending the baseline over 53 years.
The Hvar \ubv\ photometry represents probably the longest series
of homogeneous and well-calibrated observations.
Especially for hot stars, they also provide the monitoring
of the ultraviolet part of the stellar flux,
where the radiation of hot stars dominates.
Our measurements can be conveniently combined with
the spectroscopic monitoring of emission-line stars,
provided in the Be Star spectra (BeSS) database \citep{neiner2011}.

Our paper is organised as follows. In Sect.~\ref{sec:reduction},
we introduce a new reduction of the Hvar \ubv\ photometry, which leads to a significant improvement.
In Sect.~\ref{sec:types},
we present a classification of our sample of Be stars according to their variability patterns.
Then, in Sect.~\ref{sec:results}, we discuss the individual classes
of Be stars;
comments on selected individual objects are provided in
Appendix~\ref{sec:comments}.
In Sect.~\ref{sec:conclusions},
we summarise the main results from the Hvar programme,
emphasising the importance of the 53-year Hvar UBV archive.

\section{New reduction of Hvar \ubv\ photometry with variable extinction}\label{sec:reduction}

\subsection{Standard stars}

Since our motivation has been to ensure an accurate and stable transformation
to the standard Johnson system \citep{john53,john54}, especially for the dominant Be-star programme,
we soon realised that a set of non-linear transformation formulae
is required for the conversion of extinction-free instrumental magnitudes
to the standard Johnson magnitudes.
It is also necessary to derive the coefficients
of these transformations from all observations of standard stars
in each observing season \citep{hhj94}.

The strategy behind Hvar observations was first developed for the Be star programme 
and later used for all observing programmes. First, we compiled a list of all known 
bright Be stars north of the declination
of $-20^\circ$, included in the Bright Star Catalogue
\citep{hoffleit82}, along with a few fainter ones.
The total number of objects obtained in this way is 105 (Table~\ref{tab03}).
For all of them, we selected suitable comparison and check stars.
If more Be stars were found close
to each other in the sky, we observed
them all together, using the same comparison and check stars.
Since most of our observations dealt with early-type blue
stars, we also selected the so-called red standards for each group to
facilitate a good seasonal transformation.
Therefore, a typical observing sequence is
comparison–check–red-variable, repeated three or five times,
and ending with comparison–red-variable–check–comparison. Each star was observed 
with integration times of a few tens of seconds in each \ubvror\ filter.

\subsection{Comparison stars}

Whenever possible, we chose the comparison stars
with available \ubv\ photometry from Johnson and
his collaborators. Our primary source was the summary
publication by \citet{john66}. It was, of course, a `must' for
the red standards and it also allowed us to save observing time. Comparison stars,
check stars, and red standard stars with known \ubv\ magnitudes could be used
for both the monitoring of atmospheric extinction over the observing nights and for seasonal
transformation to the standard \ubv\ system.
However, it turned out that the stars with only a few original
Johnson's observations suffered from rather large accidental errors.
Our task was therefore to improve the original Johnson values, but to preserve
the original Johnson system over the whole range of the Hertzsprung--Russell
(HR) diagram.

\subsection{New reduction equations}

We used the following equations to reduce data
for all observing seasons since 1972.%
\footnote{The program suite for data
reduction, sorting, and archiving, several auxiliary programs, manual,
test examples, and data files are available as one compressed file
phot.tar.gz at the CDS via
\url{https://cdsarc.cds.unistra.fr/viz-bin/cat/J/A+A/712/A226}.}
For each star and the time of observation, the zenith distance, $z$, is
calculated from its equatorial coordinates recalculated for the time of
observation and the accurate air mass,~$X$, is estimated following
\citet{bemporad}
\begin{eqnarray}
X = \sec z-0.0018167Q-0.02875Q^2-0.0008083Q^3\,,\label{airok}
\end{eqnarray}
where
$Q \equiv \sec z-1$.

\begin{figure}
\centering
\includegraphics[width=7cm]{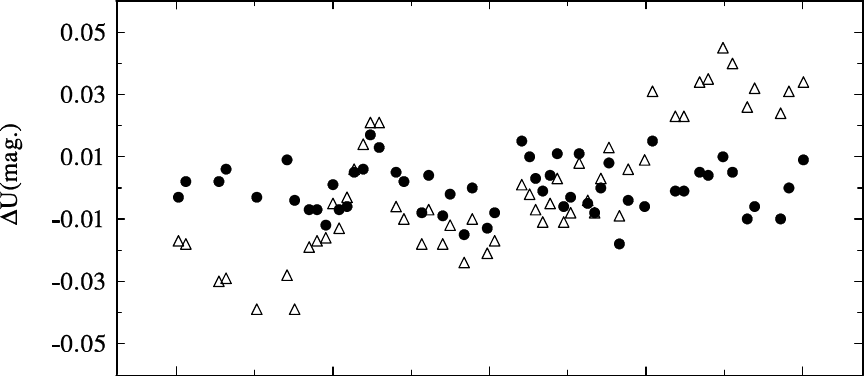}
\includegraphics[width=7cm]{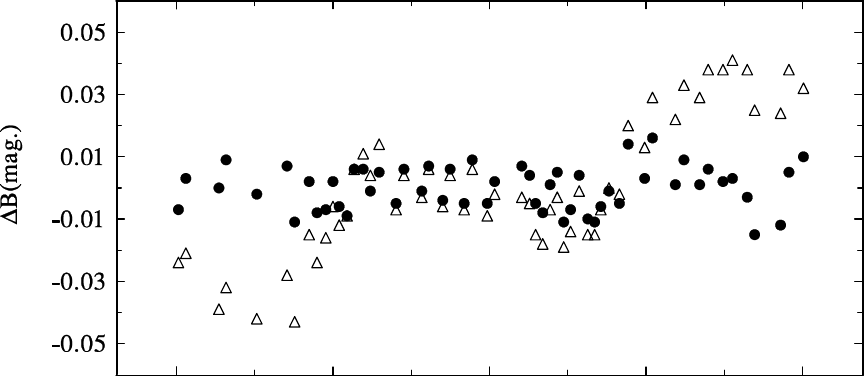}
\includegraphics[width=7cm]{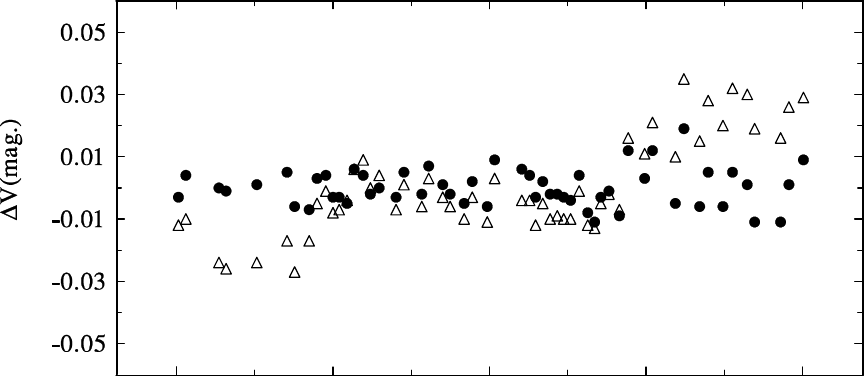}
\includegraphics[width=7cm]{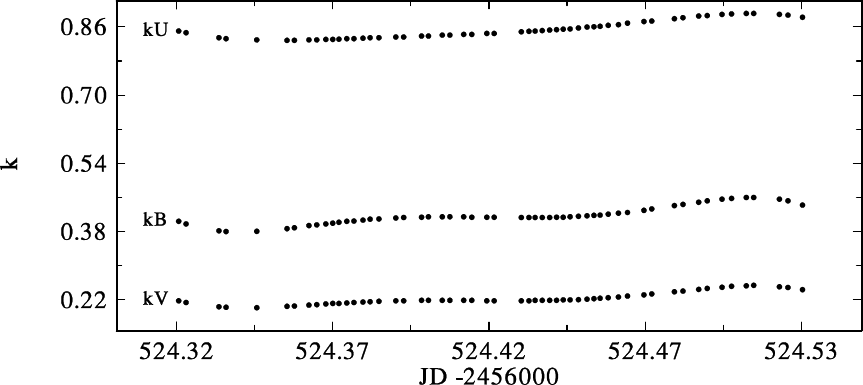}
\caption{Example of temporally variable extinction.
The fit of the residuals of the all-sky magnitudes is shown for
one long summer night
(19 Aug 2013),
using a fifth-degree polynomial in all three passbands.
The original residuals are shown by triangles;
their overall scatter is ${\pm}0.050$\,mag
The residuals after the polynomial fit are shown by circles;
their scatter is much lower, ${\pm}0.015$\,mag.
The bottom panel shows the instantaneous values of
the linear extinction coefficients
over the course of one night.
}
\label{extpol}
\end{figure}

\paragraph{Temporally variable extinction.}
Transformation coefficients $G_i$ ($i = 1,2, \dots, 28$) between observed
instrumental magnitudes, denoted by $v$, $b$, $u$, and $r$, and extinction-free
magnitudes $v_0, b_0, u_0$, and $r_0$, determined (or fixed) separately
for each night, including the first-order extinction coefficients and
a~possibility to model time variations of the linear extinction
coefficients on each night through polynomials up to the fifth degree as a function of time, $t$.
The corresponding equations are expressed as
\begin{eqnarray}
v&=&v_0+G_1\nonumber\\
 &+&(G_5+G_9\,t+G_{13}\,t^2+G_{17}\,t^3+G_{21}\,t^4+G_{25}\,t^5)\,X,\\
b&=&b_0+G_2\nonumber\\
 &+&(G_6+G_{10}\,t+G_{14}\,t^2+G_{18}\,t^3+G_{22}\,t^4+G_{26}\,t^5)\,X,\\
u&=&u_0+G_3\nonumber\\
 &+&(G_7+G_{11}\,t+G_{15}\,t^2+G_{19}\,t^3+G_{23}\,t^4+G_{27}\,t^5)\,X,\\
r&=&r_0+G_4\nonumber\\
 &+&(G_8+G_{12}\,t+G_{16}\,t^2+G_{20}\,t^3+G_{24}\,t^4+G_{28}\,t^5)X.
\end{eqnarray}
An example of how such a fit improves all-sky magnitudes over long nights is shown in Fig.~\ref{extpol}.

\paragraph{Non-linear colour transformation.}
The colour transformation equations between the Johnson standard values
for each magnitude ($V,B,U,R$) and the extinction-free instrumental
magnitudes are linear in the \ub\ index; however, they take the form of a third-degree
polynomial in \bv. This form is necessary to compensate for
the unavoidably non-linear effect of the Balmer jump on the magnitudes of
stars from mid-B to F spectral types \citep[see][]{cousins76}.
The colour extinction coefficients in the form recommended
by \citet{young91} are also included among the seasonal transformation
coefficients. It is our experience that this form of transformation
equations ensures the reproduction of the standard Johnson
system within 0.01\,mag, even in~$U$, for any standard star.
The respective equations are expressed as
\begin{eqnarray}
v_0-V&=&H_1(B-V)+H_2(U-B)\nonumber\\
     &+&H_3(B-V)^2+H_4(B-V)^3\nonumber\\
     &+&H_5\,X\,B_4(B-V+0.5\,X\,B_4)+H_6,\\
b_0-B&=&H_7(B-V)+H_8(U-B)\nonumber\\
     &+&H_9(B-V)^2+H_{10}(B-V)^3\nonumber\\
     &+&H_{11}\,X\,B_4(B-V+0.5\,X\,B_4)+H_{12},\\
u_0-U&=&H_{13}(B-V)+H_{14}(U-B)\nonumber\\
     &+&H_{15}(B-V)^2+H_{16}(B-V)^3\nonumber\\
     &+&H_{17}\,X\,B_5(U-B+0.5\,X\,B_5)+H_{18},\\
r_0-R&=&H_{19}(B-V)+H_{20}(U-B)\nonumber\\
     &+&H_{21}(B-V)^2+H_{22}(B-V)^3\nonumber\\
     &+&H_{23}\,X\,B_5(U-B+0.5\,X\,B_5)+H_{24},
\end{eqnarray}
where $B_4 \equiv G_6-G_5$ and $B_5 \equiv G_7-G_6$ are
the linear extinction coefficients in the \bv\ and \ub\ colours.

We note that in our formalism, quantities with the suffix 0 are not,
in fact, really extinction-free instrumental magnitudes; rather, they are
extinction-free magnitudes uncorrected for the colour
extinction. This is only taken into account in the seasonal transformation
equations, where they actually belong, since they depend on the colour
properties of the photometer, not on the properties of the Earth's
atmosphere. The program derives the transformation coefficients iteratively,
using all nights of a given season denoted as suitable for that purpose
and starting either with the values derived during the previous seasons
or with an instrumental system; namely, choosing all $H$ coefficients
equal to zero.

Since the non-linear transformations imply that all three passbands affect
the accuracy of the determination of the standard values for each
passband, we took care to use different integration times for each star in
each passband (usually the longest for the $U$ passband) to obtain comparable
signal-to-noise ($S/N$) ratios in all of them. Here, the usual integration times are
between 10 and 25 seconds.

\paragraph{Nightly grades.}
We inspected all the individual nights of observations for a~given observing
season and assigned them one of three possible grades: excellent if the rms
of the fit to the all-sky magnitudes for the standard stars observed was 0.010\,mag
or lower; good for the rms between 0.011 and 0.015\,mag; and poor for the
rms exceeding 0.015\,mag in the \ubv\ passbands. The respective individual observations
were then treated with the weights 1.5, 1.0, and 0.5. Besides, we selected
nights with enough different standards observed and a large range of air masses,
and only for these nights, the all-sky data were archived in the all-sky
archive. The remaining nights were only used for the differential photometry
and are stored in the archive of differential observations, together with
information about the comparison star used.

\begin{figure}
\centering
\includegraphics[width=7cm]{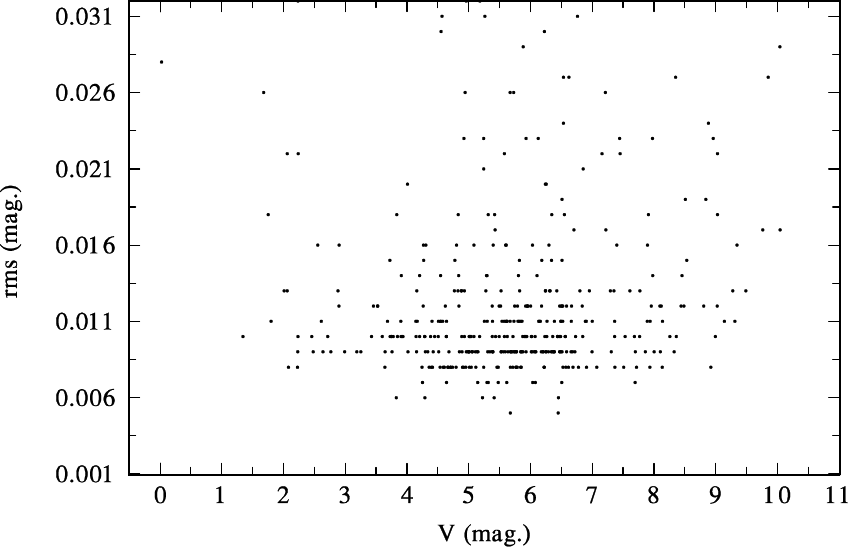}
\includegraphics[width=7cm]{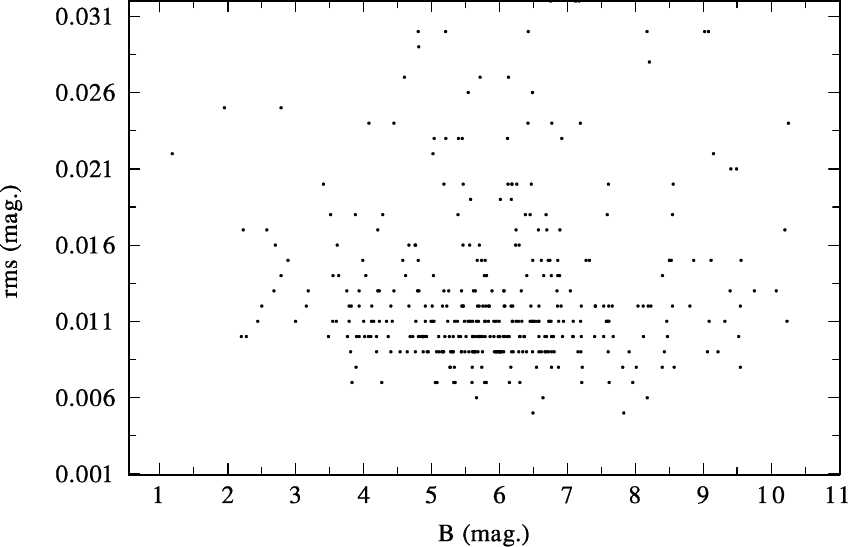}
\includegraphics[width=7cm]{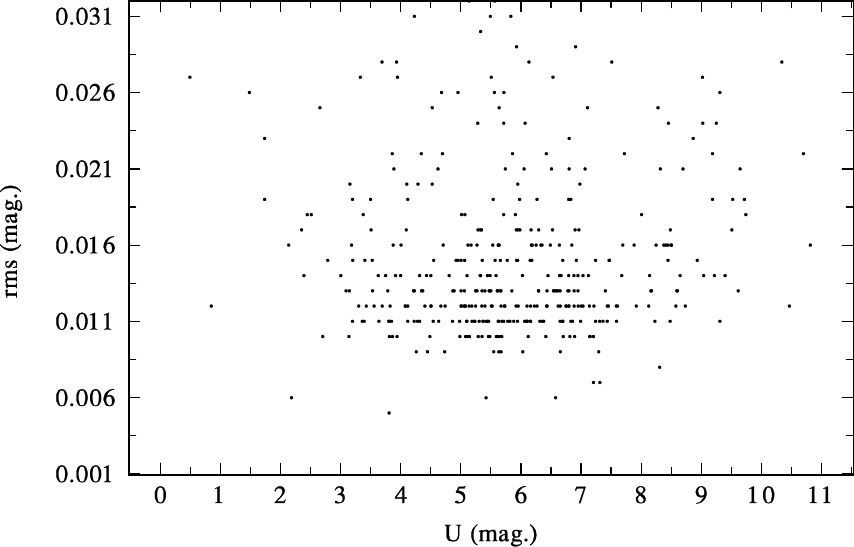}
\caption{
RMS errors of one observation
for constant, comparison stars versus their magnitude
in the U, B, V passbands.
The data were taken from the all-sky archive.
The data from the differential archive are very similar.
The precision for most stars is from 0.008 to 0.016\,mag.
}
\label{rmsmin}
\end{figure}

\begin{figure}
\centering
\includegraphics[width=8cm]{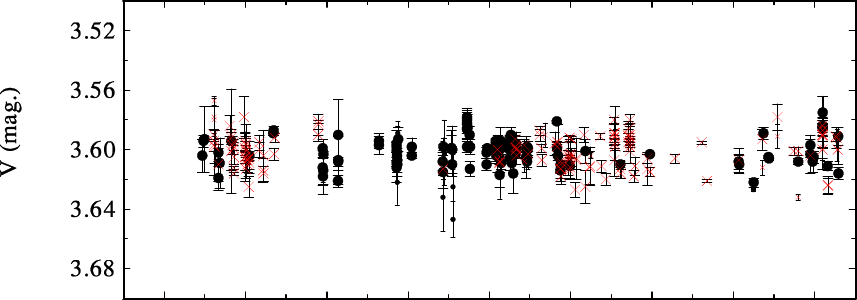}
\includegraphics[width=8cm]{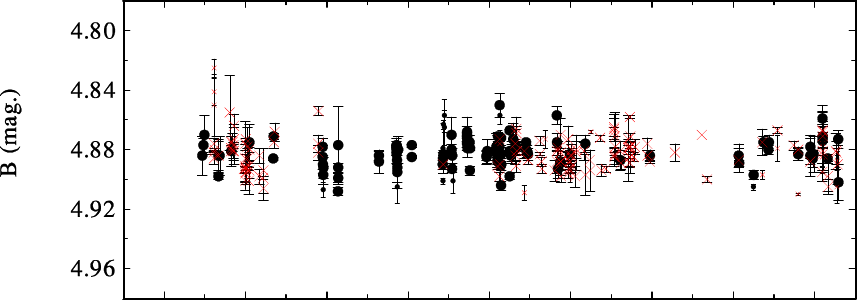}
\includegraphics[width=8cm]{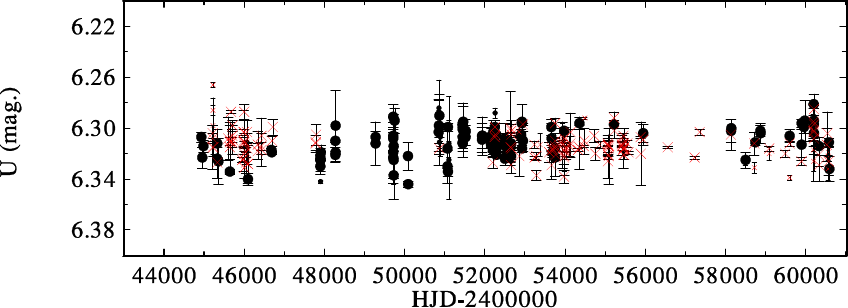}
\caption{Individual differential \ubv\ observations of
the red standard 51~And,
obtained relative to two different blue comparison stars.
Observations relative to 4~Per
($V=5.009$\,mag, $\bv =-0.073$\,mag, $\ub =-0.305$\,mag)
are shown as black circles.
Observations relative to HR~189
($V=5.674$\,mag, $\bv =-0.125$\,mag, $\ub =-0.571$\,mag)
are shown by red crosses.
Observations from nights of poor quality are shown by small symbols.
Our standard all-sky values for 51~And are
$V=3.600(10)$\,mag,
$B=4.881(10)$\,mag, and
$U=6.312(11)$\,mag.
The error bars are 1-$\sigma$.
}
\label{51and}
\end{figure}

\begin{figure}
\centering
\includegraphics[width=8cm]{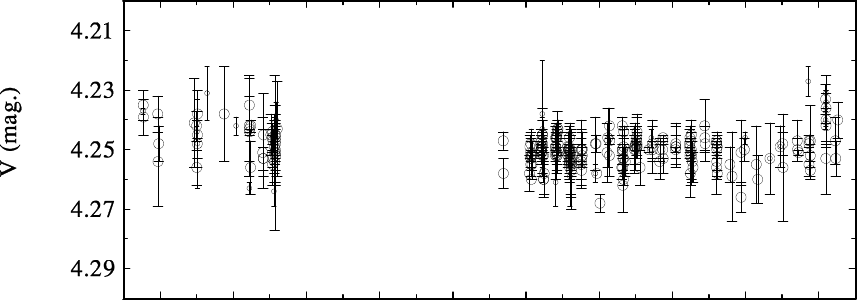}
\includegraphics[width=8cm]{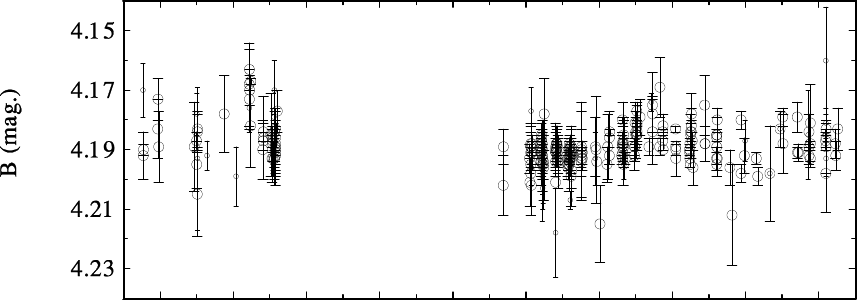}
\includegraphics[width=8cm]{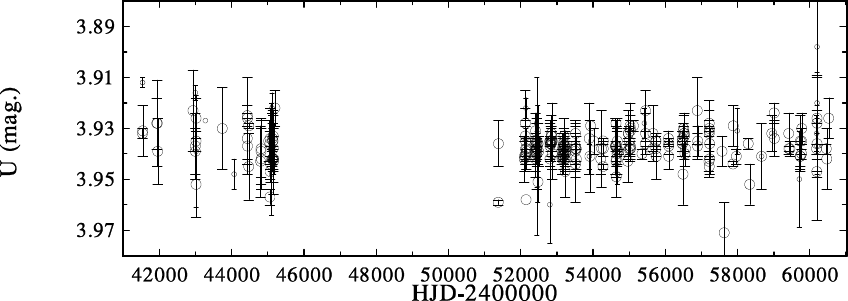}
\caption{Daily means of \ubv\ differential observations
of the check star $\varphi$~Her = HD~145389,
relative to $\upsilon$~Her,
with their rms errors per observation (1-$\sigma$).
Data from nights of poor quality are shown by small symbols.
}
\label{phiher}
\end{figure}

\begin{figure}
\centering
\includegraphics[width=8cm]{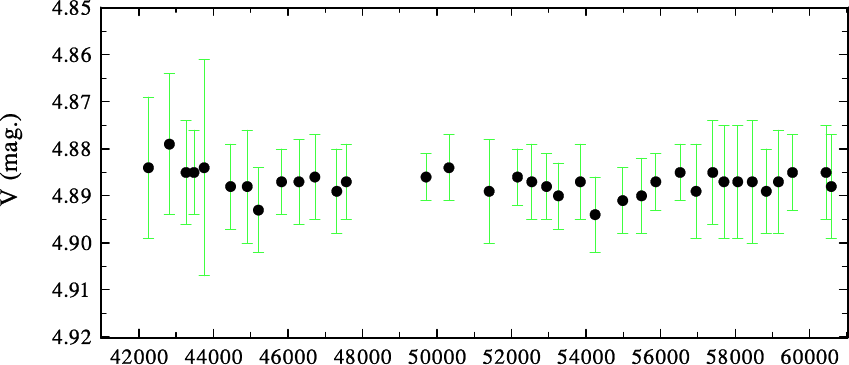}
\includegraphics[width=8cm]{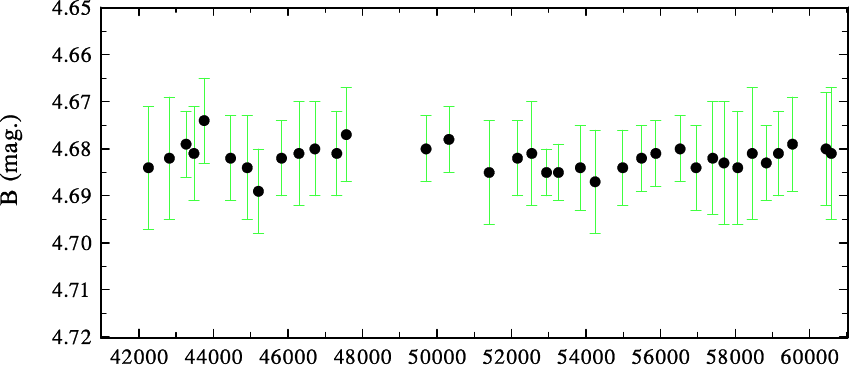}
\includegraphics[width=8cm]{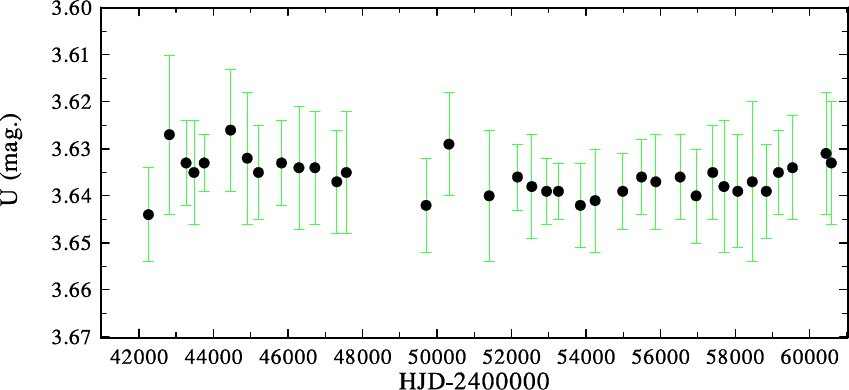}
\caption{Seasonal means of \ubv\ all-sky normal points
of the Johnson primary standard 10 Lac,
with their rms errors per observation (1-$\sigma$).
The overall mean values and the uncertainties of the means are
$V = 4.8845(14)$\,mag,
$\bv = -0.2045(12)$\,mag, and
$\ub = -1.0464(14)$\,mag.
}
\label{10lac}
\end{figure}

\subsection{Improvements with respect to previous reductions}

At the Hvar observatory, located on an island at an altitude
of only ${\sim}200\,{\rm m}$ above sea level, the variations of linear extinction
over the course of the night are usually quite large, especially during the summer seasons.
Previously, we had to split many observing nights into shorter
time segments and derive the extinction for each segment separately
\citep{hec94}.
This was not satisfactory because standard stars from a number of otherwise
good and long observing nights could not be used for seasonal
transformations. This was the primary reason why we decided to undertake
the tedious task to re-reduce all observing seasons.

The secondary reason was that we should base our standard system
on frequent observations of standard stars, recommended by Johnson and
his collaborators. Several years ago, we derived robust mean
\ubvr\ values for all individual observations of the recommended Johnson
primary standards, for which the extinction was measured, from \citet{john66}
study and used them in all consecutive observations at Hvar on good nights,
suitable for all-sky photometry.
Additionally, we did the same for several well-observed stars
from the list of 108 recommended Johnson secondary standards \citep{john54}.
For convenience, these robust \ubvr\ values for Johnson standards
are provided in Tabs.~A.3 and~A.4.

All the selected stars were carefully checked for variability with the help
of the Hipparcos satellite \hp\ photometry \citep{esa97} and
the SIMBAD bibliography. The Johnson standards were then systematically observed
several times on each good observing night. Care was taken
to observe them in high and low air masses to `bracket' the observations
of all other stars on each observing night. We also carried out dedicated 
observations of all our comparison, check, and red-standard stars, for which good Johnson values were missing.
This effort has continued till today and allowed us to get better standard
Johnson values for comparison, check, and red stars used in the early observing
seasons.

Subsequently, we repeated the reductions of all seasons since 1972. In the
vast majority of cases, this resulted in much better seasonal transformations
and a significant decrease in mean standard errors compared to previous reductions.
For most constant, comparison stars, the precision ranges from 0.008 to 0.016\,mag
(Fig.~\ref{rmsmin}).
To our satisfaction, the final mean values of the standard all-sky \ubv\ magnitudes
of standard stars observed frequently enough throughout multiple seasons
differ only by a few thousands of a magnitude
from the values derived during the first homogenisation
(for comparison, see Table~4 in \citealt{hhj94}).

Our improved standard Johnson \ubv\ magnitudes for all the
observed comparison, check, and red standard stars
are provided in Tables~A.5 and A.6
for both frequently and less frequently observed stars.
We emphasise that the less frequently observed stars
were not used as our transformation standards,
since their standard \ubv\ values are naturally less accurate.

\subsection{Accuracy and stability of our standard system}

An overview of the accuracy of Hvar photometry is shown in
Fig.~\ref{rmsmin}, where we plot
the dependence of the rms errors per observation of unit weight
versus the magnitude of the objects.
This characterises the measurement accuracy for non-variable objects.
It is encouraging to see that it is quite comparable for all three passbands,
thanks to the somewhat longer integration times used for the $U$ filter.

The accuracy of Hvar's observations of individual stars can also be estimated
with the help of the check stars used for each observed program star.
We underline that the check stars were always observed as frequently as
the corresponding programme stars.
The HD numbers of the check stars can be found in Table~\ref{tab03}
and their rms errors per one observation can be found
in Table~A.5.
For instance, the check star HR~289 = HD~6114 has the rms errors
0.009, 0.009, and 0.010\,mag
in $V$, $B$, and $U$, respectively.

In some published studies, we have already shown that our homogenised
magnitudes lead to a very good reproduction of observed magnitudes
obtained at different observing sites, even from mountain observatories
such as San Pedro M\'artir. This is well illustrated, for instance,
in Table~2 of \citet{bozic2007}, where the differential
\ubv\ magnitudes of the check star $\varphi$~Her derived for
different seasons at the Hvar, San Pedro M\'artir, and Tubitak stations
are compared. The mutual agreement of the values is excellent in all cases.

A very convincing test of the quality of our transformations into the
standard Johnson system is the case of the red standard 51~And.
This star was alternatively observed as the red standard in
two different groups, in both cases relative to blue and much fainter
comparison stars, 4~Per and HR~189. In Fig.~\ref{51and}, we show that the individual
values relative to both comparison stars are stable in time
and close to our standard all-sky value for 51~And.

We also plot in Fig.~\ref{phiher} the nightly mean values for
$\varphi$~Her = 11~Her = HD~145389. This star was used as the check star
in observations of the Be star V839~Her and has been observed since 1972. Therefore,
it provides a good idea about the accuracy of Hvar observations over time.
It is now known to be a~565-d spectroscopic binary seen nearly pole-on,
a~rotating star with a 3.708-d period and also a low-amplitude pulsating star
\citep[see][and references therein]{kochuk2021}. As shown in Fig.~4 in their paper,
the light variability has a~full amplitude lower than 0.01\,mag.

We also investigated the secular stability of the transformation to the standard system
on the example of the Johnson primary standard 10~Lac, which has been regularly observed in Hvar
since 1974. In Fig.~\ref{10lac}, we show seasonal mean values with their rms errors.
Once again, the secular stability of the values is very satisfactory.
Thus, we believe that the standard Hvar \ubv\ values for all our standards can be trusted
and used in other observing programmes elsewhere, including the CCD observations with
Johnson filters.

\begin{figure}
\centering
\includegraphics[width=9cm]{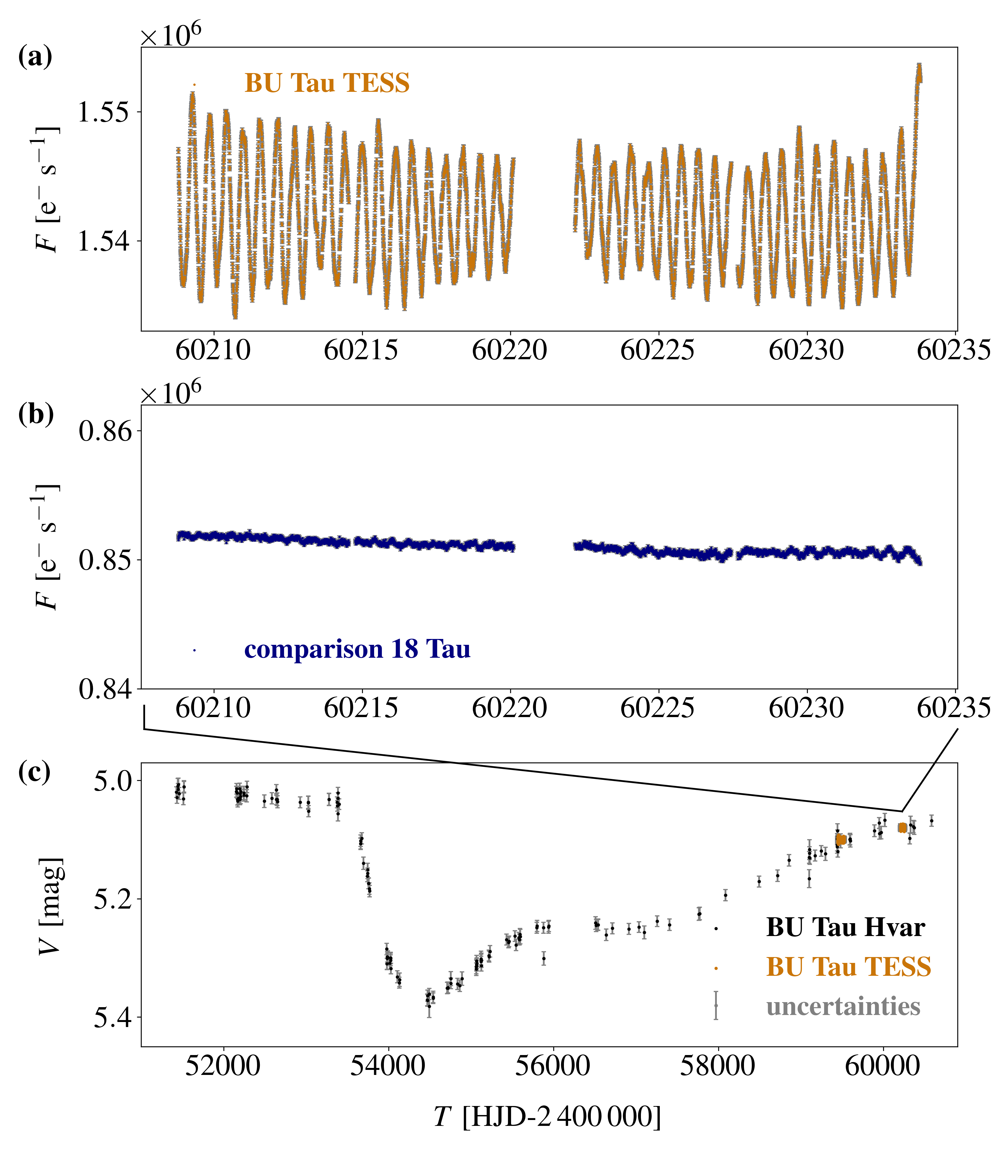}
\vskip-.2cm
\leavevmode\kern.1cm
\includegraphics[width=8.9cm]{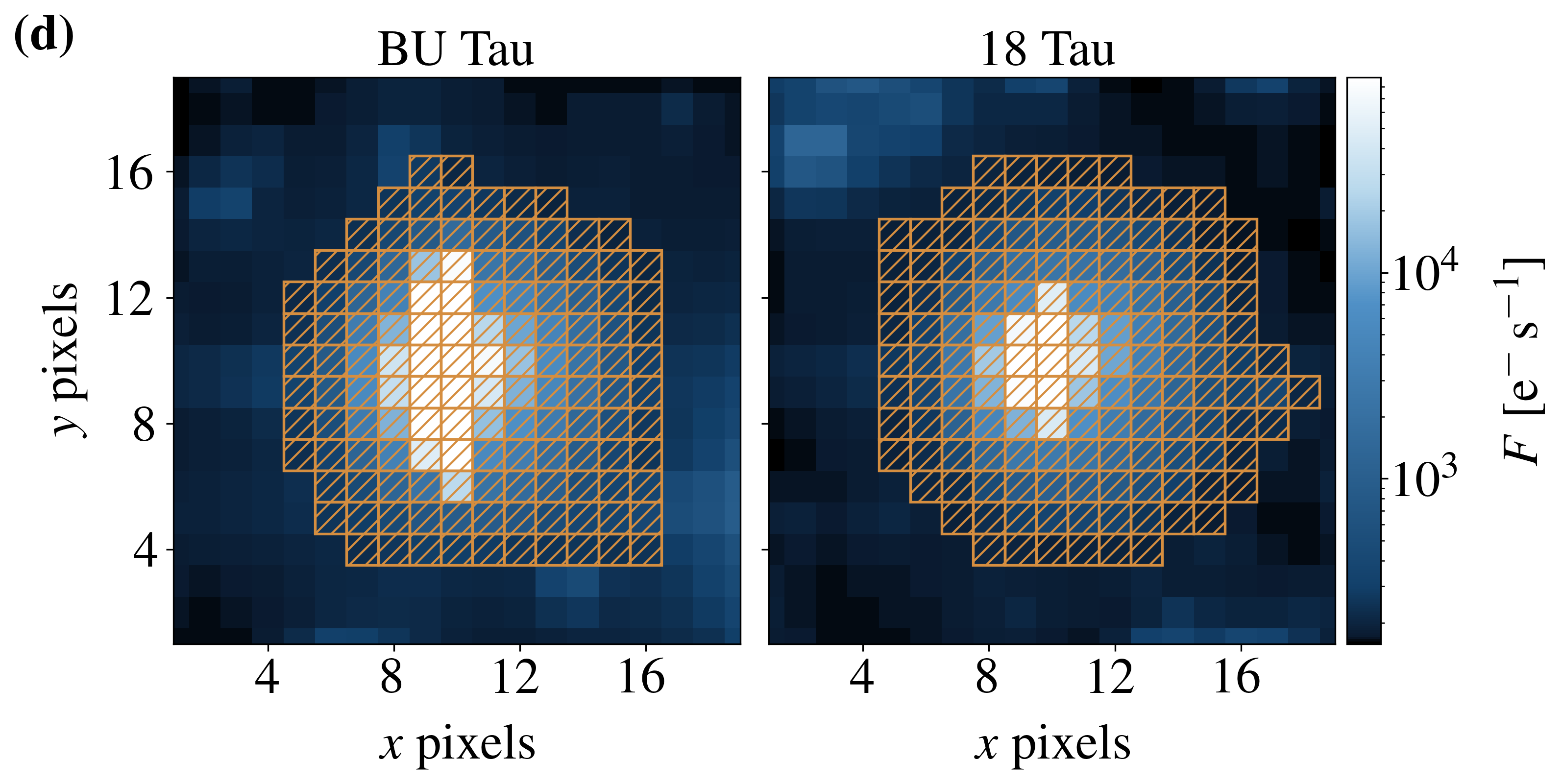}
\caption{
Comparison of ground-based photometric data from the Hvar observatory 
and space-based photometry from TESS for
BU~Tau = 28~Tau = HD~23862 = Pleione.
The TESS data sets cover five sectors (27\,d each) in 2021 and 2023,
while the Hvar \ubv\ photometry,
with a reasonable cadence,
spans the period from 1998 to 2025.
(a): The flux (in ${\rm e}^-\,{\rm s}^{-1}$) from TESS for BU~Tau (sector 70).
(b): The same for the comparison 18~Tau;
the flux range is the same as above.
(c): The Hvar $V$ photometry together with TESS (denoted by orange),
which was converted to magnitudes
and aligned with the Hvar long-term light curve.
The TESS data from 2021 were shifted by 5.10\,mag,
while those from 2023 by 5.08\,mag.
(d): For reference, we also show aperture masks
(hatched)
overlaid on median-stacked TESS observed images
of BU~Tau and 18~Tau.
}
\label{BU_Tau_TESS}
\end{figure}

\subsection{Comparison to TESS data} 

Monitoring of stellar photometric variability benefits greatly from combining
Hvar \ubv\ photometry with space-based photometry from TESS
\citep{Ricker2015JATIS...1a4003R}, 
which has observed more than 1000 Be stars. 
Together, they capture both short-term and long-term variations. 
We illustrate this synergy with the light curve of BU~Tau
(Fig.~\ref{BU_Tau_TESS}).
TESS observed the star over five sectors (three in 2021 and two in 2023), 
providing nearly continuous sector-long coverage
with a cadence of 2\,min in short-cadence mode,
or 30\,min in full-frame images,
achieving photometric precision of $\sim$0.0001\,mag
in its broad-band, red filter (600--1000\,nm).

In contrast, the Hvar observatory provides \ubv\ measurements
over a much longer temporal baseline, often spanning decades,
well-suited for tracing long-term variations,
where the seasonal and irregular sampling
does not hinder the study of secular trends. 
The uncertainty of $\pm$0.009\,mag
for the comparison star 18~Tau
represents excellent performance for ground-based photometry.

When comparing Hvar \ubv\ and TESS measurements,
the respective fluxes are normally computed from a model.
For main sequence stars, the
TESS magnitudes could be converted to $V$ or $B$ 
\citep{Eker2023MNRAS.523.2440E},
if the effective temperature is known.
However, such conversions are not advisable for Be stars,
due to circumstellar emission.
Moreover, in TESS data it is difficult to reliably 
identify trends extending beyond the duration of a single 
sector ($\sim$27 days), which limits the sensitivity to 
longer-term variability.
Since the Hvar system is stable,
we can use it to avoid detrending
and align the TESS data with the observed long-term trends.
The Hvar data thus remain essential
in coordination of ground- and space-based monitoring.

\subsection{Monitoring of atmospheric extinction at Hvar}

It was also deemed useful to present an overview of seasonal variations in
the extinction, relevant to air pollution at Hvar. Their values exceed
the astronomical interest and can be valuable for the broader community (e.g.
ecology, in particular).

\begin{figure}
\centering
\includegraphics[width=8.1cm]{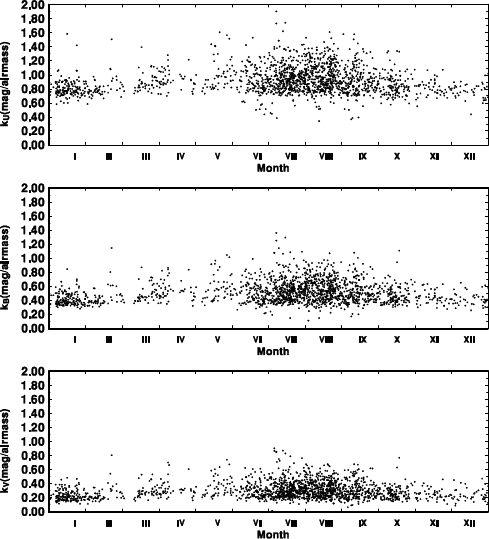}
\caption{Seasonal variations of the extinction coefficients
at the Hvar observatory.
In winter, atmospheric extinction is generally lower than in summer.
There are some values lying below the general slope
of the minimum extinction, especially for the $U$ band.
These extinction coefficients were usually determined
on poor photometric nights or the nights,
when only a few standard stars have been measured for extinction.
Of course, such nights were never used for all-sky photometry.
}
\label{extph}
\end{figure}

The seasonal variations, plotted in Fig.~\ref{extph} and
already noted by \citet{koupav82}, are clearly seen.
During the winter seasons,
atmospheric extinction is usually lower as the sea does not evaporate
too much.
During the summer seasons,
the values of the extinction coefficients exhibit a large scatter
due to local weather changes,
but on average, they are much higher than in the winter.
Another thing worth noting is that the minimum values
of the seasonal changes do not show any annual modulation.
Such behaviour has also been found at some other sites,
for instance, at La~Silla \citep{rufener86} or the Großschwabhausen
observing station \citep{reimann92}.

\begin{figure}
\centering
\includegraphics[width=8.1cm]{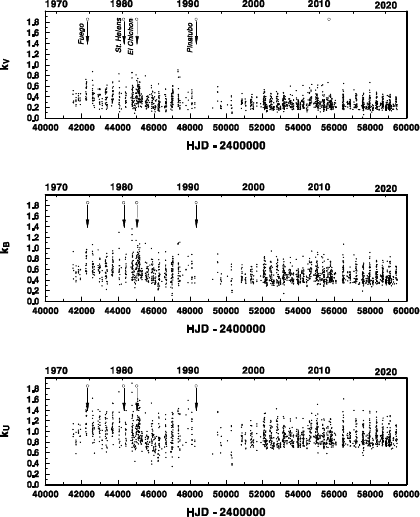}
\caption{
Long-term evolution of the extinction coefficients
at the Hvar observatory,
over the past fifty years.
The arrows (from left to right) indicate major volcanic eruptions
in the northern hemisphere:
Fuego, St.~Helens, El Chichon, Pinatubo.
}
\label{extjd}
\end{figure}

The long-term evolution of extinction coefficients
is plotted in Fig.~\ref{extjd}.
The scatter of the values is mainly caused by
seasonal changes discussed above. Nevertheless, a mild growth of
the extinction can be seen in 1975 and 1982. Such increases are often
related to large volcanic eruptions worldwide.
The first maximum, seen in 1975, may be due to the eruption of
the El Fuego volcano in Guatemala (October 1974). We note
that during the first observing seasons at Hvar, only a limited number
of nights and standard stars suitable for the accurate determination of
the atmospheric extinction were available.

The next increase in atmospheric extinction 
appeared at the time of the eruption
of the El Chichón volcano (March 1982).
The growth of extinction coefficients after this eruption
was also recorded at other observatories such as
La Silla \citep{rufener86, burki95},
Jena University Observatory \citep{reimann92}, or
Lowell Observatory \citep{loctho86}.
However, the effect of the powerful eruption of
the St. Helens volcano (March 1980)
was not obviously felt in the extinction at Hvar.
Needless to say, a very small number of observations
were carried out at that time.
It was detected only at the Lowell Observatory \citep{loctho86},
but the effect was marginal.
The strong Pinatubo eruption (April 1991)
was not covered by Hvar observations
because of the intermission due to the war in Yugoslavia.

Previously, \citet{pavlovski79} pointed out that the influence of the
molecular absorption (by O$_3$) has little effect
in the \ubv\ region.
The Rayleigh scattering on molecules (N$_2$, O$_2$)
on the other hand exhibits little variability.
The most variable component of the extinction is scattering by aerosols.
The Hvar observatory is a low-altitude station,
in the vicinity of the sea,
and is subject to significant weather changes even over a single night.
Obviously, the sea spray plays a major role in the large variations
in extinction during the summer seasons.
In addition, occasional forest fires, which are
quite frequent on the Adriatic coast,
could have a significant influence on the measured extinction.

\section{Classification according to variability patterns}\label{sec:types}

The list of all Be stars observed at Hvar, together with information
about the comparison and check stars used for their differential observations,
and the MK spectral types is in Table~\ref{tab03}.
Since the submission of the first version of this article,
one member of our team, Hrvoje Bo\v{z}i\'c, obtained 630 new individual observations
of several active Be stars discussed here during the year 2025 (JD~2460690-991).
As these observations further extend and corroborate
the existing time series,
we include them here as well,
highlighting the new data with blue symbols 
in Figs.~\ref{v442and} to \ref{omiand}. Based on our homogenised photometric observations,
we can distinguish the following patterns of light and colour variability, described below.

\begin{table*}
\footnotesize
\caption{Types of variability (as defined in Sect.~\ref{sec:types})
of 59 frequently observed hot emission-line objects at Hvar.}
\label{types}
\centering
\vskip-\smallskipamount
\begin{tabular}{@{\tiny}l@{\ \ \ }r@{\ }r@{\ }r@{\ }r@{\ \ \ }l@{\ }l@{\ }l@{\ }l@{\ }l@{\ }l@{\ }l@{}}
\hline
\hline
\noalign{\smallskip}
\# & Star            &   &    HD/BD & No.  & LTE & LTC & BIN & RLA       & TESS $\mathrm{rms}$ &  LTQ & Figs. \\
   &                 &   &          &      &     &     &     &           & [mmag]              &      &       \\
\hline
\noalign{\smallskip}           
1  & $\omicron$ Cas  & * &     4180 &  895 &   p &  ?  &  EL & yes       & 2.6--6.1 &   no & \\ 
2  & $\gamma$ Cas    & \ &     5394 &  137 &   p & yes &  ?  & yes       & 5.5      &  inc & \\
3  & V442 And        & * &     6226 &  964 &   p &  no &  no & no?       & 1.4      &   no & \ref{v442and} \\ 
4  & $\varphi$ And   & * &     6811 &  504 &  no & no? &  no & no        & 0.5      &  dec & \ref{phiand} \\
5  & $\varphi$ Per   & * &    10516 &  678 &   p & yes & yes & yes       & 9.7      &  dec & \ref{phiper} \\
6  & V554 Per        & \ &    14818 &  146 &  no & yes &  no & yes?      & 2.4      &   no & \\
7  & HR 894          & \ &    18552 &  118 &  no &  no &  no & no?       & 1.7      &   no & \\ 
8  & RX Cas          & \ & +67\,244 &  361 &  no &  no &  EB & yes       & 6.1      &   no & \\ 
9  & 13 Tau          & \ &    23016 &  114 &  no &  no &  no & yes?      & 2.7      &   no & \\
10 & 17 Tau          & * &    23302 &  367 &  no & yes &  no & no        & 0.3      &   no & \ref{17tau} \\ 
11 & V971 Tau        & \ &    23480 &  265 &  no & yes &  no & no?       & 1.3      &   no & \\
12 & $\eta$ Tau      & \ &    23630 &  205 &  no & yes &  no & no        & 0.3      &  no? & \\
13 & BU Tau          & * &    23862 &  469 &   i &  no &  no & no        & 2.8      &   no & \ref{BU_Tau_TESS}, \ref{butau} \\ 
14 & V960 Tau        & * &    36576 &  392 &   p & yes &  no & yes       & 13.7     &   no & \ref{v960tau} \\
15 & $\zeta$ Tau     & * &    37202 & 1100 & p+i & yes & yes & yes       & 9.8      &   no & \ref{zettau} \\
16 & $\omega$ Ori    & \ &    37490 &   90 &   p &  no &  no & yes       & 10.2     &   no & \\
17 & V731 Tau        & \ &    37967 &  107 &  no &  no &  no & yes       & 4.0      &   no & \\
18 & V696 Mon        & * &    41335 &  318 &  no &  no &  no & yes       & 5.8      &  dec & \\
19 & HR 2418         & \ &    47054 &   63 &  no &  no &  no &  no       & 1.1      & dec? & \\
20 & OT Gem          & \ &    58050 &  437 &   p &  no &  no & yes       & 4.3      &   no & \\
21 & $\beta$ CMi     & \ &    58715 &  168 &  no &  no &  no &  no       & 0.1      &   no & \\ 
22 & BR CMi          & \ &    61273 &  103 &  no &  no &  EL & yes       & 4.1      &   no & \\ 
23 & UX Mon          & \ &    65607 &  152 &  no &  no &  EB & yes       & 8.7      &   no & \\
24 & HD 81357        & \ &    81357 &   93 &  no &  no &  EL & no        & 0.8      &   no & \\
25 & $\kappa$ Dra    & * &   109387 &  429 &   p &  no &  no & yes?      & 2.4      &  dec & \ref{kapdra} \\
26 & $\vartheta$ CrB & \ &   138749 &  158 &  no &  no &  no & yes?      & 3.1      &   no & \\
27 & V839 Her        & * &   142926 &  685 &   i &  no & yes & no        & 2.6      &   no & \ref{v839her} \\
28 & $\delta$ Sco    & \ &   143275 &   98 &   p &  no &  no & no?       & 0.6      & inc? & \\
29 & $\zeta$ Oph     & \ &   149757 &  168 &   p &  no &  no & yes       & 5.2      &   no & \\
30 & V744 Her        & * &   162732 & 1449 &   i &  no &  no & no        & 2.1      &  dec & \ref{v744her}, \ref{v744her_cycles} \\
31 & V2048 Oph       & \ &   164284 &  164 &   p &  no &  no & yes       & 4.4      &  dec & \\
32 & V974 Her        & \ &   164447 &  183 &   p &  no &  no & no        & 0.8      &   no & \\
33 & $\omicron$ Her  & \ &   166014 &  182 &  no &  no &  no & no        & 0.1      &   no & \\
34 & NW Ser          & \ &   168797 &  327 &  no &  no &  no & yes       & 6.5      &   no & \\
35 & CX Dra          & * &   174237 & 1167 &   p &  no & yes & yes       & 3.4      &   no & \ref{cxdra} \\
36 & $\beta$ Lyr     & \ &   174638 &  544 &  no & yes &  EB & no?       & 0.5      &   no & \\
37 & 7 Vul           & \ &   183537 &  131 &   p &  no &  no & yes       & 5.1      &   no & \\
38 & V923 Aql        & * &   183656 & 1620 &   i & yes &  no & yes       & 10.8     &   no & \ref{v923aql_cycles} \\ 
39 & V1294 Aql       & * &   184279 & 1709 & p+i & yes & yes & yes       & 9.8      &  dec & \ref{v1294aql_rectification} \\
40 & V1507 Cyg       & \ &   187399 &  193 &  no &  no & yes & yes       & 6.6      &   no & \\ 
41 & V1746 Cyg       & \ &   189687 &  209 &  i? &  no &  no & yes?      & 2.7      &   no & \\
42 & V1624 Cyg       & \ &   191610 &  510 &   p &  no &   ? & yes       & 9.3      &   no & \\
43 & 20 Vul          & \ &   192044 &  130 &  no &  no &  no & no?       & 1.6      &   no & \\
44 & QR Vul          & \ &   192685 &  191 &   p &  no &  no & yes       & 4.5      &   no & \\
45 & P Cyg           & \ &   193237 &  122 &  no &  no &  no & no?       & 1.4      &   no & \\
46 & 25 Vul          & \ &   193911 &  124 &   ? &  no &  no & no?       & 1.6      &   no & \\
47 & V2119 Cyg       & \ &   194335 &  140 &   ? &  no &  no & yes       & 11.4     &   no & \\
48 & HR 7843         & \ &   195554 &  161 &  i? &  no &  no & no        & 0.3      &   no & \\
49 & V1661 Cyg       & \ &   198478 &  288 &  no &  no &  no & no?       & 1.3      &   no & \\
50 & V832 Cyg        & * &   200120 &  849 &   p & yes & yes & yes?      & 3.2      &  inc & \ref{v832cyg}, \ref{v832cyg_fou} \\
51 & V1931 Cyg       & \ &   200310 &  889 &   p &  no &  no & yes       & 10.9     &   no & \\
52 & 8 Lac\,A        & \ &   214168 &  151 &  no &  no &  no & no?       & 1.8      &   no & \\
53 & V360 Lac        & \ &   216200 &  424 &  no & yes &  EL & yes       & 3.7      &   no & \\
54 & EW Lac          & * &   217050 & 1281 &   p &  no &  no & yes       & 10.3     &  inc & \ref{ewlac} \\ 
55 & V378 And        & \ &   217543 &  267 &   p &  no &  no & yes       & 4.0      &   no & \\
56 & $\omicron$ And  & * &   217675 & 1636 &  no & yes &  no & yes       & 2.4--15.4&   no & \ref{omiand}, \ref{omiand_orbit} \\ 
57 & KX And          & * &   218393 & 1210 &  no &  no & yes & yes       & 5.0      &   no & \\
58 & KY And          & \ &   218674 &  970 &  no &  no &  no & yes       & 19.6     &   no & \\
59 & LQ And          & \ &   224559 &  590 &  no &  no &  no & yes?      & 2.6      &   no & \\
\hline
\end{tabular}
\tablefoot{
* denotes objects discussed in Appendix~\ref{sec:comments},
the last column links the respective figures; 
p = positive correlation between the brightness and emission-line strength,
i = inverse correlation;
EB = eclipsing binary;
EL = ellipsoidal variations;
yes in the BIN column means either mild sinusoidal variation with the known orbital period,
or some peculiar orbital changes;
no in the RLA columns means that the variability is not detectable from Hvar, but it was detected from TESS;\ 
 $\mathrm{rms}$ denotes the root mean square from TESS photometry (in mmag);
the corresponding peak-to-peak variability is commonly ${\sim}3$ times larger;
dec = secular decrease of brightness between different quiescent phases;
inc = secular increase.
}
\end{table*}

\begin{enumerate}
\item Long-term envelope (LTE) variability,
occurring on a timescale of several years to decades, which
is recurrent, but not periodic,
with amplitudes ${>}0.1\,{\rm mag}$ (in~V),
and correlated (or anti-correlated) with colours
(see e.g. Fig.~\ref{butau}).

\item Long-term cyclic (LTC) variability,
occurring on a timescale of several years,
which is multi-periodic or quasi-periodic,
with amplitudes ${>}0.016\,{\rm mag}$
(Fig.~\ref{omiand}).

\item Binarity (BIN)
is periodic variability
with observed periods from $6$ up to ${\sim}1000\,{\rm d}$
and amplitudes ${>}0.016\,{\rm mag}$
(Fig.~\ref{kxu}).

\item Rapid low-amplitude (RLA) variability
is periodic or multi-periodic,
with periods from 0.1 to 4\,d
and amplitudes ${>}0.016\,{\rm mag}$
(Fig.~\ref{v960tau}).
Hvar observations are not very suitable
for analyses of such rapid changes;
space-based photometry, such as TESS, should be used
instead.
Nevertheless, 
these rapid variations are present in the stars observed 
at Hvar and may contribute to the observed scatter and light-curve 
morphology. Therefore, Table~\ref{tab03}
indicates whether the RLA amplitudes are large enough to be 
detectable in the Hvar data (RLA~yes).
This flag serves as a warning to users that rapid 
variability may be present in the observations and should be 
considered when interpreting the Hvar photometry.

\item Long-term variability of quiescence (LTQ),
occurring on the timescale of years to decades,
as a linear trend from one quiescent phase to another
(i.e. not taking the LTE changes into account),
at rates of the order of ${\pm}1.0\,{\rm mmag}\,{\rm yr}^{-1}$
(Figs.~\ref{v744her}, \ref{v832cyg}).
\end{enumerate}

We determined the types of variability for 59 hot emission-line objects,
which were more frequently observed at the Hvar observatory
(see Table~\ref{types}).
\vskip\baselineskip

\paragraph{Physical interpretations.}
The LTE variability is known to be related to
the strength and/or extent of the Be envelope. It has two distinct alternatives, 
depending on the inclination of the Be disc plane with respect to the observer:
pole-on-positive, when the brightenings are accompanied by the rise of 
the emission-line strength, and edge-on-inverse,
when the light decreases. Both phases are followed by an emission-line rise
\citep{hec83, hec94, hec2000, sigut2013, Vieira_2017MNRAS.464.3071V}. 
We note that in cases of intermediate inclination,
a~given object can exhibit either
a~positive or an inverse correlation in various epochs of observations,
depending on the extent of the Be envelope \citep[see, e.g.][]{hec2022}.

The LTC variability is often correlated with $V/R$ variations
of the relative strength of the $V$ and $R$ peaks
of the double Balmer emission lines,
inferred from spectroscopy.
Both are related to non-axisymmetric geometrical
and/or dynamical changes of the structure of the Be discs.
In the past,
they were interpreted as a gradual revolution of an elongated disc
\citep[see, e.g.][]{balle89}. Since the studies by \citet{okazaki91}
and \citet{okazaki97}, they are usually understood as global
one-armed oscillations of the discs
and can also manifest themselves by the corresponding light changes
\citep[see the detailed discussion by][]{mennick97}.
V923~Aql provides a recent example, as presented by
\citet{wolf2021}.
The problem is that the LTC sometimes occur simultaneously
with the LTE changes and it is not easy to separate both effects
(see e.g. V1294~Aql; \citealt{hec2022}).

The BIN variability generally follows
the orbital periods of the Be binaries in question,
but they are not easy to detect
due to the simultaneous presence of variations on other timescales.
In addition to clear cases of binary eclipses, they can also manifest themselves
as a combination of ellipticity and reflection, or occasional eclipses
of the outer parts of the discs. Numerous examples of such changes were
discussed, with corresponding references, by \citet{bozic2013}
and are documented by Hvar's systematic photometry. We discuss the individual Be stars below.

The RLA variability is interpreted as pulsations or rotational modulation.
Since the typical observing strategy at Hvar was focused on monitoring
variations on longer time scales, Hvar observations are not very suitable
for analyses of such rapid changes, with the exception of several
dedicated observing campaigns on selected Be stars, for instance,
$\zeta$ Tau \citep{bozic88}, $\omicron$~And, KX~And, KY~And, LQ~And,
EW~Lac \citep{stagg88}, or $\omega$~Ori \citep{balona2001}.

\begin{figure*}
\sidecaption
\includegraphics[width=12cm]{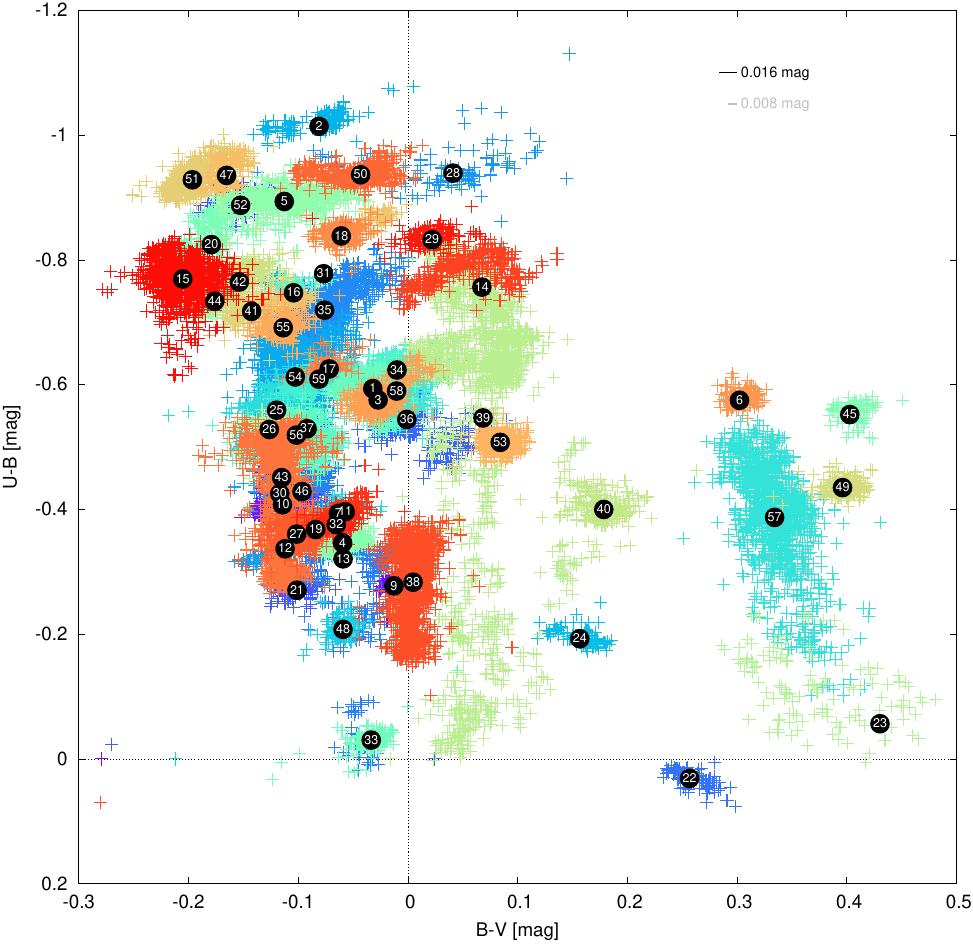}
\caption{
Colour-colour diagram \bv\ versus \ub\
for 59 studied hot emission-line stars.
Each star is plotted with a different colour.
Typical uncertainties of \ubv\ magnitudes
(0.008 to 0.016\,mag)
approximately correspond to the size of symbols.
See Fig.~\ref{Fig10} for the individual types of variability.
The numbers refer to values marked as \# in Table~\ref{types}.}
\label{Fig9}
\end{figure*}

The LTQ variability was probably first noted
for $\omega$~CMa by \citet{hec98} and empirically modelled in terms of
the viscous decretion disc model \citep{lee91,carciofi2009}
by \citet{ghore2021,ghore2023}.
Several distinct examples of such behaviour were presented
by \citet{hec2022}.
Secular brightness changes are systematically monitored
by the long-term Hvar photometry
and represent a well-established observational fact.
However, the physical interpretation of LTQ variability remains uncertain. 
One option could be changes in the circumstellar disc alone,
resulting from timings of LTE events,
with the consequence that the system does not return to the same state. 
Another option could be that the variability is connected
with long-term evolution of the pseudo-photosphere
\citep{hec2022}.
In several objects,
the Balmer emission nearly disappears during such phases,
suggesting substantial structural changes.
Apparent secular changes of $v \sin i$,
reported for $\omega$~CMa \citep{hec98}
or Achernar \citep{Souza2014A&A...569A..10D,Rivinius2016ASPC..506...17R}, 
provide additional support for this interpretation.
If the observed long-term brightening corresponds to
a gradual strengthening of the pseudophotosphere,
as in EW~Lac,
then long-term fading episodes would indicate the opposite behaviour.

\section{Results}\label{sec:results}

Our systematic observational effort and monitoring of a~representative sample
of 59 bright early-type emission-line stars from the northern hemisphere
(down to declinations of about $-20^\circ$)
had previously been helpful in determining that virtually
all Be stars are light- and colour-variable
(Fig.~\ref{Fig9}), leading to the recognition of their characteristic timescales.
We admit our sample is strictly speaking neither magnitude-
nor volume-limited.
On one hand, this is a serious limitation;
on the other hand, it contains stars of various MK spectral types,
albeit located at different distances.

\paragraph{Long-term envelope variability.}

We detected the LTE variability in $(44\pm 9)\%$ of all monitored objects.
More specifically, 20 objects show a positive correlation
between the brightness and emission-line strength,
while 6 show an inverse correlation.
We thus confirm that the positive correlation is more common
\citep{deWit_2006A&A...456.1027D}.
In two cases, we found alternation between the two types of correlation
($\zeta$~Tau and V1294~Aql).

We constructed the corresponding colour-colour
diagrams (Fig.~\ref{Fig10}),
for the various types of variability.
We can see that the objects with positive versus inverse correlations are
bluer (${\lesssim}{-}0.5$\,mag) and redder (${\gtrsim}{-}0.5$\,mag)
in the \ub\ index,
because of the pole-on versus edge-on disc geometry.
V1294~Aql with both correlations spans the entire range,
likely because of the obscuration of the star
when the disc is thick.
On the contrary, $\zeta$~Tau only exhibited inverse correlation
in H$_\alpha$ and $V$
\citep{zarf26};
otherwise, it is located among positive-correlation stars.
If no LTE variability is present,
the index is often intermediate (${\sim}{-}0.5$\,mag).

We also constructed the HR diagrams
(Fig.~\ref{Fig11}),
with the absolute magnitudes, $M_V$, derived from the known parallaxes
\citep{leeuw2007b,Vallenari_2023A&A...674A...1G}
and the interstellar absorption, $A_V$, from the reddening maps
\citep{Green_2018MNRAS.478..651G}.
We can see that the objects with positive and inverse correlations are
brighter ($M_V\lesssim{-}1$\,mag) and fainter ($M_V\gtrsim{-}1$\,mag),
respectively.
V1294~Aql, $\zeta$~Tau with both correlations do not span the entire range,
but are brighter (${\sim}{-}3$\,mag).
This also corresponds to the geometry,
because the disc attenuates more visible light if it is edge-on.%
\footnote{
We note that binaries observed close to the edge-on configuration
also exhibit varying colours similar to the LTE's inverse variability.
However, this occurs on a short, orbital timescale,
obviously due to projection effects of circumstellar matter.
This is the case of KX~And, UX~Mon, or RX~Cas.
}

\paragraph{Ultraviolet variability.}

We point out that models of Be star discs
based only on space-based photometry
with a very broad-band filter
are inherently limited.
Cases of such stars as
BU~Tau,
V744~Her,
V923~Aql, or
V1294~Aql
demonstrate that their variations in the
yellow, blue, and ultraviolet parts
of the stellar spectrum can be different. This can be seen in Fig.~\ref{Fig10},
where these stars exhibit excursions in the \ub\ index (for the temporal dependence, see
Figs.~\ref{butau}, \ref{v744her}).
Specifically,
the BU~Tau light curve reached a minimum in $V$ (by 0.3\,mag)
at around 2455000\,d, while the
\bv\ index was in phase.
However, the deepest minimum in \ub\ (by 0.45\,mag)
was 1500\,d later, at 2457500\,d.

All such differences are `smeared' or ignored
when photometry with a very broad-band filter is used.
This might be the case for
MOST \citep{walker2003},
TESS \citep{Ricker2015JATIS...1a4003R}, and
Gaia \citep{Vallenari_2023A&A...674A...1G},
where the respective filters are approximately
350--700,
600--1000, and
330--1050\,nm.

\paragraph{Long-term cyclic variability.}

We detected the photometric LTC variability in $(24\pm 6)\%$ of objects.
In some of them, it was verified by spectroscopy
(e.g. \citealt{wolf2021,hec2022})
that it is related to the $V/R$ line-profile variability.
These objects tend to be bluer
($\bv \lesssim {+}0.1$\,mag);
although similar objects have no LTC's.
Similarly, they tend to be brighter
($M_V\lesssim {-}1$\,mag).
None of the intrinsically faint
($M_V\gtrsim {-}1$\,mag)
objects show this kind of variability.

Notably, the distribution of LTC objects in the HR diagram
is similar to the LTE-positive objects
(Fig.~\ref{Fig11});
despite the fact that these represent two different types of variability.
Regarding their possible relation, we computed
the following contingency table (see Table~\ref{contingency_tab}).

\renewcommand{\arraystretch}{1.2} 
\begin{table}[h!]
\centering
\caption{Contingency table for the LTC and LTE variability.}
\label{contingency_tab}
\begin{tabular}{|c|c|c|}
\hline
\diagbox{LTE}{LTC} & no & yes \\
\hline
not positive & 22 & 10  \\
\hline
positive     & 16 & 4   \\
\hline
\end{tabular}
\end{table}

\noindent
This indicates that, excluding 22 unremarkable objects,
16 and 10 objects exhibit either LTE or LTC,
but rarely both.
These two types of variability seem to be almost mutually exclusive.

\vskip\baselineskip
\label{p9}
This might suggest the following mechanism.
On one hand,
if the LTE variability is present,
the disc can extend to larger radii
due to prolonged build-up episodes
and can reach the secondary
\citep{klement2024},
or even extend beyond it at low surface density
\citep{Panoglou_2018MNRAS.473.3039P}.
The secondary then perturbs the outer disc
and creates two spiral arms ($m=2$;
\citealt{Lindblad_1964ApNr....9..103L,Rubio_2025A&A...698A.309R}),
which are visible on both sides of the disc,
from the observer's point of view.
On the other hand,
if no significant LTE variability is present,
this disc can remain more confined
due to more rapid growth$\,\leftrightarrow\,$dissipation cycles,
and the disc cannot reach the secondary.
The secondary then perturbs the inner disc,
and creates one arm ($m=1$),
which is asymmetric,
visible on one side of the disc.
If true, the LTC period
should correspond to the orbital period of the secondary
(or a~half of it).
The phasing, however, might be modulated by other types
of variability; in particular, the time-dependent extent
of the disc.

Thus, we computed periodograms for the LTC variable stars
(as in Fig.~\ref{v832cyg_fou}).
Some of the periods, $P_i$, were instrumental
(${\sim}1\,{\rm d}$)
or corresponding to long-term trends,
but in several cases, we were able to clearly recover
the respective periods (see Table~\ref{identified_P_tab}).

\begin{table}[h!]
\small
\caption{Periods identified in the periodograms.}
\label{identified_P_tab}
\begin{tabular}{|@{\ }l@{\ }|@{\ }l@{\ }|@{\ }l@{\ }|@{\ }c@{\ }|}
\hline
V554 Per       & $P_2 \simeq 71\,{\rm d}$                              &                        & ? \\ 
V960 Tau       & $P_2 = 2731\,{\rm d}$                                 &                        & ? \\ 
$\zeta$ Tau    & $P_3 = 656\,{\rm d}$, $P_5 \simeq 74\,{\rm d}$        & $P = 132.987\,{\rm d}$ & ? \\ 
$\beta$ Lyr    & $P_1 = 6.470758\,{\rm d}$, $P_2 = 12.939670\,{\rm d}$ & $P = 12.913\,{\rm d}$  & 1 \\ 
V923 Aql       & $P_1 = 2227\,{\rm d}$, $P_5 \simeq 92\,{\rm d}$       &                        & ? \\ 
V1294 Aql      & $P_2 = 97.052\,{\rm d}$                               & $P = 192.910\,{\rm d}$ & 2 \\ 
V832 Cyg       & $P_1 = 2202\,{\rm d}$, $P_6 = 28.193\,{\rm d}$        & $P = 28.187\,{\rm d}$  & 1 \\ 
V360 Lac       & $P_1 = 10.085\,{\rm d}$                               & $P = 10.085\,{\rm d}$  & 1 \\ 
$\omicron$ And & $P_1 = 2481\,{\rm d}$                                 & $P = 2525\,{\rm d}$    & 1 \\ 
\hline
\end{tabular}
\end{table}

\noindent
Comparisons with known orbital periods, $P$ (taken from Table~\ref{tab03}). These data
allowed us to determine the order, $m$ (in the last column).
Most of the cases are standard $m = 1$ oscillations.
These systems tend to have relatively short orbital periods
(with the possible exception of $\omicron$~And), possibly indicating that the proximity of the companion plays a role.
In close binaries, the strong and asymmetric perturbation of the inner disc
could favour the development of a global one-armed asymmetry
over a symmetric $m=2$ spiral structure.
The longer periods (${\approx}10^3\,{\rm d}$) of
$\zeta$~Tau and V923~Aql
could be better explained
as global disc oscillations
\citep{Ogilvie_2008MNRAS.388.1372O};
their shorter periods ($10^1\,{\rm d}$)
would be close to $m = 2$,
but the correspondence is not exact.
These two stars are also LTE variables,
likely causing phase modulation
and frequency splitting (e.g. $2f_{\rm orb} \pm f_3$).

\vskip\baselineskip
Alternatively, if the disc is extended well beyond the secondary
and the mass ratio is relatively high,
the spiral in the circumbinary region might dominate emission line features
(\citealt{Rubio_2025A&A...698A.309R}, fig.~2).
Another possibility is that unceasing disc growth and dissipation
(i.e. LTE activity)
prevents the very development of one-armed disc oscillation
(i.e. LTC activity),
which requires some time to develop; however,
gravity and pressure gradients during growth and dissipation
are not in equilibrium.

\paragraph{Binarity.}

We detected more regular BIN variability in $(25\pm 7)\%$ of the objects in the sample.
This particular percentage is affected by (i)~the selection bias because some objects 
were selected as {\em a-priori\/} binaries; and (ii)~the limiting amplitude (${\sim}0.008$\,mag) 
because we were unable to detect sdO components \citep{wang2021}.
Nevertheless, the amplitudes in \ub\ of our binaries,
tend to be relatively larger compared to singles (Fig.~\ref{Fig10}).
Among them, $\beta$~Lyr is similar in terms of out-of-eclipse colours
to classical Be stars;
other eclipsing (or ellipsoidal) binaries are offset.
The amplitudes in $M_V$ tend to be also naturally larger
(Fig.~\ref{Fig11}),
especially for eclipsing systems.
On the contrary,
ellipsoidal systems always have low amplitudes in $M_V$.

\begin{figure}
\centering
\includegraphics[width=9cm]{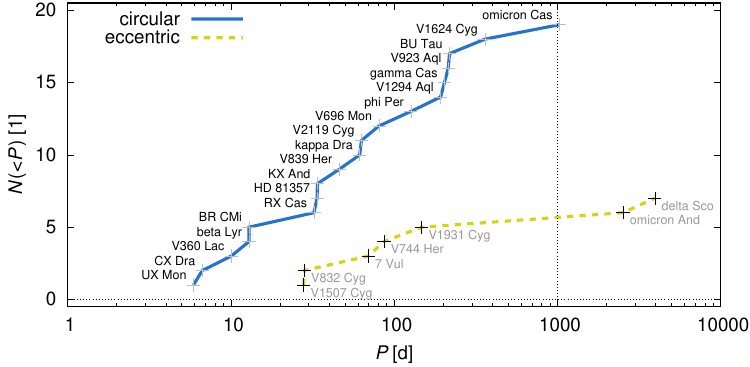}
\caption{
Cumulative distributions of periods $N({<}P)$
for circular and eccentric binaries,
according to Table~\ref{tab03}.
}
\label{fig:P}
\end{figure}

\begin{figure*}[p]
\centering
\includegraphics[width=16cm]{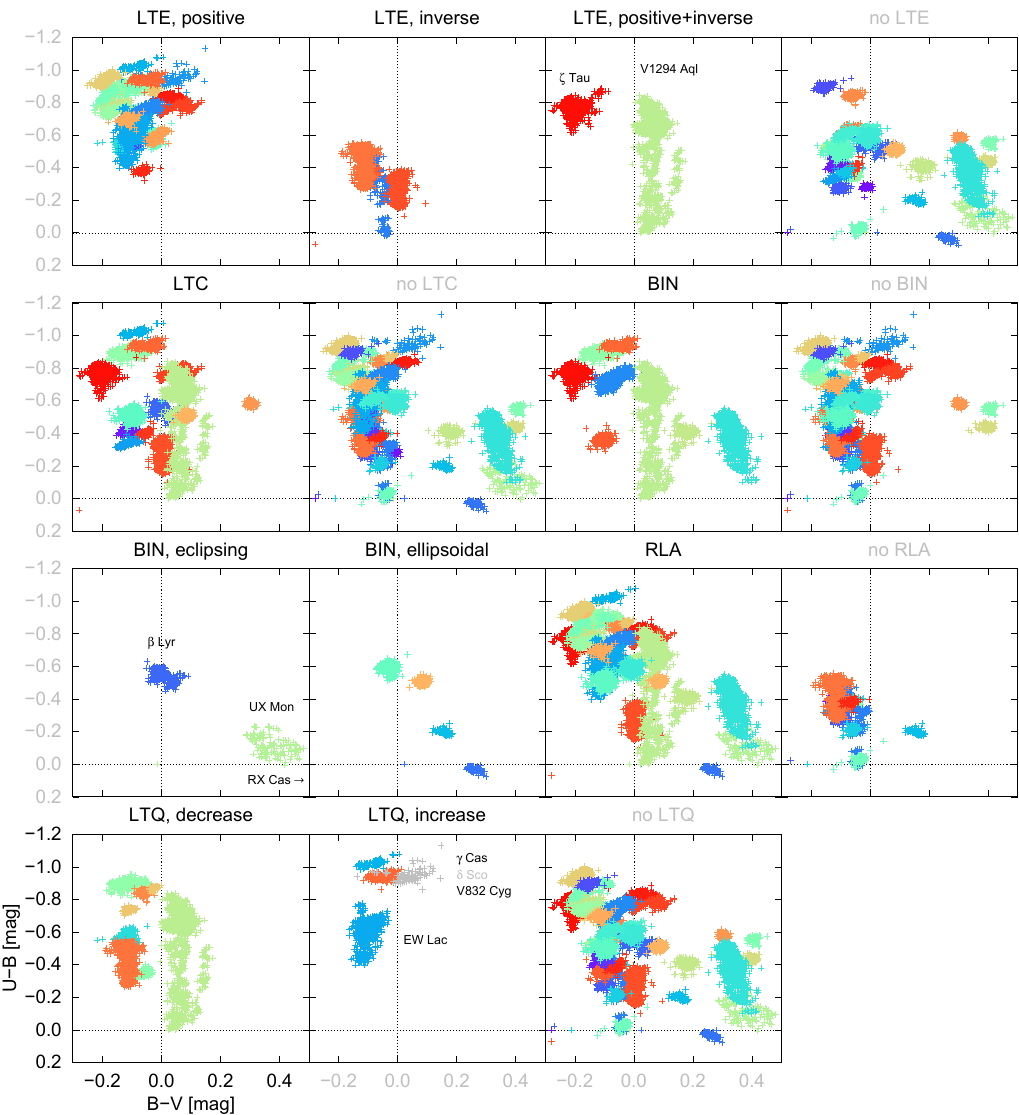}
\caption{
Same as Fig.~\ref{Fig9},
but separated according to the types of variability as follows.
LTE: long-term emission,
LTC: long-term cyclic,
BIN: binarity,
RLA: rapid low-amplitude,
LTQ: long-term quiescence.
Comparison of objects with positive and inverse correlations shows
they have lower (${\lesssim}{-}0.5$\,mag) and higher (${\gtrsim}{-}0.5$\,mag)
index \ub;
V1294~Aql with both correlations spans the entire range.
On the contrary, $\zeta$~Tau only exhibited inverse correlation
in H$_\alpha$ and $V$
\citep{zarf26},
otherwise, it is located among positive-correlation stars.
If there is no LTE variability,
the index is often intermediate (${\sim}{-}0.5$\,mag).
Objects with LTC variability tend to have lower (${\lesssim}{+}0.1$\,mag)
index \bv.
If there is more regular BIN variability,
amplitudes tend to be relatively larger.
$\beta$~Lyr is similar in terms of out-of-eclipse colours
to classical Be stars,
even though it is physically different \citep{broz2021};
other eclipsing (or ellipsoidal) binaries are offset.
All objects with very low rapid changes (i.e. no RLA)
have higher
(${\gtrsim}{-}0.5$\,mag)
index \ub.
Objects with LTQ variability,
which increase brightness in the course of time,
have the lowest (${\lesssim}{-}0.9$\,mag)
index \ub,
as if the star is not obscured.
Note: the $\gamma$~Cas objects emit hard X-rays,
arising from accretion onto the WD companion \citep{Naze_2026A&A...707A.334N}.
$\delta$~Sco was considered uncertain
(gray). Nevertheless, it is a highly eccentric binary,
where the disc has been perturbed by periastron passage
\citep{rast2024}.
}
\label{Fig10}
\end{figure*}

\begin{figure*}[p]
\centering
\includegraphics[width=16cm]{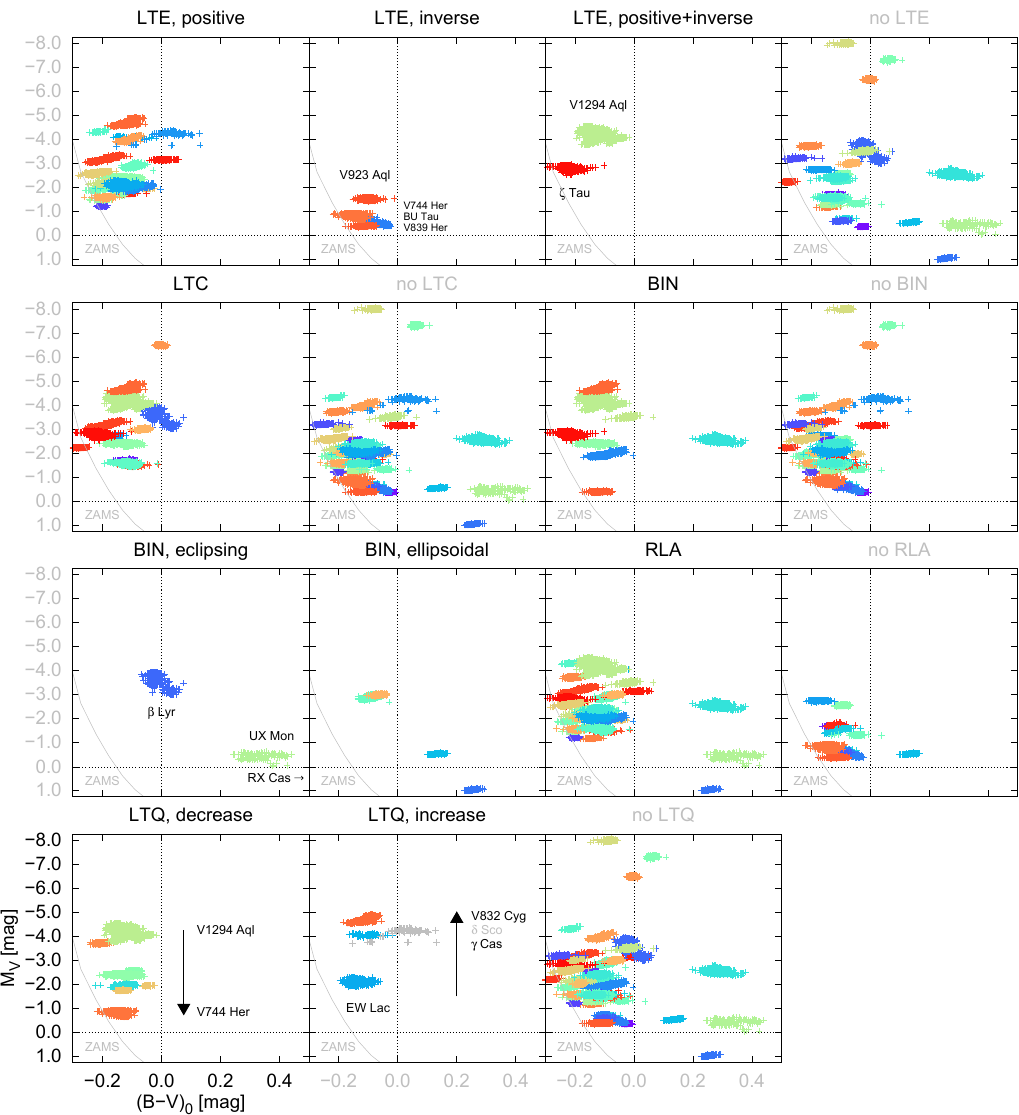}
\caption{
HR diagrams,
as the absolute magnitude $M_V$ vs $(\bv)_0$,
separated according to types of variability.
The parallaxes $\pi$ were taken from Gaia
\citep{Vallenari_2023A&A...674A...1G},
Hipparcos
\citep{leeuw2007b},
and $M_V$ was computed from the Hvar $V$ magnitudes
and reddening $E(B-V)$ \citep{Green_2018MNRAS.478..651G}, as
$A_V = 3.1\,E(B-V)$,
$M_V= V - A_V + 5\log_{10}\pi+ 5$.
The notation is the same as in Fig.~\ref{Fig10}.
The zero-age main sequence (ZAMS) is plotted
in gray.
Comparison of objects with positive and inverse correlations shows
that they have lower (${\lesssim}{-}1$\,mag) and higher (${\gtrsim}{-}1$\,mag)
absolute magnitude~$M_V$;
V1294~Aql and $\zeta$~Tau, which show both correlations, do not span the entire range, 
but have higher values (${\sim}{-}3$\,mag).
Objects with LTC variability tend to have low (${\lesssim}{-}1$\,mag)
absolute magnitude~$M_V$;
although similar objects have no LTC's.
If there is BIN variability,
amplitudes in $M_V$ tend to be naturally larger,
especially for eclipsing systems.
On the contrary,
ellipsoidal systems always have low amplitudes in $M_V$. 
Objects with LTQ variability,
which decreases brightness in the course of time,
span the entire range of $M_V$.
Their number (7 vs 4) suggests that this phase is relatively longer.
On the contrary,
only four increase brightness, and three of them are
close to $M_V \simeq -4.0$\,mag, i.e. the upper end in brightness for our Be stars 
(the three brighter stars are luminous B-type stars). This group includes
$\gamma$~Cas (a companion-origin hard X-ray source; \citealt{Naze_2026A&A...707A.334N}),
$\delta$~Sco (a highly eccentric secondary; \citealt{rast2024}).
}
\label{Fig11}
\end{figure*}

\paragraph{Eccentric orbits.}
For 26 systems,
we already know their orbital solutions from the literature,
in some cases based on Hvar data
(see the references in Table~\ref{tab03}).
Among them,
18~are circular ($e \doteq 0.0$) and
8~eccentric ($e \gg 0$).
Their cumulative distributions of periods, $N({<}P)$,
appear to be similar
(Fig.~\ref{fig:P});
the Kolmogorov--Smirnov test
indicates the distance, $d_{\rm KS} = 0.361$,
and the probability, $p_{\rm KS} = 0.379$.
Nevertheless, the two systems with the longest periods
($P\gg 1000\,{\rm d}$)
are eccentric
($\omicron$~And$, \delta$~Sco)
On the other hand,
the third system with $P\sim 1000\,{\rm d}$
is circular
($\omicron$~Cas; \citealt{Koubsky_2010A&A...517A..24K}).
This would imply that most systems have undergone circularisation,
but other systems
(i)~have remained eccentric or
(ii)~have become eccentric.
From $N({<}P)$ alone, it is not clear
which of these processes is more likely.

\paragraph{Rapid low-amplitude variability.}

We made note of the RLA variability in $(66\pm 11)\%$ of the stars
observed at Hvar.
We used the TESS data and verified that all other stars (i.e. RLA no)
also exhibit rapid changes, albeit with very low amplitudes
(${\lesssim}0.008\,{\rm mag}$). Thus, it became clear that we should 
treat these two groups of stars as having
low versus very low amplitude of rapid changes.
All objects with very low amplitude are redder
(${>}{-}0.5$\,mag)
in \ub\ index,
as if the star were obscured by the disc.
Nevertheless, this finding is also statistically consistent with
\citet{bartz2022},
who reported that later-type Be stars tend to exhibit lower RLA amplitudes,
which then simply fall below our detection limit
(i.e. 0.008 to 0.016\,mag).
This is indeed the case of
BU~Tau (Fig.~\ref{BU_Tau_TESS}).

\paragraph{Non-radial pulsations.}

A standard explanation for the RLA variability is non-radial pulsations.
The prototype $\omega$~CMa
(a star not observed at Hvar)
shows a prominent g-mode
($\ell = 2$, $m = 2$)
with high horizontal amplitude
\citep{Maintz_2003A&A...411..181M}.
In line profiles, it is seen as a wave,
sometimes having a sharp `edge',
when the high horizontal amplitude at the limb
is projected towards the observed.

Regarding stars observed at Hvar,
these non-radial pulsations were previously discussed in detail, for instance, for
V2048~Oph, V1624~Cyg, or $\omega$~Ori
\citep{Rivinius_2003A&A...411..229R,baade2018,Neiner_2002A&A...388..899N}.
However, an unambiguous detection of the modes is difficult
from ground-based photometry alone.
For comparison, \citet{bartz2022} reported this sort of variability
in the space-based TESS photometry for up to 98\% of stars.

\paragraph{Long-term variability of quiescence.}

\label{p10}
We detected the LTQ variability in $(19\pm 6)$\% of stars.
Objects with the LTQ variability,
which increase brightness in the course of time,
have the lowest (${\lesssim}{-}0.9$\,mag)
index \ub,
as if the star is not obscured.
Among them, we can find two especially interesting objects,
$\gamma$~Cas, which emits hard X-rays
(arising from accretion onto the WD companion, 
\citealt{Naze_2026A&A...707A.334N}), and
$\delta$~Sco, which is a~highly eccentric binary,
where the disc has been perturbed by periastron passage
\citep{rast2024}.

Objects whose brightness is decreasing with time
span the entire range of $M_V$.
Their number (7 vs 4) suggests that this phase 
is relatively longer.
In contrast, the objects whose brightness is increasing with time 
appear to cluster near $M_V \simeq -4.0$\,mag,
although this trend is based on only three objects,
two of which are peculiar
($\gamma$~Cas, $\delta$~Sco).
Specifically, for EW~Lac, 
\citet{hec2022} detected an increase in brightness
from $V = 5.6$ to $\sim$5.2\,mag
over about 70\,yr.
We determined the rates for other objects, see Table~\ref{stars_inc_br_tab}.

\begin{table}[h!]
\centering
\caption{Stars with increasing brightness (in mmag\,yr$^{-1}$).}
\label{stars_inc_br_tab}
\begin{tabular}{|l|c|}
\hline
EW~Lac       & $-6.5$ \\
$\delta$~Sco & $-4.0$ \\ 
$\gamma$~Cas & $-2.3$ \\ 
V832~Cyg     & $-1.8$ \\ 
\hline
\end{tabular}
\end{table}

For $\varphi$~And, on the other hand,
we observed the decrease in brightness
from $V = 4.26$\,mag to ${\sim}4.28$\,mag
in 40\,yr
(as shown in Fig.~\ref{phiand})
We emphasise that detecting 0.02\,mag decrease over four decades
was only possible due to the exceptional stability of the Hvar \ubv\ photometry.
Thus, we determined the rates for other objects (see Table~\ref{stars_dec_br_tab}).%
\footnote{Cf. $\omega$~CMa
from \citet{Ghoreyshi_2018MNRAS.479.2214G}
has $+7.8\,{\rm mmag}\,{\rm yr}^{-1}$
and $M_V = -3.2$\,mag,
so it resembles V1294~Aql.}

\begin{table}[h!]
\centering
\caption{Stars with decreasing brightness (in mmag\,yr$^{-1}$).}
\label{stars_dec_br_tab}
\begin{tabular}{|l|c|}
\hline
V1294~Aql     & $+6.0$  \\ 
V2048~Oph     & $+4.5$  \\
$\kappa$~Dra  & $+2.1$  \\ 
$\varphi$~Per & $+1.4?$ \\ 
V696~Mon      & $+1.3$  \\ 
V744~Her      & $+1.1$  \\ 
$\varphi$~And & $+1.0$  \\
\hline
\end{tabular}
\end{table}

\vskip\baselineskip
In a forthcoming paper,
we shall investigate whether the observed LTQ variability
and the triplicity of some Be stars
(Table~\ref{triplicity};
\citealt{Hutter_2021ApJS..257...69H,Dodd_2024MNRAS.527.3076D,Kalari_2025ApJ...993..192K})
suggests the following evolutionary sequence:
\label{p11}
(i)~increase in the semi-major axis~$a$ due to mass transfer ($\beta$~Lyr);
(ii)~increase in the eccentricity~$e$ due to interaction with a tertiary;
(iii)~subsequent secondary's interaction with the outer Be star disc,
accompanied by a slow increase of brightness in quiescence (EW~Lac);
(iv)~periastron passages through the inner Be star disc ($\delta$~Sco);
(v)~strong disc--secondary interactions leading to circularisation ($\gamma$~Cas); and
(vi)~relaxation of the system,
accompanied by a slow decrease in brightness in quiescence
(V1294~Aql).

\section{Conclusions}\label{sec:conclusions}

The refinement of the reduction and standardisation of 53 years
of \ubv\ photometry at Hvar has enabled us to address questions
related to various types of variability (Table~\ref{types})
and the short-term temporal evolution of hot emission-line stars.
In Figs.~\ref{Fig10} and \ref{Fig11},
we can see that there are substantial differences among the types, 
in terms of colour (\ub, \bv) 
or absolute magnitude~($M_V$).
Apart from the known inclination dependence,
the types of variability appear to be dependent on 
intrinsic stellar properties
($M_V$, or spectral type).
For instance,
several types of variability
(i.e. LTE-positive, LTC- or LTQ-increasing)
are absent in stars fainter than $M_\mathrm{V} {\sim}{-}1\,{\rm mag}$.
It appears to be the case that
every Be disc carries the signature of its host star.

\clearpage

We propose that the LTE and the long-term cyclic LTC
variabilities are almost mutually exclusive,
possibly due to a more versus a less extended disc,
which is perturbed by the inner Lindblad resonance of the secondary
(Sec.~\ref{sec:results}, p.~\pageref{p9}). We also note LTQ as the case
where objects increase or decrease in brightness
(from $-6.5$ to $+6.0\,{\rm mmag}\,{\rm yr}^{-1}$). However, it remains unclear 
whether this is related to the existence of a tertiary component
and its possible interaction with a secondary
embedded in a disc around the primary
(Sec.~\ref{sec:results}, p.~\pageref{p11}),
or whether there are other processes involved.

This would have further interesting implications.
If the secondary (sdO) were indeed found to be an efficient sink of angular momentum,
for instance, due to the Lindblad resonance
\citep{Lindblad_1964ApNr....9..103L,Okazaki_2002MNRAS.337..967O}
creating the inner spiral arm. This would help accrete the disc back onto the primary 
(Be) star and might be related to the differing values
of the viscosity parameter,~$\alpha$
\citep{Ghoreyshi_2018MNRAS.479.2214G,ghore2021},
because the coupling of the disc and the secondary
depends on the extent of the disc.
Moreover, some of the secondaries could be misaligned
with the primary's equator (and its disc),
due to the von Zeipel–Lidov-Kozai mechanism
\citep{vonZeipel_1910AN....183..345V,Lidov_1961,Kozai_1962AJ.....67..591K}
induced by a tertiary.
This might relate to the differing orientations of the $\gamma$~Cas disc
\citep{Hummel_1998A&A...330..243H,Baade_2023A&A...678A..47B},
because a misaligned secondary (${\sim}1\,M_\odot$)
induces precession, $\dot\Omega$, of the order of
$10\,^\circ\,{\rm yr}^{-1}$
and `aligns' a~low-mass, $10^{-7}\,M_\odot$ disc with itself.

\paragraph{Archive of Hvar observations.}
In general, 53 years of UBV monitoring at the Hvar Observatory
have revealed that Be stars, as a class:
i)~exhibit diverse types of long-term variability,
which have been systematically documented;
ii)~display characteristic behaviour in colour–colour diagrams,
where their motion during variability cycles
exhibits either positive or inverse correlation;
iii) have such photometric signature because of disc inclination with respect to observer,
which is a geometrical property, not intrinsic to discs \citep{hec83};
iv)~in several cases undergo orbital modulation of photometric brightness
(e.g. V832~Cyg, $\varphi$~Per, CX~Dra, KX~And, V360~Lac,
BR CMi, HD~81357),
which has led to the detection of several semidetached interacting binaries
among known Be stars,
although such systems remain underabundant among classical Be stars;
v)~in numerous cases exhibit decades-long evolution of photometric brightness,
as described by accurate Johnson magnitudes,
very useful in multi-technique modelling,
where such constraints contribute to the determination
of physical and orbital parameters.
 
In fact, the Hvar photometry constitutes the longest homogeneous
ground-based Johnson \ubvror\ time-series dataset
obtained within a single observing programme.
The quality of the measurements,
based on very well-defined magnitudes of comparison, check, and red stars, 
is exceptional owing to the reduction procedure, 
especially considering the observatory’s low altitude of only 200\,m above sea level.
Although space-based photometric missions provide superior cadence, superior precision, and higher multiplex,
the Hvar photometry offers multi-band coverage.
If combined in multi-dataset modelling,
the accurately calibrated Hvar data help to establish 
the photometric zero-point of space-based surveys (if the observations were contemporaneous).

Since a significant fraction of classical Be stars show
slow, large-amplitude photometric variability,
it can be used as a (non-exclusive) diagnostic 
for their identification
\citep{Eyer_2019A&A...623A.110G,Gao2025A&A...694A.131G}.
To identify Be stars in other surveys,
we can also use colour--colour, HR diagrams, and light curves
(Figs.~\ref{Fig10}, \ref{Fig11}, Appendix~\ref{sec:comments})
of Hvar's Be stars as templates,
including their types of variability.
We can either perform photometric transformations
to construct equivalent diagnostic diagrams in other filter systems
or use the observed magnitudes of Hvar's Be stars
in other passbands and obtain similar templates.
Such templates might then be applied to space- or ground-based surveys
such as 
Gaia \citep{gaia1},
Sloan \citep{Eisenstein_2011}, 
Vera Rubin \citep{Ivezic_2019ApJ...873..111I}, 
Pan-STARRS1 \citep{Chambers_2016arXiv161205560C},
Dark Energy \citep{DES}, 
SkyMapper \citep{Keller_2007PASA...24....1K}, or
VISTA \citep{Sutherland_2015A&A...575A..25S}. 

\paragraph{Predictions and synthetic light curves.}
Moreover, the long time span enables predictions
of more regular brightening events
and the most interesting phases of disc evolution.
These information is useful for planning proposals for spectroscopic or interferometric
(imaging) observations. For instance, we can predict for V744~Her
that the fifth minimum in $V$ will probably occur around 2060
(Fig.~\ref{v744her}),
if its activity is extrapolated assuming no change in behaviour.
On the contrary, for V923~Aql, one can expect the next \ub\ maximum
coming in just several months
(Fig.~\ref{v923aql_cycles}).
For the subset of Be stars showing the LTQ variability,
continuous monitoring will allow us to falsify the null hypothesis
that the observed increase or decrease in brightness continues as a linear trend; 
alternatively, it is perhaps more likely that it does not continue and remains episodic.

To further interpret the observations,
we have to compute synthetic UBV light curves.
Specifically, in time-dependent viscous decretion disc models,
the commonly fitted parameters include 
the \citet{Shakura_1973A&A....24..337S} viscosity parameter,~$\alpha$,
the density normalisation,~$\rho_0$,
the density exponent,~$n$ (where $\rho(r)\propto r^{-n}$),
the outer truncation radius,~$R_\mathrm{out}$, and
the observer-dependent inclination,~$i$.
Extended time span of observations,
such as the long-term UBV monitoring from Hvar, 
provide the key information on the temporal evolution
or constancy of these parameters
(in particular,~$\alpha$),
as they trace multiple events of disc formation (brightenings)
and dissipation (fadings).
The \ub\ colour is particularly important in probing
the photospheres (if not obscured) and inner Be discs, as it is sensitive to the Balmer jump region.
In conclusion, the archive of Hvar observations provides essential empirical material for modelling Be stars.

\paragraph{Data availability.}\label{data}

Complete all-sky and differential archives of Hvar observations
from the years 1972--2024,
including Tables A.3--A.7,
are available in electronic form at the CDS
via anonymous ftp to cdsarc.u-strasbg.fr (130.79.128.5) or via
\url{https://cdsarc.cds.unistra.fr/viz-bin/cat/J/A+A/712/A226}.
Potential users of these observations should credit this article
in their future publications.
A compressed file with our photometric reduction software `phot.tar.gz'
is available at the CDS via
\url{https://cdsarc.cds.unistra.fr/viz-bin/cat/J/A+A/712/A226}.

\begin{acknowledgements}
H. Bo\v{z}i\'c acknowledges financial support from the Croatian Science Foundation under
the project 6212 `Solar and Stellar Variability'.
P.~Harmanec, M.~Wolf, A.~Opli\v{s}tilov\'a, J.~Jon\'ak, and K.~Vitovsk\'y
were supported by grant GA19-01995S of the Czech Science Foundation.
M.~Bro\v{z} was supported by grant 25-16507 of the Czech Science Foundation.
Research of P.~Hadrava was supported by project RVO~67985815.
Our colleagues J.~Grygar and K.~Pavlovski were associated with the whole
program and carried out a number of observations.
Moreover, K.~Pavlovski helped us to check and correct some of his observations.
A~large number of observations have been secured by J.~Nemravov\'a-Bene\v{s}ov\'a.
These three colleagues informed us that they do not want to be listed
among the co-authors, which we have to respect.
We credit them for their significant contributions.
Our thanks go to our colleagues and students
A.~Arena, J.~Arsenijevi\'c,
R.~Braj\v{s}a, R.~Bro\v{z}, P.~Chadima, D.~\v{C}ikoti\'c, E.~Frle\v{z},
K.~Ho\v{n}kov\'a, Z.~Ivanovi\'c, \v{Z}.~Ivezi\'c, J.~Jan\'\i k,
J.~Jury\v{s}ek, B.~Jovanovi\'c, G.~Lazin, M.~Malari\'c, A.~Muli\'c,
M.~Muminovi\'c, M.~Netopil, D.~Ond\v{r}ich,
D.~Pla\v{c}ko-Vr\v{s}nak, P.~Polechov\'a, \v{Z}.~Ru\v{z}i\'c,
I.~Vince, B.~Vr\v{s}nak, and M.~Zejda
who participated in observations at Hvar;
Our special thanks go to J.~Vacek for his design,
maintenance, and regular servicing of the Hvar photoelectric photometer.
J.R.~Percy kindly put the archive of his \ubv\ observations of bright
Be stars at our disposal.
Detailed comments of the referee, Dietrich Baade, helped us to
re-think and restructure parts of this study.
The following internet-based resources were consulted:
the SIMBAD database and the VizieR service operated at
CDS, Strasbourg, France; and the NASA's Astrophysics Data System
Bibliographic Services.
\end{acknowledgements}

\bibliographystyle{aa}
\bibliography{hvar6}

\begin{appendix}

\onecolumn

\section{Supplementary tables}

\renewcommand{\arraystretch}{1.0}

\begingroup
\footnotesize

\begin{longtable}{r@{\ }r@{\ }rrllllllll}
\caption{
Complete list of 105 early-type emission-line stars observed at
the Hvar Observatory.
}
\label{tab03}
\\
\hline\hline\noalign{\smallskip}
Star            &   & HD/BD    & No.     & Comp.         & Check         & Spectral      & $P$       & $e$   & Ref. & $\pi$ & $A_V$ \\
name            &   & number   & of obs. & HD/BD         & HD            & type          & [days]    & [1]     & --   & [mas]   & [mag]   \\
\hline\noalign{\smallskip}
\endfirsthead
\caption{continued}\\
\hline\hline\noalign{\smallskip}
Star            &   & HD/BD    & No.     & Comp.         & Check         & Spectral      & $P$       & $e$   & Ref. \\
name            &   & number   & of obs. & HD/BD         & HD            & type          & [days]    & [1]     & --   \\
\hline\noalign{\smallskip}
\endhead
\hline
\endfoot
$\omicron$ Cas  & * &     4180 &  895 &      4142        &   6114        & B5III-IVe     & 1031.55   & 0.0   & 2  & 3.7   & 0.192 \\
$\gamma$ Cas    & \ &     5394 &  137 &      2626        &   2011        & B0.5IVe       & 203.52    & 0.0  & 3  & 5.94  & 0.071 \\
V442 And        & * &     6226 &  964 &      4142        &   6114        & B3.5IIIe      & -         & -     & -  & 0.815 & 0.290 \\
$\varphi$ And   & * &     6811 &  504 &      4142        &   6114        & B7IIIe        & -         & -     & -  & 4.55  & 0.111 \\
$\varphi$ Per   & * &    10516 &  678 &     12303        &  11291        & B0.5IVe+sdOB  & 126.6731  & 0.0   & 7  & 5.404 & 0.065 \\
V554 Per        & \ &    14818 &  146 &     12303        &  11291        & B2Iae         & -         & -     & -  & 0.442 & 0.952 \\
HR 894          & \ &    18552 &  118 &     18411        &  19736        & B7IVe         & -         & -     & -  & 4.462 & 0.069 \\
RX Cas          & \ & +67\,244 &  361 &     18962        &  19556,19193  & B3IIIe+K1III  & 32.3301   & 0.0   & 6  & 1.817 & 0.653 \\
13 Tau          & \ &    23016 &  114 &     23324        &  23288        & B7Ve          & -         & -     & -  & 6.275 & 0.042 \\
17 Tau          & * &    23302 &  367 &     23324        &  23288        & B6IIIe        & -         & -     & -  & 8.345 & 0.029 \\
V971 Tau        & \ &    23480 &  265 &     23324        &  23288        & B6IVe         & -         & -     & -  & 7.067 & 0.675 \\
$\eta$ Tau      & \ &    23630 &  205 &     23324        &  23288        & B7IIIe        & -         & -     & -  & 8.09  & 0.179 \\
BU Tau          & * &    23862 &  469 &     23324        &  23288        & B8Ve          & 218.34    & 0.0   & 8  & 7.241 & 0.066 \\
V960 Tau        & * &    36576 &  392 &     36589        &  37711,36819  & B2IV-Ve       & -         & -     & -  & 2.540 & 0.749 \\
$\zeta$ Tau     & * &    37202 & 1100 &     36589        &  37711,36819  & B2IVe         & 132.987   & -     & 9  & 7.33  & 0.043 \\
$\omega$ Ori    & \ &    37490 &   90 &     36591        &  36351        & B3Ve          & -         & -     & -  & 1.976 & 0.342 \\
V731 Tau        & \ &    37967 &  107 &     36589        &  37711        & B4Ve          & -         & -     & -  & 3.619 & 0.212 \\
V696 Mon        & * &    41335 &  318 &     42690        &  45546,43023  & B1Ve          & 80.913    & 0.0   & 10 & 2.000 & 0.461 \\
HR 2418         & \ &    47054 &   63 &     42690        &  45546,43023  & B8IVe         & -         & -     & -  & 4.376 & 0.145 \\
OT Gem          & \ &    58050 &  437 &     58187        &  59059        & B2Ve          & -         & -     & -  & 2.212 & 0.089 \\
$\beta$ CMi     & \ &    58715 &  168 &     58187        &  59059        & B8Ve          & -         & -     & -  & 20.17 & 0.001 \\
BR CMi          & \ &    61273 &  103 &     58187        &  61341        & B9.5e+G8III   & 12.919    & 0.0   & 13 & 5.847 & 0.018 \\
UX Mon          & \ &    65607 &  152 &     65199        &  65005        & A5IIIe+G2III  & 5.9044365 & 0.0   & 14 & 1.798 & 0.127 \\
HD 81357        & \ &    81357 &   93 &     82861        &  77692,81772  & B8e           & 33.77458  & 0.0   & 15 & 1.708 & 0.055 \\
$\kappa$ Dra    & * &   109387 &  429 &    107193        & 115612,104316 & B6IIIe        & 61.5549   & 0.0   & 16 & 7.009 & 0.021 \\
$\vartheta$ CrB & \ &   138749 &  158 &    138341        & 136849        & B6Vnne        & -         & -     & -  & 8.213 & 0.017 \\
V839 Her        & * &   142926 &  685 &    142926        & 145389,141930 & B7e           & 46.1921   & 0.0   & 17 & 6.042 & 0.042 \\
$\delta$ Sco    & \ &   143275 &   98 &    144470        & 142096        & B0.3IVe       & 3950.8    & 0.938 & 18 & 6.64  & 0.049 \\
$\zeta$ Oph     & \ &   149757 &  168 &    148367        & 147550        & O9.2IVnne     & -         & -     & -  & 7.408 & 0.041 \\
V744 Her        & * &   162732 & 1449 &    158414,162132 & 162579        & B6IIIe        & 86.7221   & 0.16  & 20 & 3.152 & 0.085 \\
V2048 Oph       & \ &   164284 &  164 &    164432        & 163641        & B2Ve          & -         & -     & -  & 4.904 & 0.071 \\
V974 Her        & \ &   164447 &  183 &    166182        & 166230        & B8IVe         & -         & -     & -  & 2.396 & 0.149 \\
$\omicron$ Her  & \ &   166014 &  182 &    166182        & 166230        & B9.5IIIe      & -         & -     & -  & 9.348 & 0.021 \\
NW Ser          & \ &   168797 &  327 &    170200        & 169578        & B2Vne         & -         & -     & -  & 2.388 & 0.292 \\
CX Dra          & * &   174237 & 1167 &    173664        & 172883        & B3e+F5III     & 6.696     & 0.0   & 21 & 2.826 & 0.087 \\
$\beta$ Lyr     & \ &   174638 &  544 &    176437        & 174602,172044 & B8II+B0e::    & 12.913779 & 0.0   & 22 & 3.598 & 0.047 \\
7 Vul           & \ &   183537 &  131 &    188260        & 184606        & B4-5IVe       & 69.4212   & 0.113 & 23 & 3.582 & 0.342 \\
V923 Aql        & * &   183656 & 1620 &    183227        & 184663        & B6IIIe        & 214.716   & 0.0   & 24 & 3.545 & 0.333 \\
V1294 Aql       & * &   184279 & 1709 &    183227        & 184663        & B0.5IVe       & 192.91    & 0.0   & 25 & 0.705 & 0.587 \\
V1507 Cyg       & \ &   187399 &  193 &    188170        & 186357        & B8III+Be      & 27.9705   & 0.389 & 26 & 1.057 & 0.615 \\
V1746 Cyg       & \ &   189687 &  209 &    188892        & 193369        & B3IVe         & -         & -     & -  & 2.442 & 0.155 \\
V1624 Cyg       & \ &   191610 &  510 &    188892        & 193369        & B2.5Ve+sdO    & 359.98    & 0.0   & 27 & 3.892 & 0.097 \\
20 Vul          & \ &   192044 &  130 &    190993        & 191747        & B7Ve          & -         & -     & -  & 3.187 & 0.063 \\
QR Vul          & \ &   192685 &  191 &    190993        & 191747        & B3Ve          & -         & -     & -  & 5.691 & 0.053 \\
P Cyg           & \ &   193237 &  122 &    188892        & 193369        & B1-2Iae       & -         & -     & -  & 0.625 & 1.061 \\
25 Vul          & \ &   193911 &  124 &    190993        & 191747        & B6IVe         & -         & -     & -  & 2.597 & 0.148 \\
V2119 Cyg       & \ &   194335 &  140 &    188892        & 193369        & B2IIIe        & 63.146    & 0.0   & 28 & 2.712 & 0.060 \\
HR 7843         & \ &   195554 &  161 &    194668        & 197618        & B8.5Ve        & -         & -     & -  & 3.271 & 0.071 \\
V1661 Cyg       & \ &   198478 &  288 &    203245        & 199311        & B2.5Iae       & -         & -     & -  & 0.543 & 1.485 \\
V832 Cyg        & * &   200120 &  849 &    203245        & 199311        & B1.5Ve+sdOB   & 28.1871   & 0.144 & 34 & 1.468 & 0.212 \\
V1931 Cyg       & \ &   200310 &  889 &    203245        & 199311        & B1Ve          & 147.617   & 0.20  & 5  & 2.662 & 0.098 \\
8 Lac\,A        & \ &   214168 &  151 &    217101        & 214680        & B1Vne         & -         & -     & -  & 1.891 & 0.304 \\
V360 Lac        & \ &   216200 &  424 &    217101        & 214680        & B3IIIe+F9IV   & 10.0854   & 0.0   & 30 & 2.026 & 0.458 \\
EW Lac          & * &   217050 & 1281 &    218470        & 219080,212593 & B4IIIpe       & -         & -     & -  & 3.479 & 0.093 \\
V378 And        & \ &   217543 &  267 &    217101        & 214680        & B2.5Vne       & -         & -     & -  & 2.686 & 0.252 \\
$\omicron$ And  & * &   217675 & 1636 &    217101        & 214680        & B6Ve          & 2525      & 0.24  & 35 & 9.305 & 0.045 \\
KX And          & * &   218393 & 1210 &    218470        & 219080,212593 & B0.5e+K1III   & 38.919    & 0.0   & 31 & 1.317 & 0.160 \\
KY And          & \ &   218674 &  970 &    218470        & 219080,212593 & B4Ve          & -         & -     & -  & 1.713 & 0.343 \\
LQ And          & \ &   224559 &  590 &    223229        & 222439,224342 & B4Ve          & -         & -     & 32 & 2.571 & 0.162 \\
\hline\noalign{\smallskip}
10 Cas          & \ &      144 &   18 &      2626        &   2011        & B9IIIe        & -         & -     & -  \\
V742 Cas        & \ &      698 &    6 &      2626        &   2011        & B5II-IIIe+BVI & 55.9233   & 0.0   & 33 \\
$\kappa$ Cas    & \ &     2905 &   17 &      2626        &   2011        & B1Iae         & -         & -     & -  \\
HD 9709         & \ &     9709 &    7 &      4142        &   6114        & B8Ve          & -         & -     & -  \\
V777 Cas        & \ &    11606 &   11 &     12303        &  11291        & B2Ve          & -         & -     & -  \\
V780 Cas        & \ &    12302 &    9 &     12303        &  11291        & B1:V:pe+sdOB  & -         & -     & 5  \\
HR 654          & \ &    13854 &   13 &     12303        &  11291        & B1Iabe        & -         & -     & -  \\
HR 1051         & \ &    21551 &   30 &     21278        &  24546        & B8Ve          & -         & -     & -  \\
$\psi$ Per      & \ &    22192 &   29 &     21278        &  24546        & B5Ve          & 126.6982  & -     & 7  \\
HR 1113         & \ &    22780 &   41 &     21856        &  23193        & B7Ve          & -         & -     & -  \\
HR 1160         & \ &    23551 &   17 &     21278        &  24546        & B8Ve          & -         & -     & -  \\
MX Per          & \ &    25940 &   32 &     21278        &  24546        & B3Ve          & -         & -     & -  \\
HR 1500         & \ &    29866 &   30 &     33641        &  29722        & B8Ve          & -         & -     & -  \\
BV Cam          & \ &    32343 &    9 &     39283        &  34787        & B3Ve          & -         & -     & -  \\
$\lambda$ Eri   & \ &    33328 &   24 &     32249        &  33224        & B2IIIep       & -         & -     & -  \\
HR 2231         & \ &    43285 &   36 &     44783        &  43526        & B6Ve          & -         & -     & -  \\
AX Mon          & \ &    45910 &   40 &     44783        &  43526        & B2IIIe+K2II   & 232.499   & 0.0   & 6  \\
HR 2370         & \ &    45995 &    7 &     44783        &  43526        & B1.5Vne       & 103.1     & 0.0   & 11 \\
HD 46150        & \ &    46150 &    3 &     44783        &  43526        & O5V((f))z     & -         & -     & -  \\
$\psi^9$ Aur    & \ &    50658 &   29 &     49949        &  52860,50860  & B8IIIe        & -         & -     & -  \\
AU Mon          & \ &    50846 &   15 &     50109        &  50169        & B5V+F0        & 11.113    & 0.0   & 12 \\
HR 3135         & \ &    65875 &   24 &     63975        &  71155        & B2Ve          & -         & -     & -  \\
$\chi$ Oph      & \ &   148184 &   11 &    144470        & 142096        & B1.5Ve        & 34.121    & 0.26  & 19 \\
V2315 Oph       & \ &   161261 &    5 &    161677,169420 & ----          & hB8VekA0      & -         & -     & -  \\
V4024 Sgr       & \ &   178175 &    1 &    177817        & 182678,182645 & B2Ve          & -         & -     & -  \\
$\upsilon$ Sgr  & \ &   181616 &   46 &    177817        & 182678,182645 & B2Vpe         & 137.9343  & 0.0   & 1  \\
HD 183261       & \ &   183261 &    1 &    184606        & 188260        & B3IIe         & -         & -     & -  \\
$\beta$2 Cyg    & \ &   183914 &   14 &    188260        & 184606        & B8Ve          & -         & -     & -  \\
HR 7482         & \ &   185859 &   38 &    188260        & 184606        & B3I           & -         & -     & -  \\
V395 Vul        & \ &   187811 &   39 &    188260        & 184606        & B2.5Ve        & -         & -     & -  \\
V2120 Cyg       & \ &   194883 &   74 &    194668        & 197618        & B2Ve          & -         & -     & -  \\
HR 7983         & \ &   198625 &   78 &    203245        & 199311        & B4Ve          & -         & -     & -  \\
V2140 Cyg       & \ &   199478 &   90 &    203245        & 199311        & B7Iae         & -         & -     & -  \\
HR 8103         & \ &   201733 &   71 &    203245        & 199311        & B4IVe         & -         & -     & -  \\
$\upsilon$ Cyg  & \ &   202904 &    6 &    202349        & 204403        & B2Ve          & -         & -     & -  \\
HR 8153         & \ &   203025 &    8 &    208218        & 202214        & B2IIIe        & -         & -     & -  \\
V382 Cep        & \ &   203467 &   82 &    208218        & 202214        & B3IVe         & -         & -     & -  \\
HR 8259         & \ &   205551 &   21 &    207330        & 206259,207793 & B7IIIe        & -         & -     & -  \\
$\epsilon$ Cap  & \ &   205637 &    2 &    144206,213420 & ---           & B3IIIe        & 128.3     & 0.0   & 29 \\
HD 206773       & \ &   206673 &   57 &    208218        & 202214        & B0.5V:pe      & -         & -     & -  \\
EM Cep          & \ &   208392 &   10 &    208218        & 202214        & B1Vne         & -         & -     & -  \\
HR 8375         & \ &   208682 &    8 &    208218        & 202214        & B2Ve          & -         & -     & -  \\
HR 8682         & \ &   216057 &  205 &    218470        & 219080,212593 & B7Ve          & -         & -     & -  \\
V639 Cas        & \ &   225094 &   22 &      2626        &   2011        & B2.9Iabe      & -         & -     & -  \\
MWC 327         & \ &   227611 &    6 & +35\,3955        & 190919        & B1:III/Ve     & -         & -     & -  \\
V1322 Cyg       & \ &   229221 &    2 &    229234        & 229238        & B0.2IIIe      & -         & -     & -  \\
\end{longtable}
\tablefoot{
Columns are the numbers of stars' observations,
the comparison and check stars used,
the MK types \citep{Wenger_2000A&AS..143....9W},
the orbital period~$P$ and
the eccentricity~$e$ (if known)
with a reference to the orbital solution,
the parallax~$\pi$ \citep{Vallenari_2023A&A...674A...1G}, and
the extinction~$A_V$ \citep{Green_2018MNRAS.478..651G}.
Column `References':
1...\citet{zarf25}; 2...\citet{zarf28}; 3...\citet{zarf29}; 4...\citet{bozic95}; 
5...\citet{klement2024}; 6...\citet{hec2001}; 7...\citet{mourard2015}; 8...\citet{zarf27}; 
9...\citet{zarf26}; 10...\citet{peters2016}; 11...\citet{naze2022}; 12...\citet{sahade82}; 
13...\citet{zarf30}; 14...\citet{sudar2011}; 
15...\citet{zarf31}; 16...\citet{zarf16}; 17...\citet{zarf18}; 18...\citet{tycner2011}; 
19...\citet{hec87c}; 20...\citet{zarf4}; 21...\citet{cxdrahorn}; 22...\citet{mourard2018}; 
23...\citet{hec2020}; 24...\citet{wolf2021}; 25...\citet{hec2022}; 26...\citet{hutchings73}; 
27...\citet{hec2025}; 28...\citet{klement2022a};  
29...\citet{rivi2006}; 30...\citet{zarf24}; 31...\citet{floquet89}; 32...\citet{jaymie91}; 
33...\citet{rivi2025}; 34...\citet{peters2013}; 35...\citet{grant88}.
* denotes objects discussed in Appendix~\ref{sec:comments}. The last two columns are filled
in for the stars discussed in this paper (above the horizontal line).
}
\endgroup

\twocolumn

\begin{table}[h!]
\caption{
Hvar programme stars that have a third component.
}
\label{triplicity}
\centering
\small
\begin{tabular}{|@{\ }l@{\ }|@{\ }l@{\ }|@{\ }l@{\ }|}
\hline
$\gamma$ Cas    & 00567+6043 & Aa, Ab (0.019), B (2.1) \\ 
BU Tau          & 03492+2408 & Aa, Ab (0.2), F (4.7) \\
$\beta$ Lyr     & 18501+3322 & Aa1, Aa2 (0.0008), Ab (0.54), B (46) \\ 
V832 Cyg        & 20598+4731 & Aa1, Aa2 (0.0011), Ab (0.16) \\
V1931 Cyg       & 21012+4609 & Aa, Ab (<0.2), B (2.9) \\ 
$\omicron$~And  & 23019+4220 & Aa, Ab (0.05), Ba+Bb (0.33) \\ 
\hline
\end{tabular}
\tablefoot{
The Washington Double Star (WDS) catalogue
\citep{Mason_2001AJ....122.3466M}
designations and angular separations (in~$''$).
Out of 59 Be stars,
at least 6 have a more distant tertiary component.
Their presence is important because of possible
interactions within the system
(e.g. \citealt{Broz_2026arXiv260522545B}).
We note that
\citet{Hutter_2021ApJS..257...69H} and \citet{Kalari_2025ApJ...993..192K}
confirmed the triplicity of $\beta$~Lyr, V1931 Cyg,
as well as of other Be stars not observed at Hvar.
}
\end{table}

\vskip\baselineskip
\noindent
The following tables are only available at the CDS:

\smallskip
\noindent
Table A.3:
Archive of all \ubv\ observations of 106 hot emission-line stars
and binaries discussed in this paper.

\smallskip
\noindent
Table A.4:
Robust mean values of the $V$ magnitude
and the \bv, \ub, and \vr\ indices
for the ten primary Johnson standards
\citep{john66}.

\smallskip
\noindent
Table A.5:
Robust mean values of the $V$ magnitude
and the \bv, \ub, and \vr\ indices
of seven selected secondary standards
\citep{john54}.

\smallskip
\noindent
Table A.6:
Standard Johnson \ubv\ magnitudes of all
comparison, check, and red standard stars used at Hvar,
which have at least 20 observations
secured in two or more seasons.

\smallskip
\noindent
Table A.7:
Standard Johnson \ubv\ magnitudes of less frequently observed
comparison, check, and red standard stars used at Hvar.

\newdimen\tmpdim
\tmpdim=8cm

\section{Comments on individual stars}\label{sec:comments}

In the following, we comment on selected stars
that were observed more systematically at Hvar
and exhibit interesting long-term temporal variability.
Because the observations at Hvar were interrupted over 1990-1992,
we complemented the Hvar $V$ series in the $V$ magnitude
time plots with the Hipparcos \hp\ photometry (red triangles),
transformed to Johnson $V$ magnitude after \citet{hpbv}.
For several particularly interesting objects,
we show new, preliminary reduced data obtained in 2025
(blue crosses).

\paragraph{$\omicron$~Cas = 22~Cas = HD~4180.}
This bright Be star is
the primary component A of a wide visual system AB WDS~J00447+4817.
\citet{zarf28} confirmed that it is also a spectroscopic binary with a~period
of 1031\fd6 and a rather large RV semi-amplitude of 22\,\ks, which implies
a~large mass function. They also obtained a visual orbit with an~inclination
of 115$^\circ$ and since they did not detect any lines of the secondary,
they suggested that the secondary is actually a close pair of two sharp-lined stars
(See also \citealt{grund2007,tou2013, hutter2021}.)
Recently, \cite{Harmanec_2026A&A...709A.167H} found that the true
period of the inner system is 11\fd66 and demonstrated that the
object offers a unique opportunity to derive a very
accurate mass of the Be star purely on the dynamical grounds.

In the Hvar archive,
we publish an extended series of photometric observations.
It records two complete LTE cycles
which were $\sim$18.8 and 16.2\,yr long;
the third cycle
(starting at 2\,457\,500)
is poorly covered.
The overall amplitude in $V$ is 0\m20.
The RLA variability typically reaches $\sim$0\m08
before the maximum light
and only 0\m02
after the maximum light.

\begin{figure}
\centering
\includegraphics[width=\tmpdim]{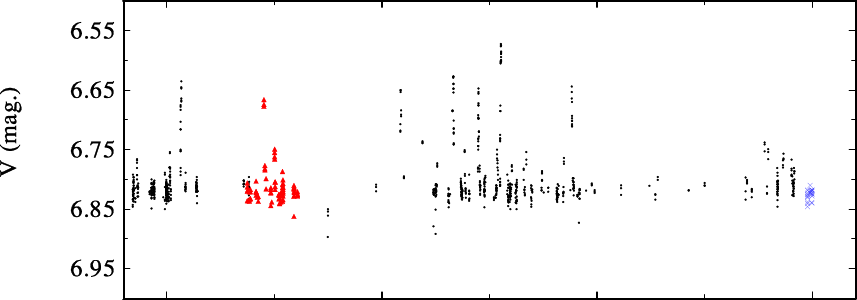}
\includegraphics[width=\tmpdim]{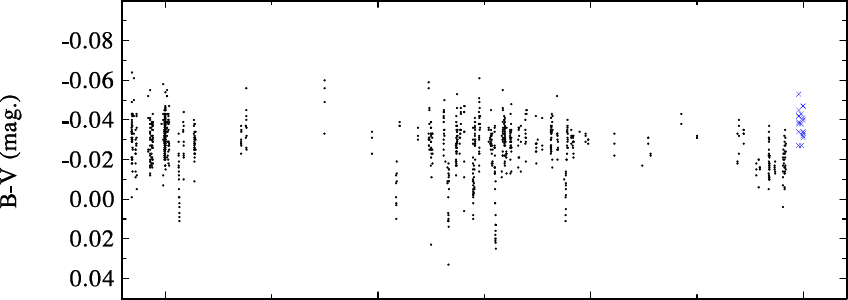}
\includegraphics[width=\tmpdim]{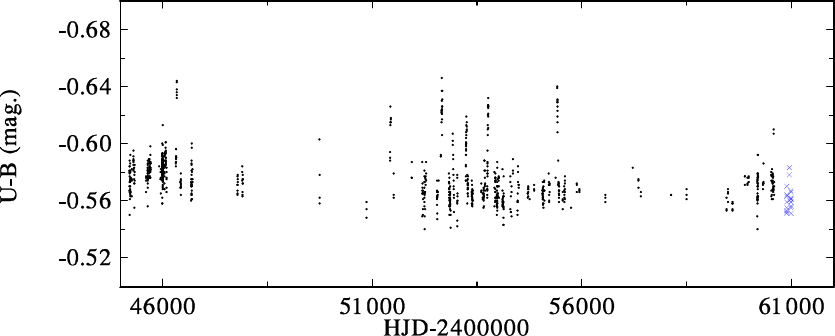}
\caption{\ubv\ time variations (LTEp) of V442~And. 
The \textcolor{red}{red} triangles in this
as well as in all following plots,
denote the Hipparcos \hp\ observations,
transformed to Johnson $V$ magnitude \citep{hpbv}.
The \textcolor{blue}{blue} crosses indicate new data obtained in 2025.
As explained in Sect.~2,
the accuracy of these observations can be estimated from the rms errors
per one observation of the respective check star,
which are provided in Table~A.6.
For the check star HR~4142 observed with V442~And, they are
0\m008, 0\m009, and 0\m011
for the $V$, $B$, and $U$ passbands, respectively.
Usually, the observations were repeated three times 
over an interval of several tens of minutes.
They can be seen in the plots as vertical `streaks'.
They reflect both a measurement uncertainty (0.008--0.016\,mag) and 
intrinsic RLA variability known from published studies.
The correlated redder-when-brighter behaviour
corresponds to a nearly pole-on inclination.
This particular light curve is characterised by intermittent brightenings,
occurring at integer multiples of $P_3$
(see Fig.~\ref{v442and_rectification}).
The disc growth is fast (${\sim}\,{\rm d}$)
and its viscous dissipation is also fast,
occurring on the timescale shorter than the basic cycle period $P_3$.
Otherwise, the system seems to be in steady state.
}
\label{v442and}
\end{figure}

\paragraph{V442~And = HD~6226.}
This star was originally used as one of the check stars
in our Be star observing programme,
and this led to the discovery of its light variability
in the form of occasional brightenings,
accompanied by reddening in \bv\ and bluing in the \ub\ index.
This finding was published by \citet{hrvoje98},
who suggested that the object could be a Be star.
This was confirmed by a detailed spectroscopic and photometric study
of \citet{zarf22},
who found that V442~And is a pole-on Be star
with a positive correlation between brightness and emission-line strength and pronounced line-profile variations,
reminiscent of an archetype Be star $\omega$~CMa.
They demonstrated that every brightening was 
accompanied by a steep rise of the Balmer emission.
Analysing RV measurements of the outer wings and
sharp core of the \ion{He}{i}~6678\,\AA\ absorption,
they found they vary in anti-phase and with a strict
2\fd61507 period, stable over 6~years of observations.
\citet{rich2021} published a~detailed study,
based on numerous spectra and on the KELT and TESS photometries.
They interpreted the 2\fd615 period of line-profile changes as a~non-radial pulsation.
Analysing the \ha and \hb equivalent widths,
they found intermittent brightenings showing two periods, 87\fd0 and 211\fd9.
The former seems to be related to the beating of two close oscillations
with periods of 1\fd41269 and 1\fd39013,
which have low amplitudes (0\m002 to 0\m003),
but sometimes may lead to large-amplitude brightenings.

\begin{figure}
\centering
\includegraphics[width=9cm]{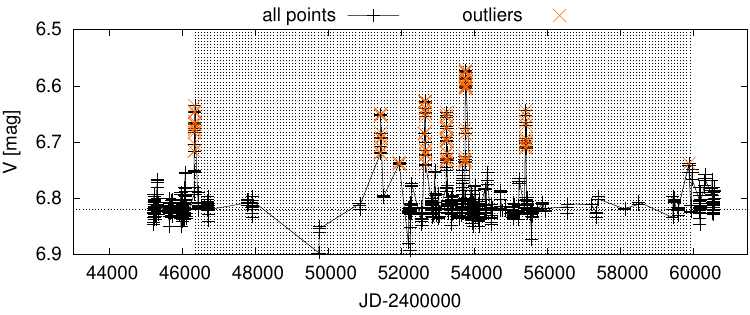}
\caption{
Light curve of V442~And,
with all points (black) and outliers (\textcolor{orange}{orange}),
which were less than 0\m08 from the mean 6\m82.
The vertical dotted grid is plotted at zero phases
of the period $P_3 = 86.477\,{\rm d}$,
corresponding to the brightenings
occurring at integer multiples of $P_3$.
}
\label{v442and_rectification}
\end{figure}

The light curve of V442~And secured at Hvar is shown in Fig.~\ref{v442and}.
Since the \citet{zarf22} publication,
we obtained 20 years of new photometric data (2005-2025).
It turned out that the more recent activity of the star is lower,
and no brightenings ${>}0\m05$ were recorded.
Over the whole time span of 40 years,
the undisturbed brightness level has remained constant.
The prominent photometric period,
$P_3 \simeq 86\fd4$,
is very similar to the spectroscopic one
and is seen also in TWS observations, studied by \citet{rich2021}.
Additionally, the brightenings ${>}0\m05$ occur always
at integer {\em multiples\/} of~$P_3$
(Fig.~\ref{v442and_rectification}).
Therefore, $P_3$ is likely the orbital,
or in Richardson's view, cycle period.
The RV of putative orbital motion of the Be star
with such a period for the pole-on inclination
is below the current detection limit.

\paragraph{$\varphi$~And = 42~And = HD~6811.} 
This Be star is obviously observed pole-on,
shows a~single-peaked \ha emission profile,
little light variability, and
was alternatively classified as B7Ve and B5IIIe \citep{jones2011, barnsley2013}.
It is also a member of the multiple visual system WDS~J01095+4715
for which \citet{muter2010} published an uncertain visual orbit
with a period of about 550\,yr.
Given its angular separation ($0.5''$),
it should be probably interpreted as a tertiary.

\begin{figure}
\centering
\includegraphics[width=\tmpdim]{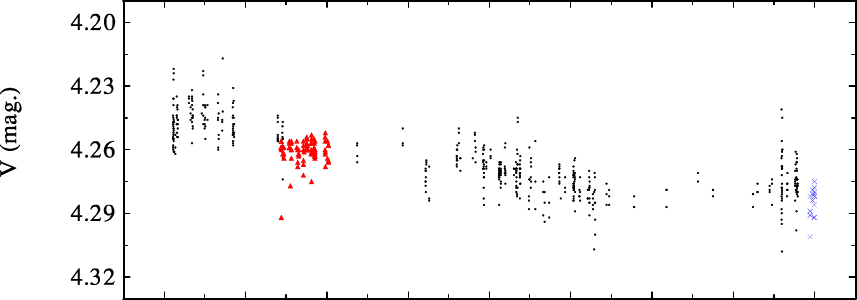}
\includegraphics[width=\tmpdim]{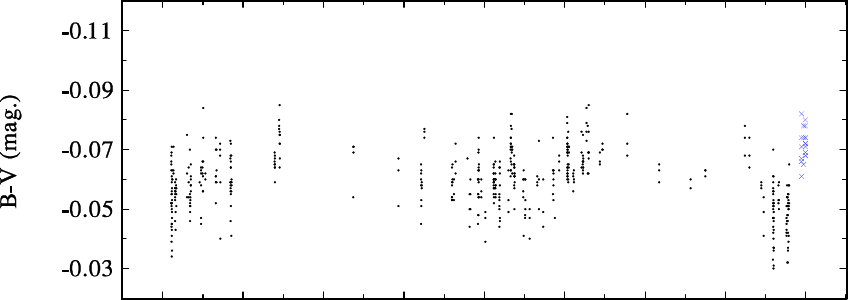}
\includegraphics[width=\tmpdim]{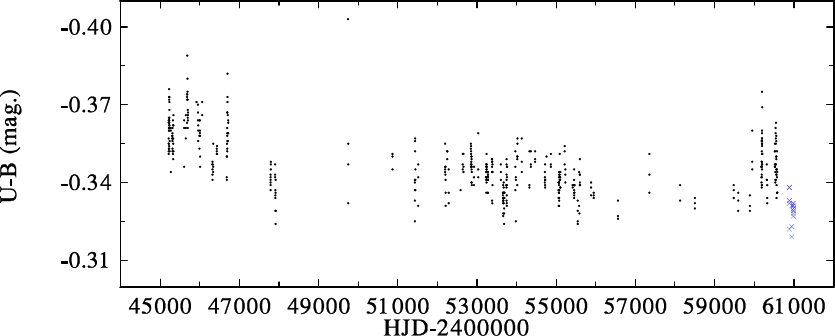}
\caption{\ubv\ time variations (LTQdec) of $\varphi$~And.
The system exhibits a steady slow decrease of brightness, possibly 
because of the long-term changes in the pseudophotosphere.
}
\label{phiand}
\end{figure}

Light and colour variability of $\varphi$~And
would probably go unnoticed
without very systematic monitoring and careful reduction;
see Fig.~\ref{phiand}.
Apart from uncertain RLA variability,
it is characterised by a steady slow light decrease
(by $1.0\,{\rm mmag}\,{\rm y}^{-1}$)
on the longest timescale (LTQ),
with no known episodes of brightenings or fadings.

\paragraph{$\varphi$~Per = 54~And = HD~10516.} 
This object is a B0.5e+O6\,VI binary with a 126\fd7 period 
\citep[see][and references therein]
{gies98}.
\citet{stefl2000} and \citet{hummel2001}
interpreted that the emission in \ion{He}{i}~6678\,\AA\ 
arises in the part of primary's disc region
facing the hotter secondary.
\citet{bozic2013} reported small sinusoidal variation in the $B$ 
magnitude with the orbital period.
Using optical interferometry, \citet{mourard2015} resolved the secondary 
and derived a detailed stellar model, obtaining component masses of 
9.6 and 1.2\,\ms.
The secondary (sdO) is overluminous
due to a short-lived He-shell burning phase \citep{Schootemeijer_2018A&A...615A..30S}.

\begin{figure}
\centering
\includegraphics[width=\tmpdim]{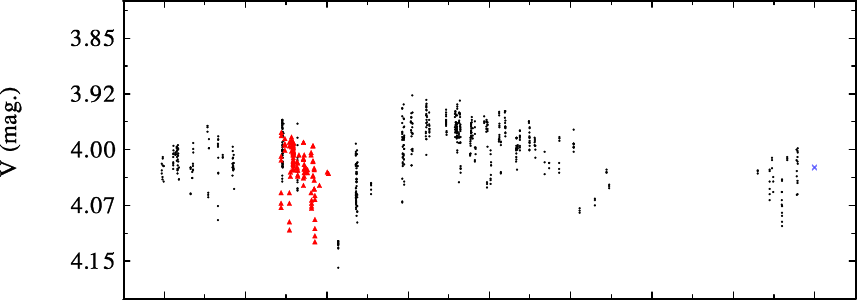}
\includegraphics[width=\tmpdim]{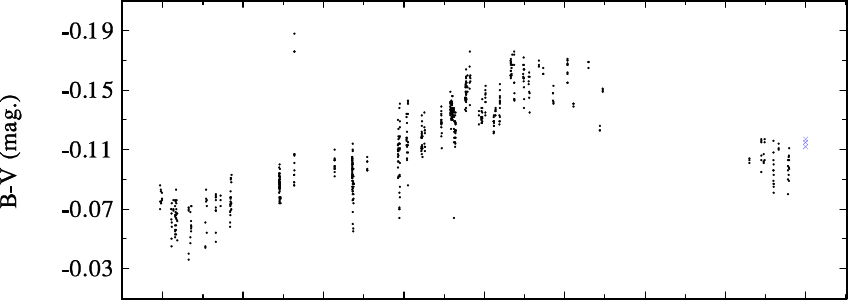}
\includegraphics[width=\tmpdim]{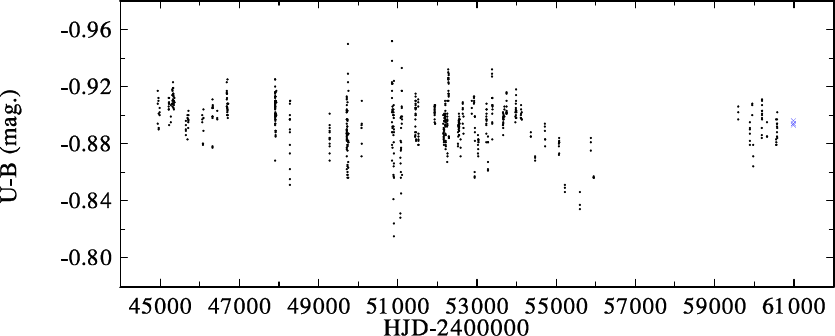}
\caption{\ubv\ time variations (LTEp, LTC, 126\fd7 BIN, LTQdec) 
of $\varphi$~Per.
The most prominent variation is LTE,
evident in $V$ and especially in the \bv\ index.
The disc evolves continuously and relatively slowly,
in terms of the density.
However, the temperature of the visible layers
(or the pseudophotosphere)
seems to be decoupled
(cf.~\bv).
}
\label{phiper}
\end{figure}

The complete Hvar photometry is shown in Fig.~\ref{phiper}.
The light variations are dominated by slow changes
on the LTE timescale.
New data suggest that the $V$ magnitude varies in a cycle of more than 10 000\,d
while the \bv\ index after gradual brightening
(by 0\m10 over the first $\sim$25\,yr),
now gets redder.
This Be star is rather exceptional,
alongside V832~Cyg,
since its \ub\ index is much more stable than \bv.

\paragraph{17~Tau = HD~23302.} 
This bright member of the Pleiades cluster,
also known as Electra, was reported to be a~single-line spectroscopic binary
with an orbital period of 100\fd46 by \citet{abt65}.
A faint companion ($0.2''$), detected from occultations,
has been interpreted as a tertiary
\citep{Richichi_1996A&A...309..163R}.
\citet{torres2020} derived accurate RVs over about 800 days and concluded no evidence of RV variability.
Photometric studies by \citet{breger72}, \citet{mcnamara85}, and Hipparcos found 17~Tau to be constant down to the millimagnitude level.

\begin{figure}
\centering
\includegraphics[width=\tmpdim]{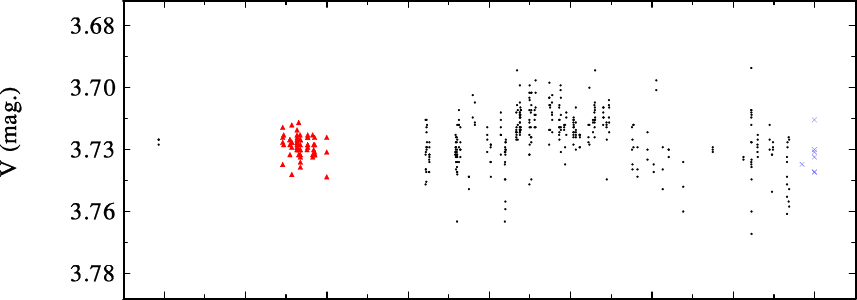}
\includegraphics[width=\tmpdim]{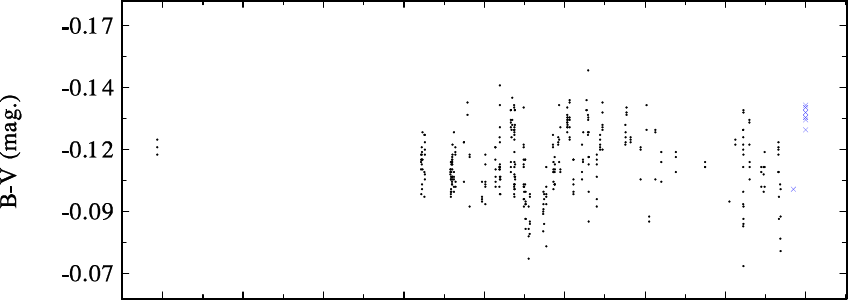}
\includegraphics[width=\tmpdim]{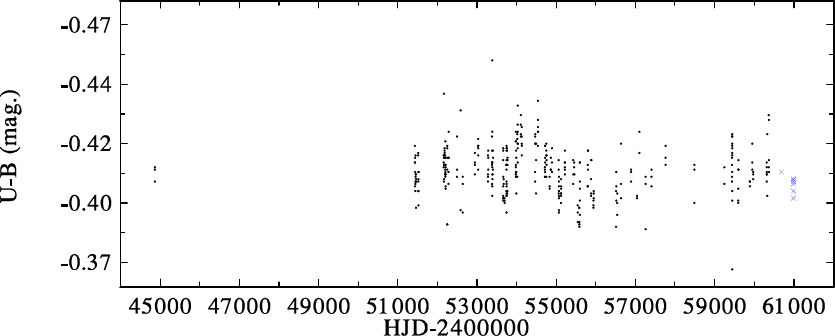}
\caption{\ubv\ time variations (LTC) of 17~Tau.
The long-term variation of $V$ is only mild;
more pronounced variability is in \bv\ and \ub\,
on intermediate time scales.
The disc evolution is thus mild,
probably accompanied by less regular disc oscillations.
}
\label{17tau}
\end{figure}

In contrast,
the Hvar photometry presented in Fig.~\ref{17tau} reveals
mild long-term variations in brightness and colours
at the level of a few hundredths of a~magnitude,
possibly with a cycle of ${\sim}9200\,{\rm d}$
present in all three filters.
Additional variability is seen on intermediate (LTC) time scales. Repeated TESS observations also reveal cyclic variability
with a period of 1\fd1 and a full amplitude of 0\m0006--0\m0008,
about 10 times below the Hvar sensitivity limit.
However, the offsets between individual TESS sectors are comparable to variability detected at Hvar.

\paragraph{BU~Tau = 28~Tau = HD~23862 = Pleione.}  
This Be star is one of only a few of its type whose brightness and colour variations have been recorded systematically over a long time, mainly due to the efforts of Sharov and Lyutyj
\citep[see][and references therein]{sharov72,sharov97}. 
It is the primary
component of a binary with a highly eccentric orbit and an orbital period
of 218\fd03 \citep{zarf27}. The orbit was not resolved
interferometrically by \citet{klement2024}.
The spectroscopic behaviour of this star is characterised by
remarkable changes between the normal Be state to Be-shell and Be
phase \citep{hirkog76, hirata95}. After a long
B phase that lasted until 1937, the star passed through 3
cycles of photometric and spectral variation.
The period of the complete cycle lasts 34-36 years \citep{hirkog76, hec82}.
Using speckle interferometry \citet{mcalis89} found a third companion at
a separation of $0.22''$. \citet{hec82} and \citet{Gies1990} (and some
others) considered a causal connection between the motion of that distant
companion and the occurrence of the shell phases. \citet{robert07}
recorded also the fourth companion of BU~Tau at a distance of
$4.66''$.
\citet{iliev2025} clearly demonstrated that large changes in the strength
of the \ha profile occur shortly after each periastron passage
in the 218\fd03 orbit.
According to \citet{Marr_2022ApJ...928..145M},
historical light and \ha observations can be interpreted
as precession and `tearing' of the outer disc.

\begin{figure}
\centering
\includegraphics[width=\tmpdim]{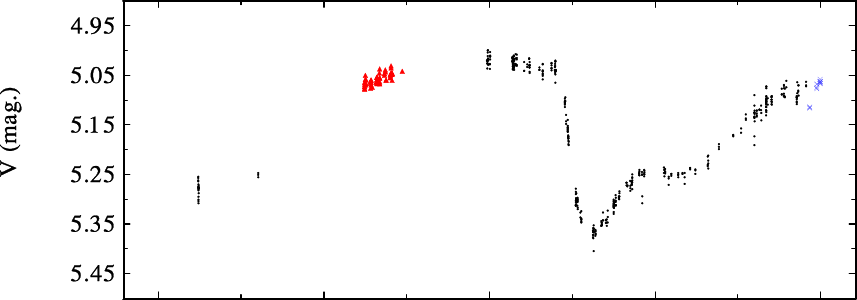}
\includegraphics[width=\tmpdim]{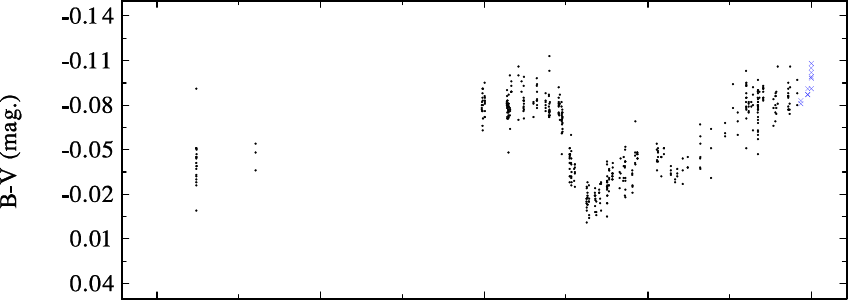}
\includegraphics[width=\tmpdim]{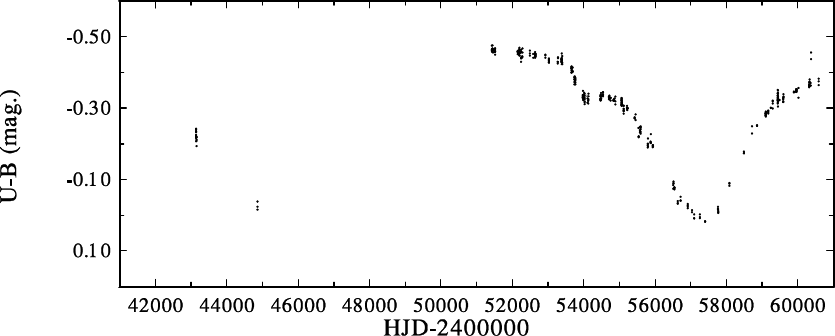}
\caption{
\ubv\ time variations (LTEi) of BU~Tau.
The light curve covers one nearly complete cycle
of disc growth and dissipation.
The brighter-when-bluer behaviour corresponds to
obscuration of the Be star by the disc during the shell phase,
indicating edge-on disc geometry.
At 51000, the star is in a bright Be phase,
unobscured by the disc.
Around 53000, a build-up of the disc begins,
causing a gradual fading by 0\m4 in $V$,
accompanied by reddening
and lasting about three years.
After the light minimum near 54000,
the system gradually brightens again towards its initial level.
This brightening
is interrupted by a mild secondary dip in $V$ between 56000--57000,
associated with the development of a~metallic shell
\citep{hec82}.
This phase corresponds to the strongest reddening in \ub.
The disc then continues to dissipate towards a new cycle.
}
\label{butau}
\end{figure}

Apart from a~few observations from the season 1976/77,
the Hvar data cover almost the entire last cycle of photometric changes.
BU Tau shows the inverse correlation between the brightness and
emission-line strength.
A secondary minimum is visible on the ascending part of the curve,
the depth of which is the most prominent at shorter wavelengths;
see Fig.~\ref{butau}.
A~detailed study of this object is under preparation.

\paragraph{V960~Tau = 120~Tau = HD~36576.}
The light variability of this star was discovered at Hvar
and was first reported by \citet{pavlovski82}.
\citet{Gies_1986ApJS...61..419G} studied a series of RVs and
concluded that its RV changes,
for which they found a possible period of 3\fd399,
is due to pulsations of photospheric variability
and questioned its runaway status.
Rapid periodic or multi-periodic light variations were reported by several authors
\citep[see, e.g][]{Bossi_1989IBVS.3348....1B,Percy_2002PASP..114..551P,bartz2025},
with periods near one day
(for instance 0\fd922, 0\fd975 or 1\fd037)
and harmonics.

\begin{figure}
\centering
\includegraphics[width=\tmpdim]{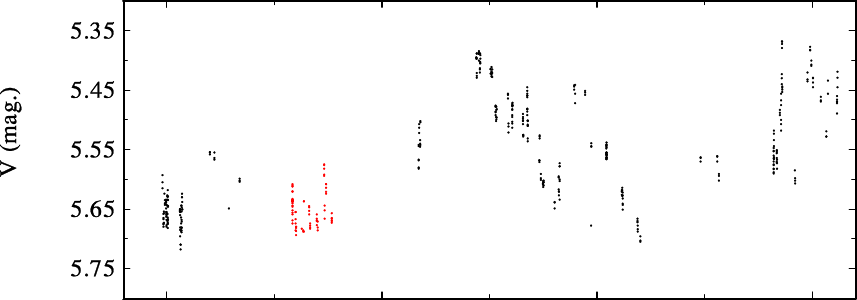}
\includegraphics[width=\tmpdim]{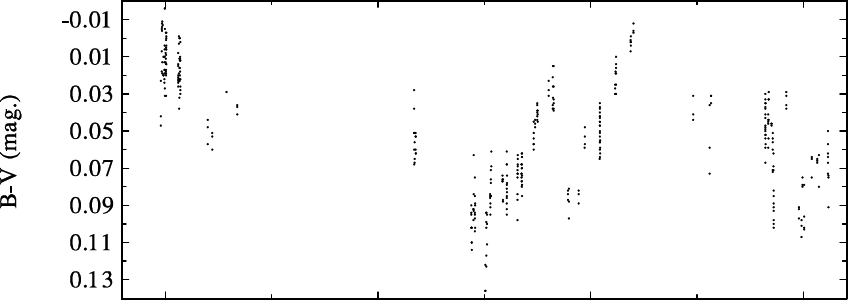}
\includegraphics[width=\tmpdim]{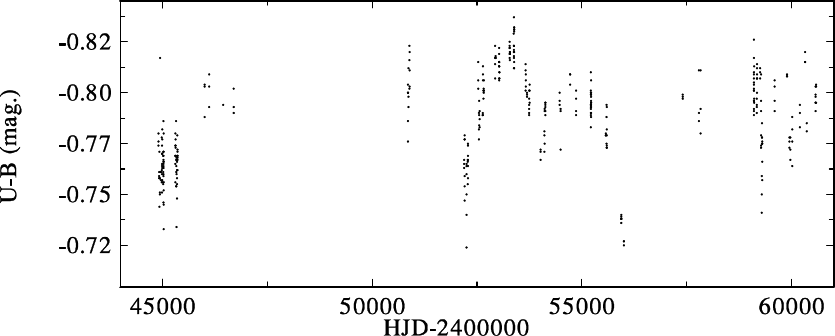}
\caption{\ubv\ time variations (LTEp, LTC) of V960~Tau.
The light curve shows mixed LTE and LTC variations
with an amplitude up to 0\m3 in $V$.
The brigther-when-redder behaviour indicates a pole-on geometry.
The disc growth episodes, which were well covered at Hvar,
started at approximately
52000, 54000, and 59000.
The growth is substantially faster than subsequent dissipation.
The minimum of brightness 5\m70 likely corresponds to
a quiescent Be star phase.
}
\label{v960tau}
\end{figure}

As shown in Fig.~\ref{v960tau}, 
the star was monitored quite systematically at Hvar.
Apart from ${\sim}1\,{\rm d}$ periods,
also seen in Hvar photometry,
there is a distinct long-periodic signal,
$P_2 = 2731\,{\rm d}$,
interpreted as the LTC variability.
A detailed study, which will also include spectroscopic observations, is under preparation.

\paragraph{$\zeta$~Tau = 123~Tau = HD~37302.}
This well-studied Be star
is a single-line spectroscopic binary with a 132\fd99 orbital period,
exhibiting correlated cyclic RV and $V/R$ variations
during some intervals
\citep[see][where also references to original
papers can be found]{delplace70,hec84,zarf26,stefl2009,carciofi2009}.
However, \citet{zarf26} have demonstrated that the light and colour behaviour
differed from cycle to cycle,
alternating between positive and inverse correlation.
Already \citet{bozic88} found
the RLA variations with a 0\fd8 period
(or a 1\fd6 period with a double-wave phase curve).
They also reported light decreases resembling atmospheric eclipses
during some orbital cycles, which were missing in the others
-- see also fig.~1 in \citet{bozic2013}.
Additionally, $\zeta$~Tau is a soft X-ray source,
containing a white dwarf companion
accreting material expelled from the central Be star
\citep{Naze_2024A&A...688A.181N,Toala_2025MNRAS.542..876T}.

\begin{figure}
\centering
\includegraphics[width=\tmpdim]{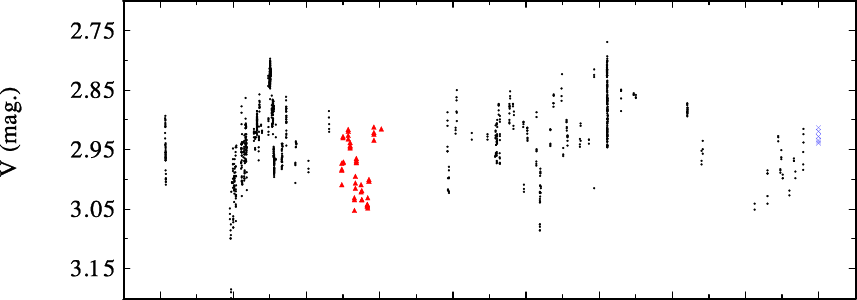}
\includegraphics[width=\tmpdim]{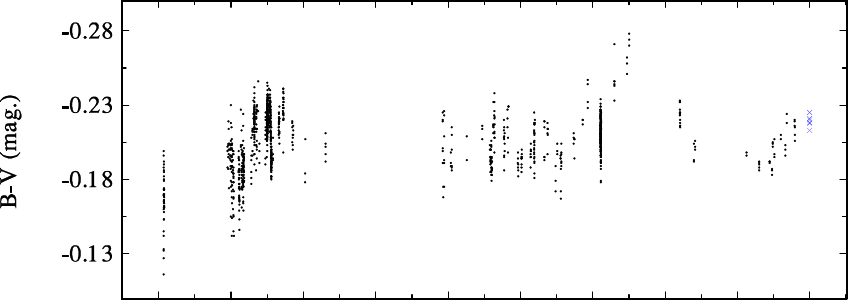}
\includegraphics[width=\tmpdim]{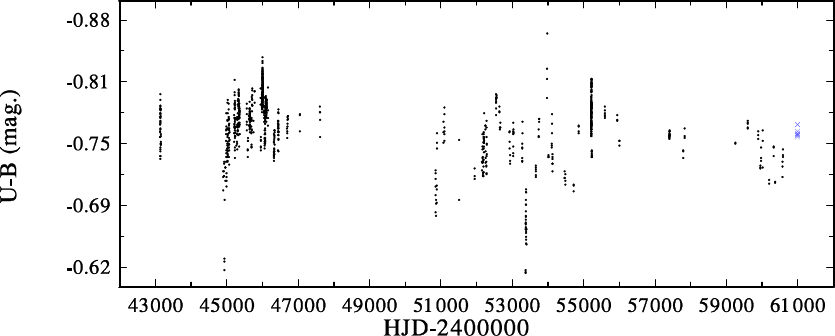}
\caption{\ubv\ time variations (LTEp+i, LTC, 132\fd99 BIN) of $\zeta$~Tau.
This unique light curve exhibits a combination of
several types of variability,
including both positive and inverse correlations.
The disc is thus seen at an intermediate inclination.
The variations on intermediate time scales
are rather related to global disc oscillations,
or to $m=2$ spiral arms due to companion,
modifying absorption (or emission)
when a dense region rotates in front of the star.
}
\label{zettau}
\end{figure}

The Hvar observations are shown in Fig.~\ref{zettau}.
Unfortunately, the interval with the inverse correlation
(from 48 100 to 48 800)
was not covered at Hvar; for that phase additional data from \citet{zarf26} should be used.
New, post-2009 observations allowed us to constrain longer photometric periods, particularly
$P_3 = 656\,{\rm d}$, $P_5 \simeq 74\,{\rm d}$. Their relation to the orbital period can be expressed as
$f_5 = 2f_{\rm orb} + f_3$,
likely due to phase modulation.
The remaining `scatter' reflects non-random RLA variability.
The periodogram from TESS is superior,
clearly showing multi-periodicity,
in particular, the periods
$4.133$,
$0.758$,
$1.119$,
$0.992$, or
$0.908\,{\rm d}$.
Since many are close to 1\,d,
which is the dominant instrumental frequency at Hvar,
we should keep it in mind when analysing the Hvar photometry.

\paragraph{V696 Mon = HR 2142 = HD 41335.}
This Be star was found to exhibit two consecutive short-lived shell phases
periodically every 80\fd85 days
\citep{peters71, peters72}.
Later, \citet{peters83} found that the object is a single-line spectroscopic binary
with the same orbital period.
Much later, \citet{peters2016} detected a weak signal in the IUE spectra
corresponding to a hot subdwarf companion,
with a mass ratio of 0.07 and \tef around 43000\,K.
The existence of shell lines close to conjunctions is still unexplained.

The photometric series published in the Hvar archive
shows a~very slow light decrease
(by $+1.3\,{\rm mmag}\,{\rm yr}^{-1}$)
on the LTQ timescale.
On the contrary, both colour indices are much more stable
(upper limit ${<}0.1\,{\rm mmag}\,{\rm yr}^{-1}$).
One could suspect also some RLA variability,
but this star is always observed at air masses larger than 1.55 at Hvar,
which implies somewhat higher scatter of individual observations.

\paragraph{$\varkappa$~Dra = 5~Dra = HD~109387.} 
This bright Be star
observed in an~unusually high northern declination, was found to be
a~single-line spectroscopic binary with a 61\fd55 orbital period and a~small
RV amplitude by \citet{zarf16}. This was confirmed by subsequent orbital
solutions by \citet{saad2021}. 
The secondary was detected from near-IR interferometry by \citet{klement2022b}. They found
a~mass ratio of $0.117\pm0.009$. 
\citet{juza94} investigated spectral, polarimetric and
photometric observations from the past 100 years, 
including their own,
and argued that all observables varied with a period of 8406 days.
\citet{zarf23} concluded that this variability (which they estimated
to $8044\pm167$~days) is probably cyclic, not strictly periodic.
\citet{balona2021} reported rotational modulation with a period of 1\fd134
from their analysis of TESS photometry.

\begin{figure}
\centering
\includegraphics[width=\tmpdim]{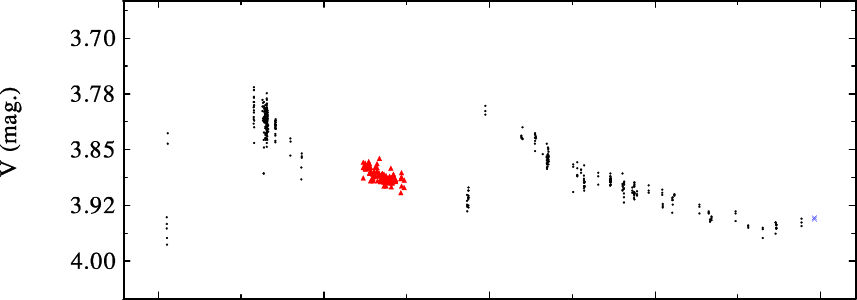}
\includegraphics[width=\tmpdim]{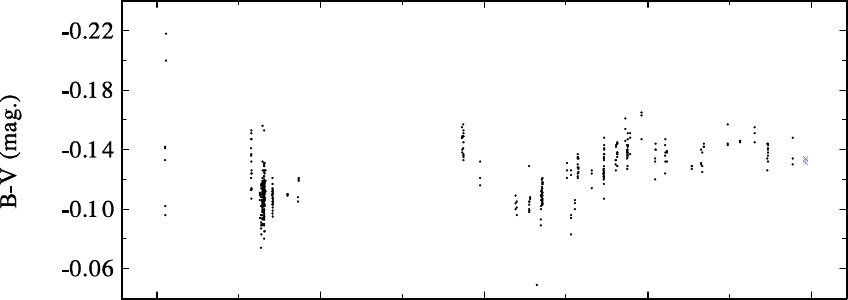}
\includegraphics[width=\tmpdim]{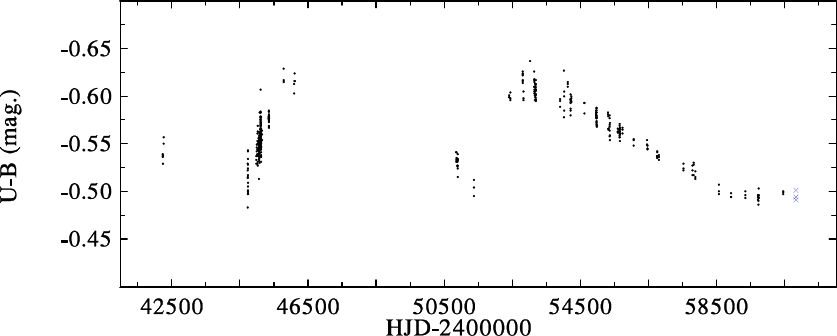}
\caption{\ubv\ time variations (LTEp, LTQdec) of $\kappa$~Dra.
The light curve shows two cycles of disc growth and dissipation.
The first cycle (44500) is relatively shorter
compared the second one (51500),
which is better covered though.
Interestingly, the time scales of dissipation are also different
(the first one being shorter).
The LTQ variability indicates the disc probably did not dissipate
completely during the first cycle,
but instead it was replenished by an onset of the second cycle.}
\label{kapdra}
\end{figure}

This is definitely confirmed by the extended series of Hvar
photometry, shown in Fig.~\ref{kapdra}, which shows a long systematic
brightness decrease since about JD~2451000 up to the present time.
This nicely correlates with the disappearance of the Balmer emission
documented over a similar time interval by \citet{klement2022b}.
A slow secular decrease on the LTQ scale seems to be present.

\paragraph{V839~Her = 4~Her = HD~142926.}
This Be star belongs to a few
Be stars systematically monitored at Hvar since 1972.
It was identified as a~single-line spectroscopic binary with a 46\fd2 period
\citep{zarf3}.
\citet{bozic2013} reported very mild light variations with the orbital period.
So far, the only large light decrease, associated with
the formation of a new shell phase, was discovered by \citet{percy97} and is
also recorded in the \hp\ photometry.
\citet{sigut2023} estimated the inclination angles of a number of Be stars,
based on the gravitational darkening and on the \ha emission line profile modelling.
For V839~Her, they found it is seen almost equator-on.

\begin{figure}
\centering
\includegraphics[width=\tmpdim]{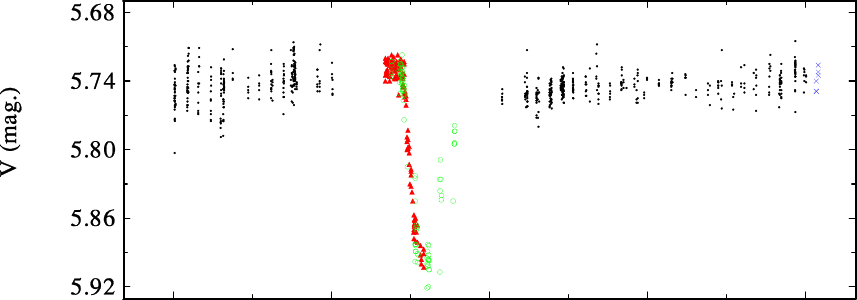}
\includegraphics[width=\tmpdim]{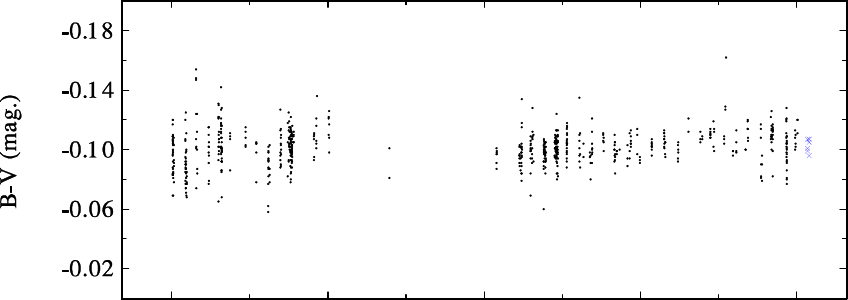}
\includegraphics[width=\tmpdim]{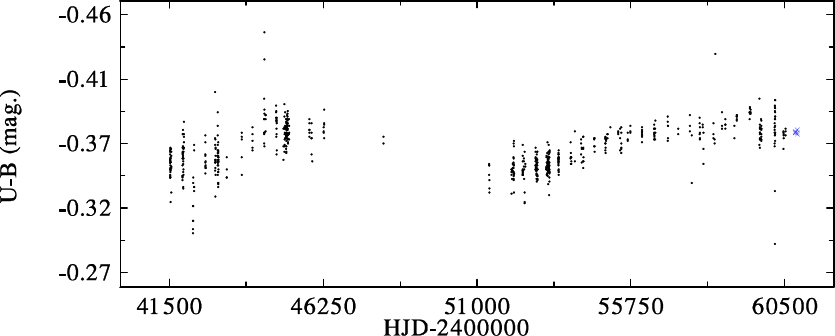}
\caption{\ubv\ time variations (LTEi, 46\D19 BIN) of V839~Her. 
The green circles denote 
observations published by \citet{percy97}.
The system was relatively stable,
but showed a strong (0.2\,mag) dimming,
centred around 50000. 
The disc growth and dissipation episode,
associated with the dimming,
lasted only several hundreds days
\citep{percy97}.
According to \ub\ and \bv\ indices,
an inverse correlation is noticeable,
confirming the edge-on viewing geometry
noted by \citet{sigut2023}.
}
\label{v839her}
\end{figure}

The time plot of all Hvar and transformed \hp\ photometric observations is shown
in Fig.~\ref{v839her}.
Since this is a very interesting case,
we made an exception and also included the observations by \citet{percy97}
in our time plot (shown by green circles).
This confirms V839~Her is an~object with the inverse correlation
between the light and the emission-line strength.

\paragraph{V744~Her = 88~Her = HD~162732.}
This Be star has also been systematically monitored at Hvar since 1972.
It was found to be a single-lined spectroscopic binary
with a period of 86\fd7
\citep{zarf1}.
The secondary star is probably a~hot, small object,
but it has never been directly observed
\citep[see, e.g.][]{wang2018}.
This Be star is a very good example of
the inverse correlation
between brightness and emission-line strength.

\begin{figure}
\centering
\includegraphics[width=\tmpdim]{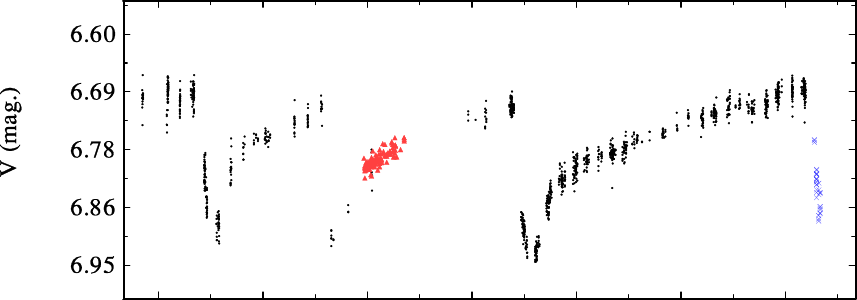}
\includegraphics[width=\tmpdim]{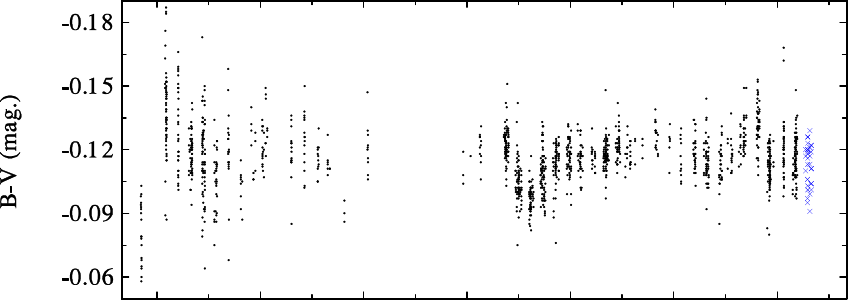}
\includegraphics[width=\tmpdim]{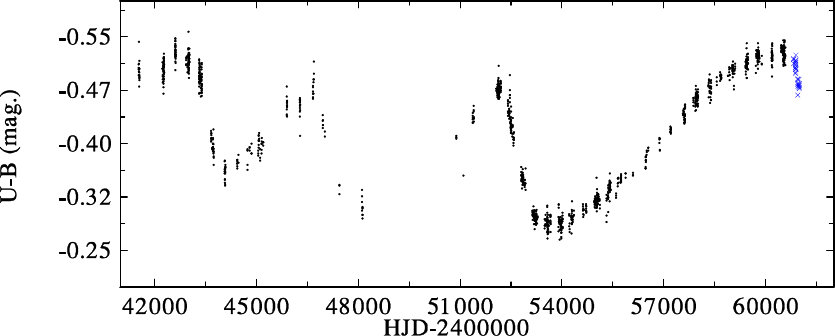}
\vskip-.2cm
\includegraphics[width=\tmpdim]{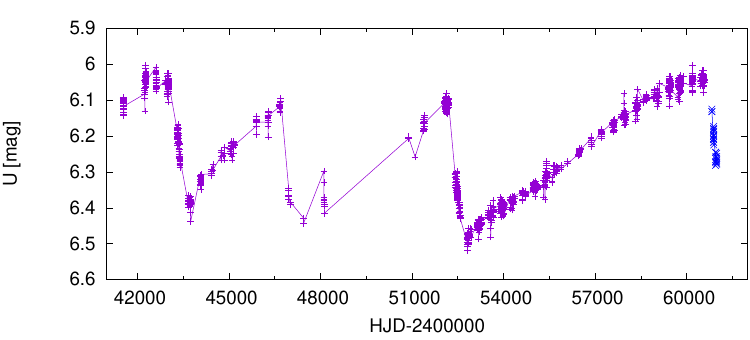}
\caption{\ubv\ time variations (LTEi, LTQdec) 
of V744~Her.
Since the system is brighter-when-bluer,
the disc viewing geometry is edge-on.
The three cycles correspond to grow and dissipation,
but the disc did not dissipate completely
and the system did not return to the quiescent (bright) state.
The shell phase characterized by the reddest \ub\ is delayed
(52500 vs 54000).
According to the $U$ band (not \ub),
the brightening between 52500 and 60000 is almost linear light curve.
Since $U$ is closely related to the Balmer jump,
i.e. the depth of hydrogen absorption lines, 
the line-of-sight integrated density of the disc should also evolve linearly.
}
\label{v744her}
\end{figure}

In Figure~\ref{v744her}, we show its light and colour changes
over 53~years of Hvar observations.
It is seen that the star underwent three episodes of secular light decreases,
accompanied by the emission and metallic shell episodes,
the first one discussed in detail by \citet{zarf8}.
The 2025 observations and recent BeSS \ha observations
demonstrate the start of the fourth episode.
There is a slow secular decline of the three light minima,
occurring on the LTQ timescale,
although the four light maxima are different from cycle to cycle.
There are almost no secular colour changes in \bv,
but very pronounced ones in \ub.

\begin{figure}
\centering
\includegraphics[width=9cm]{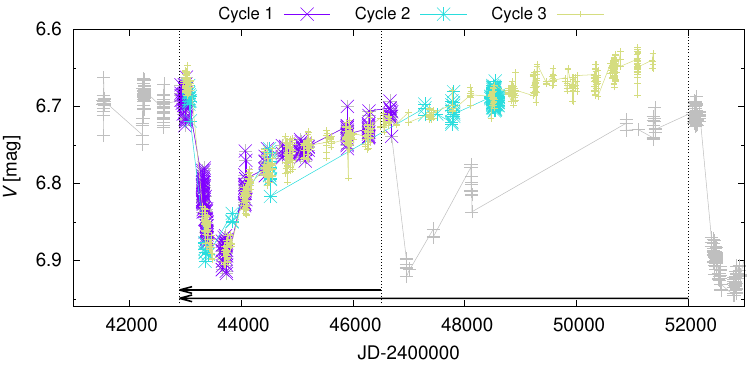}
\caption{
Three cycles in the light curve of V744~Her
have very similar shape and slope.
Cycles 2, 3 were shifted by 3600 and 9100\,d,
in order to match the beginning of Cycle~1.
}
\label{v744her_cycles}
\end{figure}

The individual cycle durations
(3700, 5400, 8400\,d, respectively)
become subsequently longer.
Their amplitude, especially in \ub, is correlated with the duration.
However, the shape and slope of light and colour curves remain very similar
(Fig.~\ref{v744her_cycles}),
suggesting very similar ionisation and/or viscosity state of the envelope --
first for outflow,
second for inflow.
The cycle-to-cycle variability should be related to something else.
It is unlikely to be just envelope mass,
because the observed variability is not random.
It is more likely the
timing of ejection from the stellar surface
driven by non-random oscillations,
or `beating' of three close, fast frequencies.

\paragraph{CX~Dra = HD~174237.}
This object is known as a B2e+F5III semi-detached binary
with a 6\fd696 period seen under an intermediate inclination of the orbit
-- see the RV studies by
\citet{cxdrakou,zarf9,cxdrahorn,cxdramr}, and references therein.
The light and colour variations show cyclic changes 
correlated with the variations of the strength of the \ha emission
on a timescale of several hundred days,
showing a positive correlation
\citep{zarf10,cxdrahvar}.
The lower envelope of low-amplitude light changes
is reminiscent of ellipticity and reflection in the undisturbed state,
but frequent brightenings are seen.

\begin{figure}
\centering
\includegraphics[width=\tmpdim]{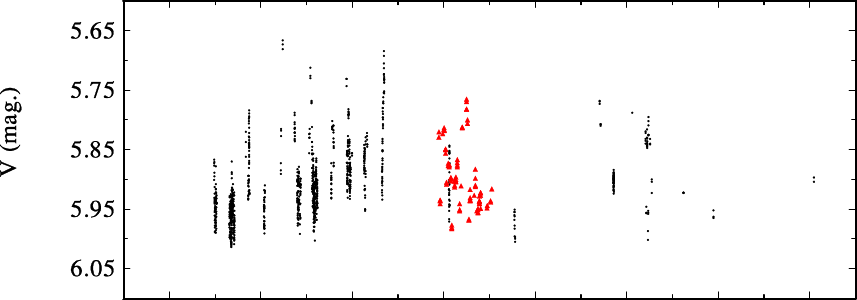}
\includegraphics[width=\tmpdim]{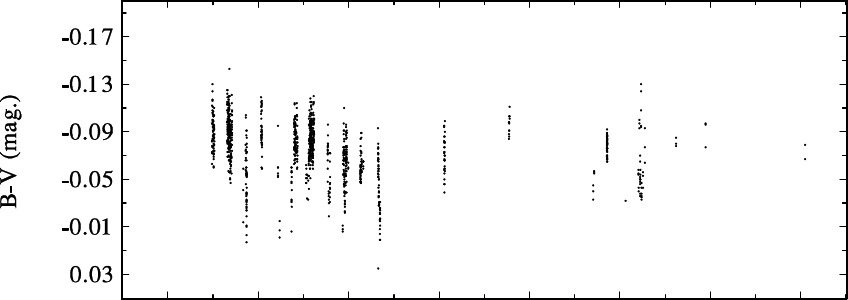}
\includegraphics[width=\tmpdim]{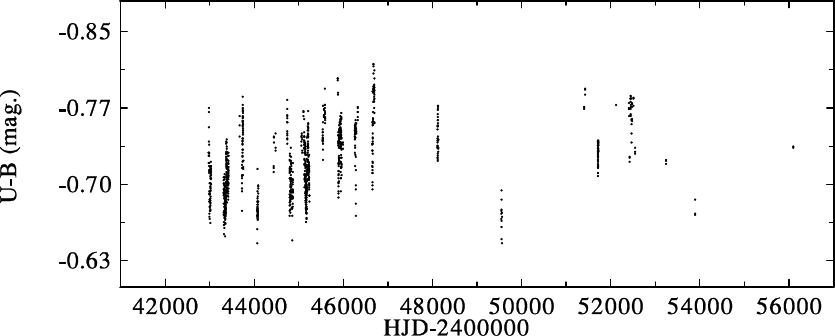}
\caption{\ubv\ time variations (LTEp, 163\D0 BIN) of CX~Dra.
The light curve is dominated by intermittent brightenings.
They are not related to the short orbital period
($6\D696$),
but rather to $P_3 = 163\D0$.
The disc also grows (and dissipates) over long, decadal time scales.
The viewing geometry is rather pole-on,
according to the brighter-when-redder behaviour.
}
\label{cxdra}
\end{figure}

We show the complete light and colour changes recorded 
at Hvar
in Fig.~\ref{cxdra}.
New, post-1998 data are sparse, but they help to constrain periods.
The observed, non-instrumental photometric periods are
$P_3 = 163\D0$,
$P_4 = 6\D696$,
where the latter corresponds to the BIN variability.
Typical duration of the brightenings ($V < 5\m82$)
estimated from well-covered events,
is 5 or 6 orbital periods.
This suggests that the circumstellar material dissipates,
but not within one orbital period,
which is among the shortest for Be stars
(Table~\ref{tab03}).

\paragraph{V923~Aql = HD~183656.}
This Be star is a primary component of a 214\fd7 binary, see
\citet{zarf13}. A detailed study based on decades of
spectral and photometric data was
published by \citet{wolf2021}. They documented its long-term,
orbital, and rapid variations. The object is a typical example of the
inverse correlation between the brightness and emission-line strength.
Although it exhibits large-amplitude cyclic RV and $V/R$ variations
with cycle lengths between about 1 800 and 3 000 days in RV, $V/R$,
$V$ magnitude, and \ub\ index;
its \bv\ index remains secularly stable and close to zero.
Rapid changes, if periodic, may follow a 0\fd8442 period.

\begin{figure}
\centering
\includegraphics[width=9cm]{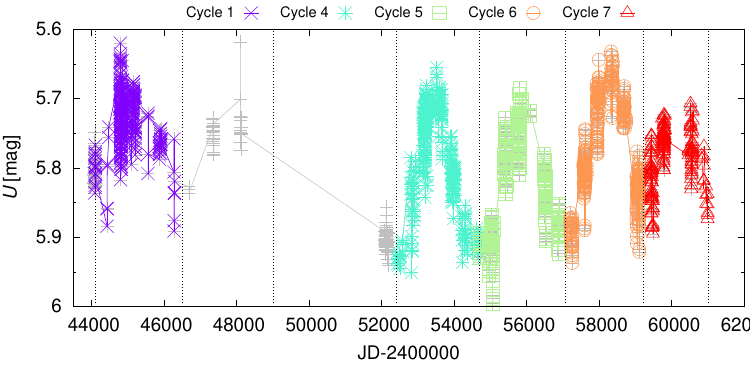}
\includegraphics[width=9cm]{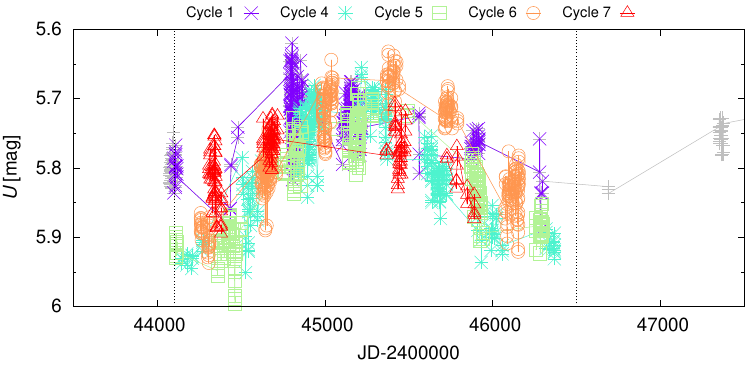}
\caption{
Five cycles in the light curve of V923~Aql
have different minima,
but similar maxima.
Their duration (minimum-to-minimum) is approximately 2400\,d.
Cycles 4, 5, 6, 7 were shifted
in order to match the beginning of Cycle~1.
}
\label{v923aql_cycles}
\end{figure}

In \citet[][fig. 2]{wolf2021}, we showed the extended time plot of
Hvar \ubv\ photometry.
It shows the large amplitude
(${\sim}0\m30$) in $U$,
compared to other Be stars,
while the constancy of $B$ and $V$,
even on the longest LTQ timescale.
It might suggest a different mechanism,
connected to a systematic shift of the Balmer jump,
when the high-order spectral lines of hydrogen
become narrower/broader,
which is reminiscent of
\citet{Divan_1982IAUS...98...53D,Moujtahid_1999A&A...349..151M}.
Five of seven cycles are sufficiently covered;
they show different minima, but similar maxima,
and their duration (minimum-to-minimum) is
approximately 2400\,{\rm d}
(Fig.~\ref{v923aql_cycles}).

\paragraph{V1294~Aql = HD~184279.}
This is one of the Be stars with the largest recorded range
of light variations. Its amazing and complicated spectral,
light, and colour variations were recently described
in the study published by \citet{hec2022},
who discovered that the object is
a spectroscopic binary with a 192\fd9 period.

In \citet[][fig. 2]{hec2022}, we showed an extended series of
photometric observations from Hvar.
It is seen that the brightness of the object had still been rising,
and reached the maximum brightness
($V = 6\m8$)
throughout the recorded history.
This maximum expressed in terms of the absolute magnitude
($M_V \simeq -4\m5$)
is actually close to the maximum over of all 
classical Be stars
(cf. Fig.~\ref{Fig11}).

\begin{figure}
\centering
\includegraphics[width=9cm]{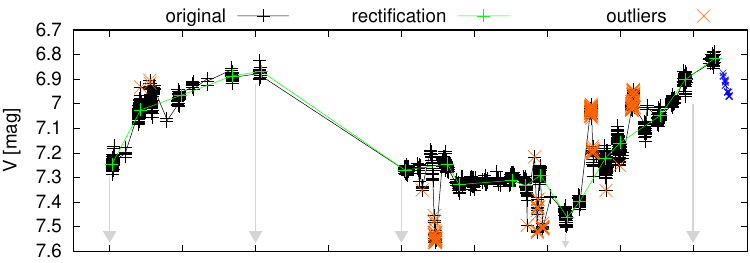}
\vskip-.1cm
\includegraphics[width=9cm]{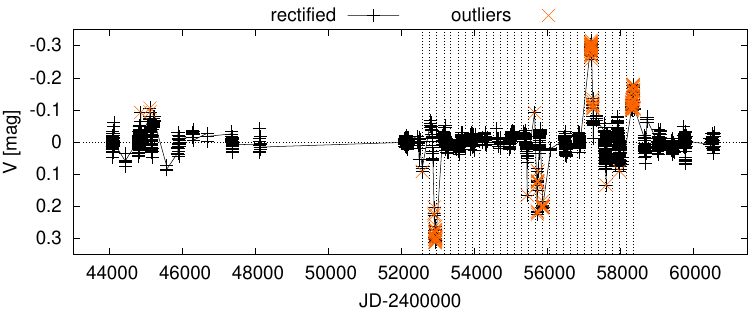}
\caption{
Rectified light curve of V1294~Aql,
without long-term trends.
Points (black) which were less than 0\m09 from zero
were analysed separately from outliers
(\textcolor{orange}{orange}).
The vertical dotted grid is plotted at zero phases of the
orbital period $192\D9$
\citep{hec2022}
corresponding to the brightenings or fadings
occurring at integer multiples of this period.
}
\label{v1294aql_rectification}
\end{figure}

However, new 2025 observations indicate a start of
another light decline.
The cycle duration turned out to be 31.5\,yr.
While onsets of the brightness were different,
both declines were similar.
This object has a `record' amplitude (0\m85) in \ub\ index
out of all Be stars.
The amplitude is much less ($\sim$0\m1) in \bv\ index,
suggesting again a connection to the Balmer jump.
On the other hand, it is not zero,
implying some projection effects,
or circumstellar matter, which sometimes hides the central star.
Apart from long-term trends and some outliers,
the light curve shows photometric periods,
$P_2 = 97\D052$,
$P_3 = 100\D251$,
where the former is very close to the half of the orbital period.
It would correspond to the LTC variability
driven by two spiral arms ($m=2$).
On the other hand, if outliers were included
(see Fig.~\ref{v1294aql_rectification}),
different periods become prominent,
$P_1 \simeq 574\,{\rm d}$,
$P_3 \simeq 373\,{\rm d}$,
which are {\em multiples\/} of the orbital period.
This would indicate a non-equilibrium variability,
occurring only once per two, three, or more orbits.

\paragraph{V832~Cyg = 59~Cyg = HD~200120.}
This object is the brightest
component of the multiple visual system ADS~14526,
with rather distant fainter components B, C, D, and E.
A close visual companion was discovered by \citet{mcalister84}, 
confirming earlier suggestions that the spectral variability and shell phases could be related to binarity \citep{hec82}.
\citet{zarf21} carried out a detailed study
of this Be star, based on spectral and photometric observations from
several observatories. They concluded that V832~Cyg is a spectroscopic
binary with a 28\fd1971 period. 
They also documented long-term light and colour changes
with a positive correlation between
the light and emission-line strength combined
with changes probably attributable to one-arm disc oscillation.
In prewhitening photometry for these long-term changes,
they found mild, but well-defined sinusoidal light changes
phase-locked with the orbital period.
\citet{peters2013} derived the RV curves of both binary components
from the far-UV spectra and proved that the secondary is a hot O-type subdwarf.

\begin{figure}
\centering
\includegraphics[width=\tmpdim]{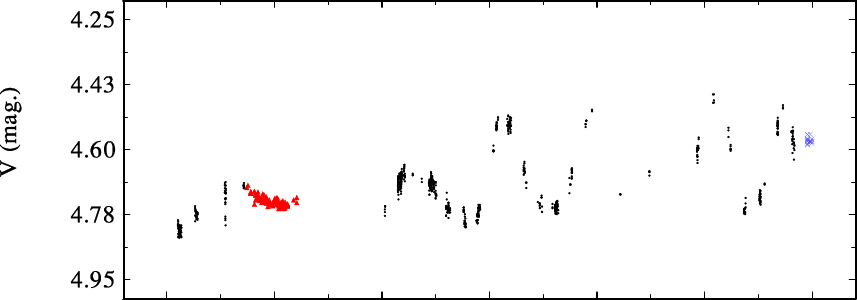}
\includegraphics[width=\tmpdim]{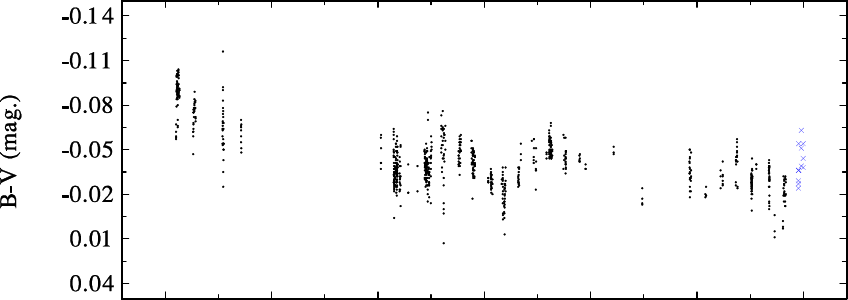}
\includegraphics[width=\tmpdim]{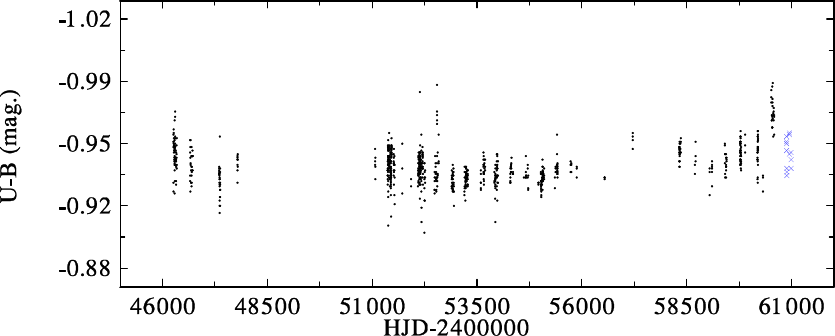}
\caption{\ubv\ time variations (LTEp, LTC, 28\D1971 BIN, LTQinc) of V832~Cyg.
The disc is viewed pole-on,
as the system becomes brighter-when-redder
when the disc grows.
The variability in $V$ is quasi-cyclic but asymmetric,
hence the disc growth is faster than its dissipation. 
The long-term modulation on the LTQ timescale also contributes
to the overall brightening,
indicating that the disc tends to grow on average.
}
\label{v832cyg}
\end{figure}

The long-term brightness and colour variations of V832~Cyg
extended to the present time are shown in Fig.~\ref{v832cyg}.
One can see a steady increase in brightness throughout the time interval
covered by the \ubv\ observations,
attributable to the variability on the LTQ timescale.
The long photometric period is approximately
$P_1 \simeq 2200\,{\rm d}$,
but it is modulated.
The \ub\ index shows less variability
compared to other Be stars.
In this context,
V832~Cyg seems to be a `mirror' of V923~Aql
(cf. their $V$ vs $U$ plots).

\begin{figure}
\includegraphics[width=9cm]{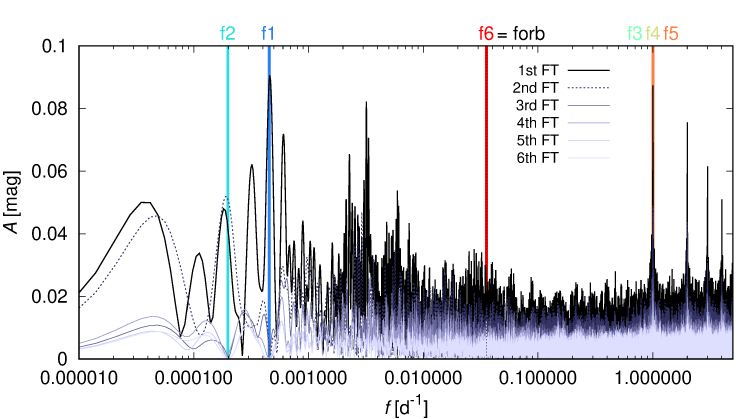}
\caption{
Periodogram of V832~Cyg,
computed with Period04
\citep{Lenz_2005CoAst.146...53L}
using six Fourier transforms (FT),
with sequential fitting of frequencies, amplitudes and phases,
and subtracting the respective model,
$\sum_i A_i \sin[2\pi(f_i t + \phi_i)]$.
The errors of frequencies are of the order of $1/\Delta$,
where $\Delta$ is the time span;
for example, the lowest peak width is $0.000050\,{\rm d}^{-1}$.
The corresponding periods $P_i$ were
$2202$,
$5054$,
$0.9966$,
$0.9952$,
$0.9986$,
and
$28.193\,{\rm d}$,
respectively.
$P_1$~corresponds to the LTC variations,
$P_6$~to the orbital period of the Aa1+Aa2 system.
Other non-instrumental periods
probably reflect the amplitude modulation of the primary signal,
not any additional orbital motion.
}
\label{v832cyg_fou}
\end{figure}

\paragraph{EW~Lac = HD~217050.}
This is a very active Be star, worthy of a detailed and complex study.
It has been studied for about 100 years. Its Balmer emission
was variable until 1926. Then a~strong emission with sharp shell lines
visible over the whole Balmer series has developed and remained stable
until 1977, when the star became active again \citep[see, e.g.][]{hec79}.
These authors also showed that the star is a classic example of the positive
correlation between the brightness and emission-line strength.
\citet{stagg88} suggested
a period of 0\fd72, but the light curve morphology changed with time. The spectra from the BeSS database
\citep{neiner2011} show that the emission gradually weakened and is virtually missing
since 2022.
It is also seen that
rapid, low-amplitude variations have declined
as the star is gradually loses its Balmer emission (disc).

\begin{figure}
\centering
\includegraphics[width=\tmpdim]{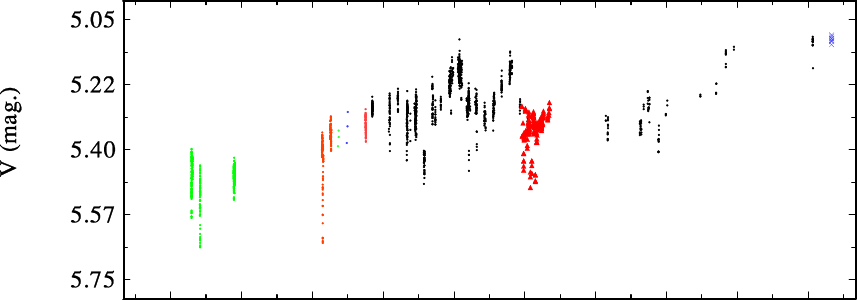}
\includegraphics[width=\tmpdim]{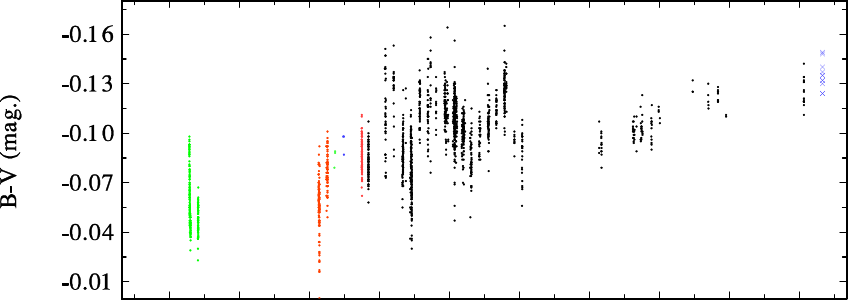}
\includegraphics[width=\tmpdim]{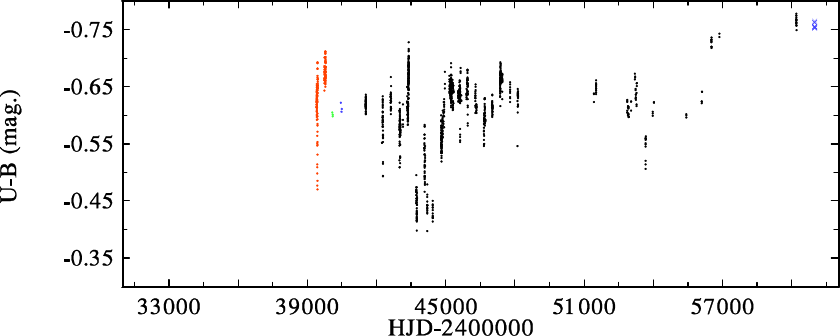}
\caption{\ubv\ time variations (LTEp, LTQinc) of EW~Lac.
The system is dominated by a long-term brightening trend,
by 0.45\,mag over 70\,yr.
The $V$ band is correlated with \bv\ and \ub\ indices.
The disc geometry should be pole-on according to the Balmer emission
\citep{hec79}.
Recent Hvar observations might indicate that the maximum brightness
has been reached,
which would imply that the disc has dissipated.
This might be closely related to the decreasing RLA variability;
if intermittent turbulence in the core does not excite
low-$m$ modes in the envelope
\citep{Neiner_2020A&A...644A...9N}
no gas outflow is possible.
}
\label{ewlac}
\end{figure}

The complete Hvar photometry shown in Fig.~\ref{ewlac},
complemented by measurements from \citet{1958AJ.....63..237W,
1953ApJ...118..481W,1974A&AS...15..311H},
indicates that the star is
an example of variability on the LTC timescale, 
with a secular brightness increase.
The long photometric period is approximately
$P_1 \simeq 2500\,{\rm d}$
but it is amplitude-modulated
by other types of variability.
Given its similarity to other Be stars
(e.g. $\omicron$~And),
this is most likely a~period of a~companion,
although none has been confirmed.

\paragraph{$\omicron$~And = HD~217675 = HR~8762.}
The rapid photometric variability of this remarkable Be star was discovered during World War I,
as revealed by early photoelectric observations reported by \citet{gutpra18}.
Beyond its individual variability, the object belongs to a complex multiple stellar system.
It serves as the brightest component of a quadruple system arranged in a 2+2 configuration
\citep[see][and references therein]{hec87,grant88,zhuchkov2010,mitro2021}.
\vskip\baselineskip

\noindent
\begin{tabular}{@{}llll@{}}
AB    & $P=118$\,yr  & $e=0.34$    & $\omega=146^\circ$ \\
Aa-Ab & $P\sim6$\,yr & $e\sim0.15$ & $\omega\sim 55^\circ$ \\
Ba-Bb & $P=33$\,d    & $e=0.24$    & $\omega=228^\circ$ \\
\end{tabular}
\vskip\baselineskip

\noindent
The orbit of the Aa-Ab pair is very uncertain, as it is based on only five interferometric observations
\citep{Mason_2001AJ....122.3466M}.
The orbital period is sensitive to $180^\circ$ ambiguities
in individual observations;
\citet{zhuchkov2010}'s solution is 5.6\,yr.
The Ba-Bb pair is observed as a 33-d spectroscopic binary. 
Since \citet{guth41} reported a 1\fd5765 period of light variations,
the star was long considered to be a contact binary
\citep[see][for the early history of the search for
the true period]{hec83}.

\begin{figure}
\centering
\includegraphics[width=\tmpdim]{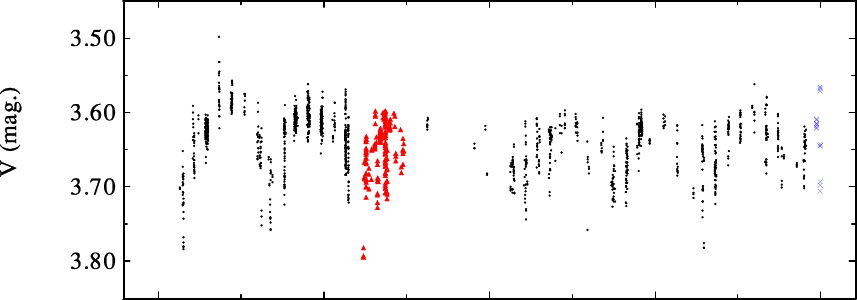}
\includegraphics[width=\tmpdim]{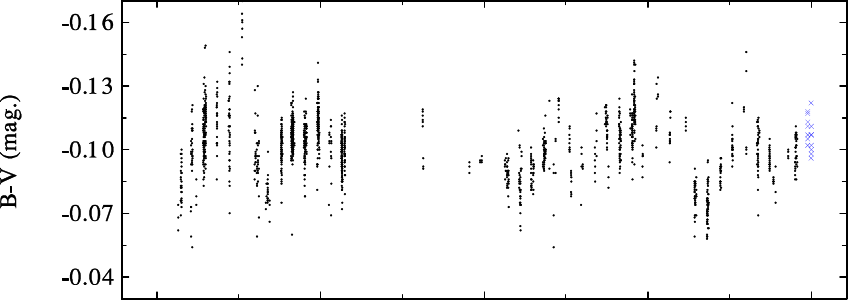}
\includegraphics[width=\tmpdim]{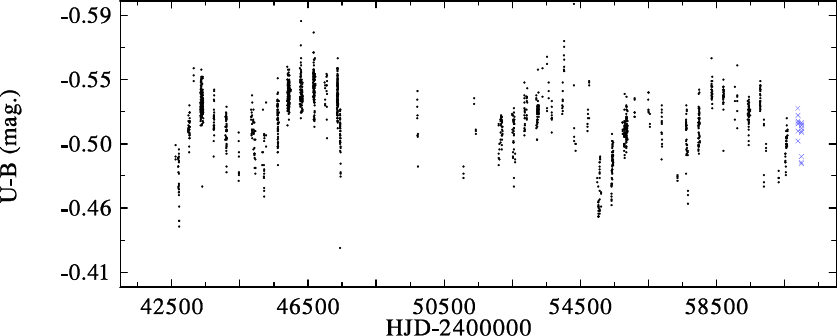}
\caption{\ubv\ time variations (LTC) of $\omicron$~And.
The variability is regular, 
on a timescale of 
approximately 6.8~years.
The disc is therefore possibly perturbed by a companion,
creating an inner spiral arm ($m = 1$)
and overdensity periodically moving around the central star.
The disc inclination should be rather pole-on,
according to the interferometric orbit
(Fig.~\ref{omiand_orbit}).
}
\label{omiand}
\end{figure}

\begin{figure}
\centering
\includegraphics[width=\tmpdim]{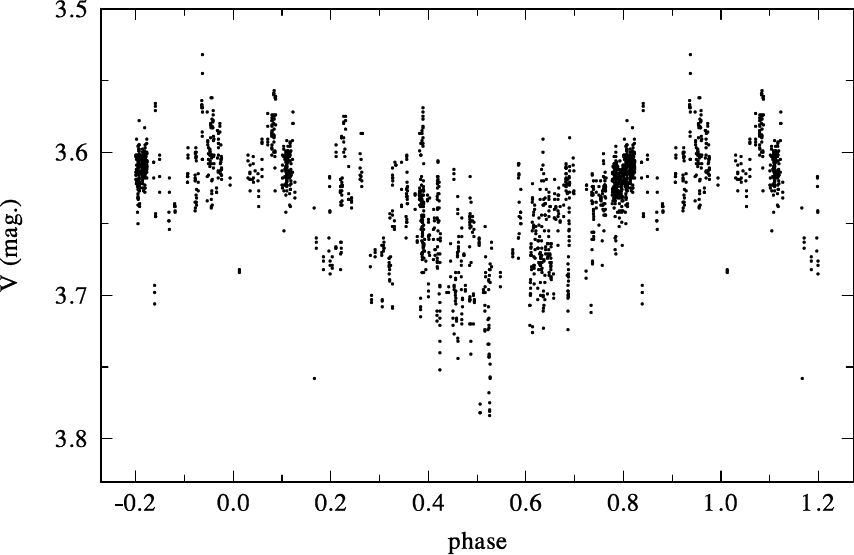}
\caption{
Phase plot of $\omicron$~And $V$ magnitude observations
for the best-fit sinusoidal period of $2498\fd3\pm4\fd1$
(see orbital solution in Fig.~\ref{omiand_orbit})
and epoch of the maximum brightness of JD~$2453894\pm12$,
which could correspond to the orbital period of the Aa+Ab system.
The semi-amplitude of the change is $0\m037\pm0\m002$.
}
\label{oand2498}
\end{figure}

\begin{figure}
\centering
\includegraphics[height=6.5cm]{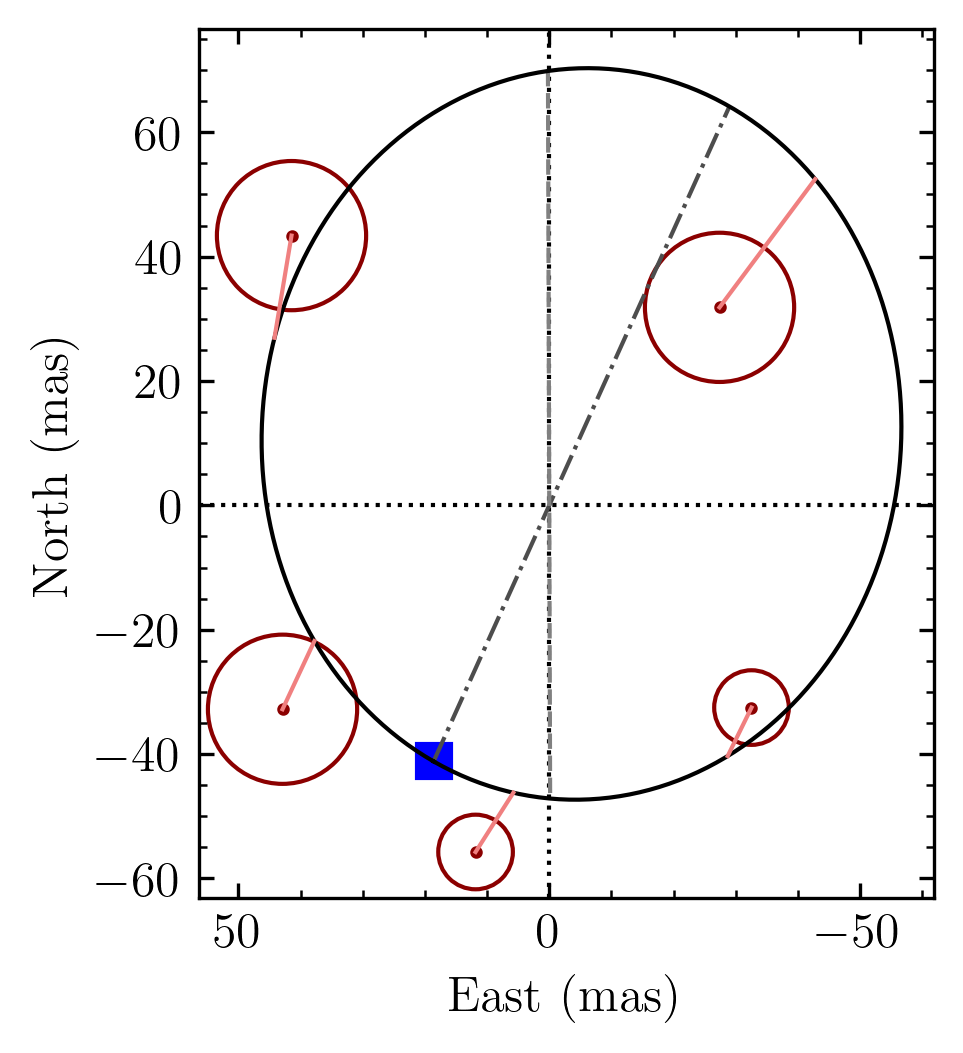}
\caption{
Orbital solution of $\omicron$~And fitting the interferometric data
\citep{Mason_2001AJ....122.3466M},
in agreement with the period derived from photometry
(see Fig.~\ref{oand2498}).
The third datum from the WDS catalogue was flipped by $180^\circ$.
It is one of the admissible solutions sampled by the MCMC method.
The corresponding orbital elements are as follows:
$P = (2491\pm 50)\,{\rm d}$,
$a = (65.1\pm 1.3)\,{\rm mas}$,
$e = 0.2^{+0.1}_{-0.2}$,
$i = (150\pm 30)\,^\circ$,
$\omega = (205\pm 30)\,^\circ$,
$\Omega = (0\pm 30)\,^\circ$, and
$T_0 = (2443000\pm150)\,{\rm d}$,
Assuming the Gaia distance of $107$\,pc,
the corresponding total mass is ${\sim}5$\,\Mnom.
The resulting $\chi^2 = 10.0$ is higher than
the number of degrees of freedom,
$\nu \equiv N-M = 4$,
which is worse than for \citet{zhuchkov2010}'s solution 5.4\,yr.
We note that \citet{olevic99}'s solution 8.9\,yr
is excluded due to the fifth datum and excessive mass.
}
\label{omiand_orbit}
\end{figure}

The Hvar photometry plotted in Fig.~\ref{omiand},
a large fraction of which is shown here for the first time, 
shows obvious cyclic changes with a possible period of about 6.8\,yr.
We calculated a~sinusoidal fit for the $V$ magnitude data,
which yielded a dominant period of $2498\fd3\pm4\fd1$;
the corresponding phase plot is shown in
Fig.~\ref{oand2498}. 
It is conceivable that this period corresponds to the true orbital period of Aa+Ab.
One of us, Jana \v{S}vr\v{c}kov\'a,
thus checked other orbital solutions,
accounting for $180^\circ$ ambiguities.
A solution with the 6.8\,yr period was found
(Fig.~\ref{omiand_orbit}).
According to this solution,
the periastron passage at JD~($2443000\pm150$)\,d (August 1976)
occurs shortly before the maximum brightness.
However, its statistical significance is worse than that of
\citet{zhuchkov2010}'s.
New high-quality interferometric observations of this
closer pair are very desirable.

\paragraph{KX~And = HD~218393.}
The light variability of this object was discovered
from Hvar photometry \citep{kxand01}. A large spectroscopic and photometric study,
based also on Hvar photometry, was published by \citet{zarf14}.
These authors found periodic RV, light and colour variations with
a 38\fd919 period and suggested tentatively that the object
could be an interacting binary. \citet{floquet95} and \citet{tolja98}
obtained a clear circular-orbit period based on the sharp lines of G8II
secondary component, with a~semi-amplitude of 86\,\ks,
proving thus definitively the binary nature of the object.

\begin{figure}
\centering
\includegraphics[width=\tmpdim]{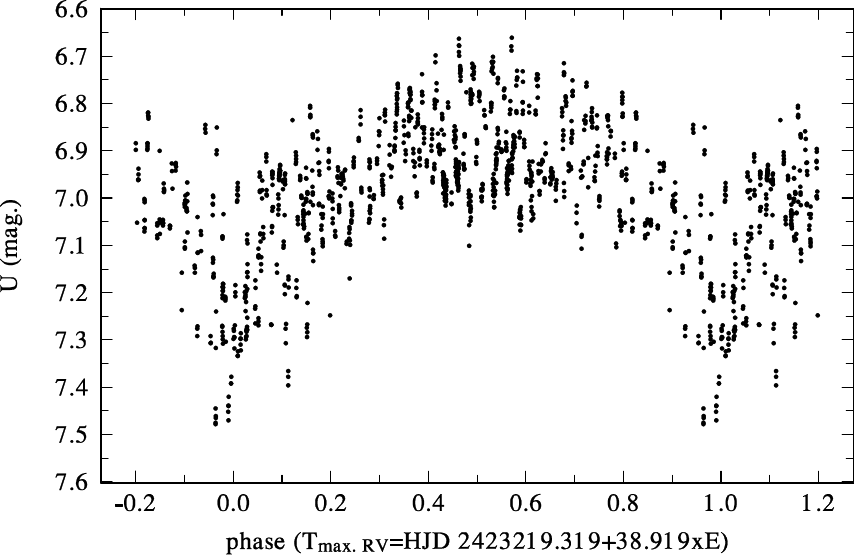}
\caption{
Phase plot of KX~And $U$ magnitude observations
for the period determined by \citet{zarf14}.
}
\label{kxu}
\end{figure}

A complete Hvar photometry is shown in Fig.~\ref{kxu}.
Observations after JD 2447000 are published here for the first time.
No secular changes are seen.
The object is notable for the fact that the inverse correlation
in the \ub\ versus \bv\ diagram
occurs on a timescale of the 38\fd9 binary orbit,
indicative of the inner spiral arm ($m=1$),
which modulates the light changes.
Moreover, the exciting discovery of extended bipolar jets by
\citet{rnaas2024}
makes KX~And similar to another strongly interacting binary $\beta$~Lyr
\citep{broz2021}.
It is conceivable that also for KX~And,
the bulk of the observed \ha emission originates in the jets,
not in the accretion disc around the primary.
This should be investigated by future modelling.

\end{appendix}

\end{document}